# Near room-temperature intrinsic exchange bias in an Fe intercalated $ZrSe_2$ spin glass

Zhizhi Kong[1], Corey J. Kaminsky[2], Catherine K. Groschner[1], Ryan A. Murphy[1], Yun Yu[1], Samra Husremović[1], Lilia S. Xie[1], Matthew P. Erodici[1], R. Soyoung Kim[3], Junko Yano[2], and D. Kwabena Bediako[1,3]*

[1] *Department of Chemistry, University of California, Berkeley, California 94720, United States*

[2]*Molecular Biophysics and Integrated Bioimaging Division, Lawrence Berkeley National Laboratory, Berkeley, CA 94720, USA*

[3]*Chemical Sciences Division, Lawrence Berkeley National Laboratory, Berkeley, CA 94720, USA*

**Correspondence to: bediako@berkeley.edu*

## Abstract

Some magnetic systems display a shift in the center of their magnetic hysteresis loop away from zero field, a phenomenon termed exchange bias. Despite the extensive use of the exchange bias effect, particularly in magnetic multilayers, for the design of spin-based memory/electronics devices, a comprehensive mechanistic understanding of this effect remains a longstanding problem. Recent work has shown that disorder-induced spin frustration might play a key role in exchange bias, suggesting new materials design approaches for spin-based electronic devices that harness this effect. Here, we design a spin glass with strong spin frustration induced by magnetic disorder by exploiting the distinctive structure of Fe intercalated $ZrSe_2$, where Fe(II) centers are shown to occupy both octahedral and tetrahedral interstitial sites and to distribute between $ZrSe_2$ layers without long-range structural order. Notably, we observe behavior consistent with a magnetically frustrated, and multi-degenerate ground state in these $Fe_{0.17}ZrSe_2$ single crystals, which persists above room temperature. Moreover, this magnetic frustration leads to a robust and tunable exchange bias up to 250 K. These results not only offer important insights into the effects of magnetic disorder and frustration in magnetic materials generally, but also highlight as design strategy the idea that a large exchange bias can arise from an inhomogeneous microscopic environment without discernible long-range magnetic order. In addition, these results show that intercalated TMDs like $Fe_{0.17}ZrSe_2$ hold potential for spintronics technologies that can achieve room temperature applications.

## Introduction

In a system that displays magnetic hysteresis, the magnetization depends on the history (sweep direction) of the external magnetic field, produces a remanent magnetization at zero-field, and a non-zero coercive field is needed to reestablish a zero-magnetization state following polarization.[1] Magnetic hysteresis loops are usually centered on zero-field. However, some systems display a shift in the center of the hysteresis loop away from zero field—usually when cooled under an external magnetic field—an effect that is termed "exchange bias".[2,3] Magnetic materials that exhibit this exchange bias effect play a crucial role in spin-based memory and electronic devices (including spin valves,[4,5] magnetic read heads,[6] and magnetic random access memory[7]), since the shifted hysteresis loop prevents random noise from inadvertently reversing the magnetization of a component. Traditionally, magnetic heterostructures and thin-films have been the primary platforms for engineering and studying exchange bias—including ferromagnetic (FM)/antiferromagnetic (AFM) bilayers[8-12] (e.g. NiFe/FeMn), FM/ferrimagnetic (FiM) architectures[13,14] (e.g. NiFe/TbCo), and FiM/AFM heterostructures[15] (e.g. $Fe_3O_4$/CoO). Yet, despite the extensive study of exchange bias within a variety of such systems over seven decades, a comprehensive mechanism behind exchange bias remains lacking. One key unresolved issue is an understanding of the critical role of disorder and/or spin frustration at the interfaces of the constituent layers.

In canonical exchange bias constructs comprised of layers of a FM on an AFM, the traditional explanation of exchange bias involves interfacial pinning of the FM layer to the adjacent AFM layer, resulting in a hysteresis loop that is polarized with the orientation of the AFM moment at the interface.[16] This simplistic picture has since been replaced by a model in which exchange bias is driven by a disordered magnetic state—a spin glass (SG)—that is incidentally formed at the interface of the FM and AFM layers. This SG, which is characterized by random and frustrated exchange interactions, is believed to stem from the interplay of structural disorder,[12,17,18] interface roughness,[9,19,20] chemical intermixing[21] and/or opposing, yet energetically similar, exchange interactions.[22] Several studies have suggested that the spin glass itself is indeed closely tied to the exchange bias phenomenon, since this effect has been observed not only in prototypical magnetic heterostructures including FM/SG bilayers,[18] but also in apparently intrinsic single phase spin glass materials, including magnetically glassy dilute metal alloys,[23] single crystals of compounds with inherently coexisting magnetic phases of AFM and SG,[24] and geometrically frustrated lattices.[25] Still, the role of magnetic disorder and frustration on the emergence of exchange bias remains unclear, though evidently critical for optimizing material and device performance.[26]

Layered transition metal dichalcogenides (TMDs) are a class of materials in which two-dimensional layers of $MCh_2$ ($M$ = transition metal; $Ch$ = S, Se, Te) are stacked via van der Waals (vdW) interactions along the crystallographic $c$-axis. The vdW interfaces allow for the intercalation of a range of chemical species, such as atoms,[27] molecules,[28] and ions.[29-31] TMDs intercalated with open-shell first-row transition metals, $T$, are versatile platforms for designing magnetic materials, where the spin density and magnetic ordering can be precisely controlled through the choice of host lattice, intercalant, and stoichiometry of the resulting $T_xMCh_2$ compound. When $x$ = 1/4 or $x$ = 1/3, it is possible for intercalants to fully order with commensurate superlattices (of size $2a \times 2a$ or $\sqrt{3}a \times \sqrt{3}a$, respectively, where $a$ is the lattice constant of the primitive $MCh_2$ lattice) and to exhibit long-range magnetic order.[32-34] However, when the extent of intercalation deviates from these stoichiometric compositions, some $T_xMCh_2$ materials have been found to exhibit spin-glass phases.[24,35,36] Analytis and colleagues recently synthesized slightly off-stoichiometric $Fe_{0.33\pm\delta}NbS_2$ ($\delta \leq 0.03$), which displayed a predominant AFM order coexisting with a minor spin-glass phase, exhibiting a large exchange bias below 40 K.[24] Interestingly, the spin glass in this material, though a very minor component, may also enable ultralow current-induced switching of the AFM order, a technology that holds promise for low power spintronics.[31,32] Intercalant site disorder coupled with the oscillatory nature of the purported Ruderman–Kittel–Kasuya–Yosida (RKKY) exchange interaction (*i.e.*, exchange mediated by conduction electrons) was proposed as the origin of the crucial spin glass phase. In turn, the coupling between an uncompensated spin glass and a highly anisotropic AFM within a single crystal was suggested as the source of the highly enhanced exchange bias effect.

To directly interrogate the role of intercalant disorder on spin-glass behavior in intercalated TMDs, we sought to explore a system possessing the spin glass itself as the predominant magnetic component. To accomplish this, we chose $Fe_xZrSe_2$ for three reasons. First, structurally, $ZrSe_2$ is a promising intercalation host lattice for constructing a predominant spin glass phase that may enable the study of an intrinsic spin glass-derived exchange bias effect. While the 2$H$ and 1$T$ polytypes of TMDs contain both octahedral and tetrahedral vacancy sites in the interlayer vdW sites, the experimentally determined crystal structures of most Fe-intercalated TMDs compounds show that intercalants occupy the octahedral sites exclusively.[37] However, Fe-intercalated Zr-based TMDs ($Fe_xZrCh_2$; $Ch$ = S, Se) are unique in that intercalants can occupy both tetrahedral and octahedral sites (**Figure 1**), with Fe atoms distributed between both sites.[38,39] The difference in coordination environment of intercalated Fe at different sites, combined with the variation in the distance between magnetic sites and the oscillatory nature of potential exchange interactions (including *p–d* hybridization for semiconductors,[40,41] RKKY for metals[42-44]), should lead to variations in sign and magnitude of the coupling between adjacent magnetic centers, potentially resulting in spin frustration that might be expected to manifest as a spin-glass phase. Second, the development of systems with high transition temperatures closer to room temperature is crucial for any potential technological application. In this regard, selenide analogues of intercalated TMDs are promising, as these compounds generally exhibit higher magnetic ordering temperatures compared to their related sulfides (likely due to the stronger spin–orbit coupling in heavier selenide compounds and/or greater orbital overlap).[38,39,45-47] Finally, $Fe_xZrSe_2$ has been reported to be semiconducting.[39] The interaction of any localized moments with conduction band electrons has a profound effect on exchange interactions. To a greater extent than metals, semiconducting materials permit the strong modulation of their charge carrier densities electrostatically[48-50] or electrochemically,[51] offering the possibility of achieving electrically controlled magnetism.

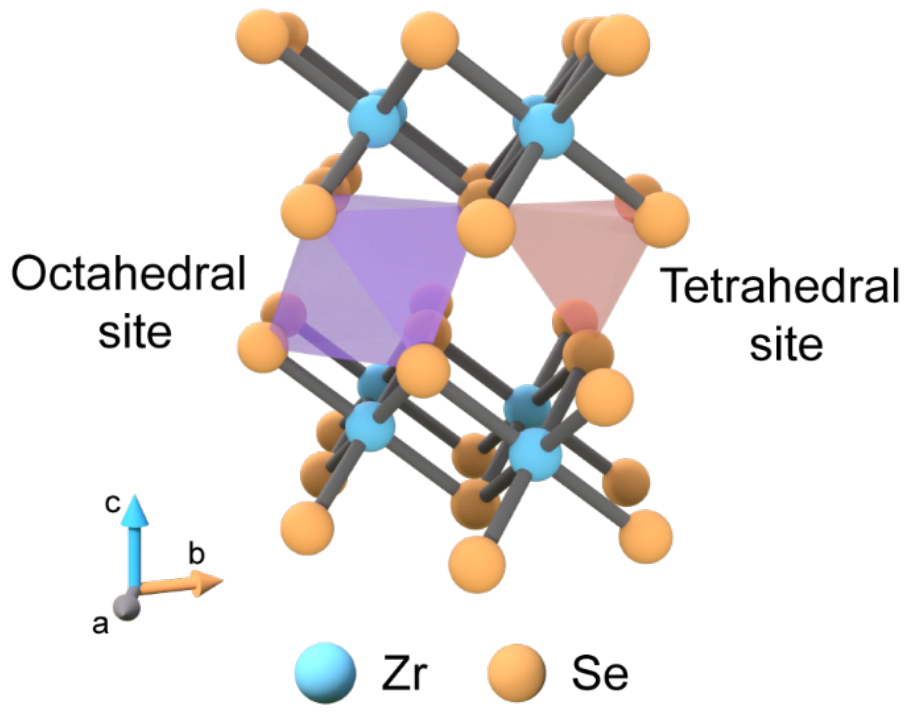

**Figure 1.** Intercalants can occupy octahedral and tetrahedral interstitial sites in 1$T$-$ZrSe_2$

As a consequence of inherent intercalant disorder, we expect $Fe_xZrSe_2$ to exhibit spin glass properties. While there have been suggestions that the material displays glassy magnetic behavior,[39] the unambiguous experimental verification of this spin glass state has not yet been reported, and the resultant manifestation of exchange bias has not been demonstrated. Here, we show that $Fe_xZrSe_2$ ($x \sim 0.17$) consists of high-spin $Fe^{2+}$ ions located without long-range order in both octahedral and tetrahedral sites between the layers of 1*T*-$ZrSe_2$. This material displays semiconducting behavior with a bandgap of about 0.4 eV, and magnetic characterization reveals a high degree of spin frustration above room temperature, much higher than the glassy transition temperatures (< 40 K) reported for other $Fe_xMCh_2$ materials. Consequently, upon magnetic field cooling, $Fe_xZrSe_2$ displays a measurable, apparently intrinsic, exchange bias for temperatures ≤ 250 K. These results indicate that the spin glass phase is closely linked to the exchange bias phenomenon and suggest a powerful strategy for designing novel exchange bias systems with near-room temperature transition temperatures, which could have potential applications in future spintronics devices.

## Results

### *Synthesis, Composition, and Structure of Fe-intercalated $ZrSe_2$*

Single crystals of $Fe_xZrSe_2$ were synthesized using chemical vapor transport (CVT). A mixture of source materials Fe, Zr, Se in a ratio of 0.5:1:2 with transport agent $I_2$ (1 mg/cm$^3$) was loaded into an evacuated quartz ampoule and placed in a two-zone furnace with temperature set points of 900 °C and 1050 °C (see Supporting Information for additional experimental details and **Figure S1** for set-up). After 15 days, hexagonal plate-shaped silver-black crystals with lateral dimensions of several millimeters were obtained (**Figure S1c**). The crystals were thoroughly rinsed with toluene to remove the excess $I_2$ and stored in an argon-filled glove box.

To determine the composition of $Fe_xZrSe_2$ crystals, scanning electron microscopy (SEM) combined with energy-dispersive X-ray spectroscopy (EDS) was performed (see Supporting Information for experimental details). Prior to SEM-EDS measurements, the as-grown crystals were cleaved with adhesive tape to expose a fresh surface. The SEM image of the as-grown crystal and the corresponding EDS spectrum and elemental mapping are shown in **Figure S2**. The elemental maps are consistent with a uniform spatial distribution of Fe, Zr, and Se over a large area of approximately 90,000 μm$^2$ with the Fe, Zr, Se in the ratio 0.17:1:2.02. This composition is consistent with the empirical formula $Fe_{0.17}ZrSe_2$ determined through single-crystal X-ray diffraction (SCXRD) refinement (**Table S1**). The main X-ray diffraction peaks can be well integrated using the trigonal space group ($P\bar{3}m1$). The lattice parameters ($a = b = 3.7662(2)$ Å, $c = 6.1236(4)$ Å) at room temperature were found to be close to those of reported structures for $ZrSe_2$,[52-55] indicating that the intercalated Fe atoms did not significantly affect the interlayer distance. **Figure 2a** depicts the unit cell of the single-crystal structure solution, where the $ZrSe_2$ host lattice is preserved in the $CdI_2$ (1*T*) structure, with Fe atoms partially occupying both tetrahedral and octahedral sites coordinated by Se atoms within the interlayer region.

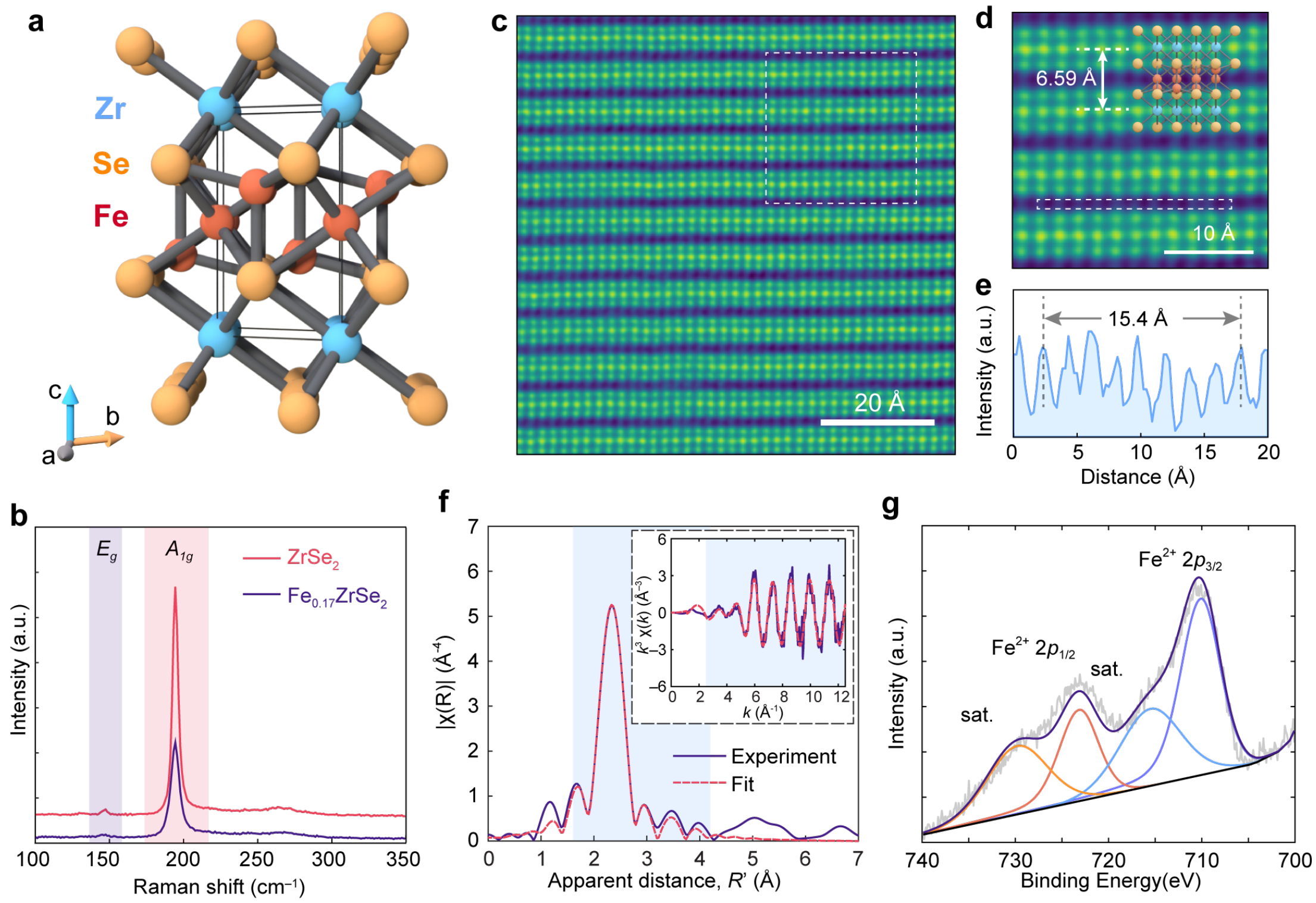

**Figure 2. Compositional and structural characterization of $Fe_{0.17}ZrSe_2$.** (a) The crystal structure of $Fe_{0.17}ZrSe_2$ is that of 1*T*-$ZrSe_2$ with iron atoms intercalated in both octahedral and tetrahedral vacancy sites in the vdW interface. The occupancies of octahedral and tetrahedral sites are 0.1 and 0.07, respectively. The unit cell is framed in solid lines. (b) Raman spectra of 1*T*-$ZrSe_2$ (purple line) and $Fe_{0.17}ZrSe_2$ (red line) flakes with 532 nm excitation. (c) Cross-sectional HAADF-STEM image of an $Fe_{0.17}ZrSe_2$ sample along the $[10\bar{1}0]$ zone axis. (d) Zoomed-in image of white dashed line area in (c) overlaid with the $Fe_{0.17}ZrSe_2$ structure. (e) Line profile integrating intensity along the white dashed box in (d) revealing presence of intercalant ions between layers 1*T*-$ZrSe_2$. (f) Fe *K*-edge EXAFS spectra obtained at room temperature in a He atmosphere. Experiments are fitted by means of ARTEMIS program. The figures show Fourier transforms |χ(R)| of the $k^3$-weighted EXAFS spectrum (not corrected for phase shift). The inset shows $k^3$-weighted background-subtracted EXAFS spectrum. Experimental data is plotted as a solid purple trace and fits are shown as dashed red lines. The regions highlighted in blue are the fitting regions. (g) Fe 2p XPS spectra in $Fe_{0.17}ZrSe_2$. The Fe $2p_{3/2}$ and Fe $2p_{1/2}$ features located at 710.09 eV (purple) and 723.15 eV (red), respectively, are attributed to $Fe^{2+}$ centers. The peaks located at 715.5 eV (blue) and 729.8 eV (orange) are attributed to the Fe satellite peaks.

**Figure 2b** presents Raman spectra of $ZrSe_2$ and $Fe_{0.17}ZrSe_2$ crystals obtained using 532 nm excitation. $ZrSe_2$ exhibits the characteristic phonon modes, out-of-plane $A_{1g}$ (194 cm$^{-1}$) and in-plane $E_g$ (147 cm$^{-1}$) vibrations, consistent with previously reported values.[56] The spectral features observed in $Fe_{0.17}ZrSe_2$ were found at the same Raman shifts as those of $ZrSe_2$, providing additional evidence that the intercalation of Fe atoms did not break the crystal symmetry of host lattice 1*T*-$ZrSe_2$.[57] No Raman peaks corresponding to a superlattice were detected in $Fe_{0.17}ZrSe_2$, suggesting that Fe centers are disordered between the layers and no long-range superstructure is formed.[57] The absence of superlattice formation is also supported by selected area electron diffraction (SAED), as only diffraction spots corresponding to the host lattice were observed (**Figure S3**). Furthermore, the decrease in peak intensities and the broadening of peak widths observed in the Raman spectrum of $Fe_{0.17}ZrSe_2$ are attributed to the site disorder induced by the disordered distribution of Fe centers in the lattice.[57,58]

To more directly establish the presence of Fe between $ZrSe_2$ layers, we performed high angle annular dark-field scanning transmission electron microscopy (HAADF-STEM) of cross-sectional samples. **Figures 2c,d** show HAADF-STEM cross-sectional images taken along the $[10\bar{1}0]$ direction, revealing high crystallinity of the $ZrSe_2$ lattice. Inspection of these HAADF-STEM images shows significant intensity between $ZrSe_2$ layers, and **Figure 2e** displays a line intensity profile along the dashed rectangle in **Figure 2d**, confirming that substantial atomic contrast is present between the layers of $ZrSe_2$, consistent with intercalated Fe centers.

The local electronic and geometric features of Fe centers in $Fe_{0.17}ZrSe_2$, were revealed with X-ray absorption spectroscopy, XAS (including X-ray near-edge spectroscopy, XANES, and extended X-ray absorption fine structure, EXAFS) and X-ray photoelectron spectroscopy (XPS) on $Fe_{0.17}ZrSe_2$ crystals. Fitting to the EXAFS of the $Fe_{0.17}ZrSe_2$ crystals (**Figure 2f**) revealed that the majority of Fe atoms occupy octahedral positions, with a minority occupying tetrahedral sites. Paths were generated from the crystal structure of $Fe_xZrSe_2$ (**Figure 2a**) where Fe atoms were placed in either tetrahedral or octahedral positions.

For each fit, paths corresponding to Fe–Se and Fe–Fe single-scattering for both tetrahedral and octahedral sites were included. There is minimal signal beyond $R$ = 2.8 Å, shorter than all Fe–Fe and Fe–Zr paths, suggesting no clustering of Fe atoms near each other. No viable fits were obtained when any Fe–Zr scattering was included. To evaluate the ratio of octahedral to tetrahedral sites, the amplitude factor in the fits for the octahedral and tetrahedral paths were weighted by a factor $x$ or $1–x$, respectively, where $x = 1$ corresponds to 100% of Fe atoms in octahedral sites and $x = 0$ corresponds to 100% in tetrahedral sites. In a given fit, $x$ was constant and a series of fits with the same parameters were performed where $x$ was varied in steps of 0.05 or 0.1 from 0 to 1. The best fit corresponds to 90% of the Fe atoms modeled in octahedral sites (see Supporting Information, Section S7 **Figure S5** for details).

XPS spectra of $Fe_{0.17}ZrSe_2$ single crystals (**Figure 2g**) are also consistent with Fe in the +2 oxidation state, displaying peaks at 710.09 eV ($2p_{3/2}$) and 723.15 eV ($2p_{1/2}$), which are consistent with what have been reported for $Fe^{2+}$ in Se-based coordination environments.[59,60] The comparison of Fe $K$-edge XANES spectra for $Fe_{0.17}ZrSe_2$ and the standard Fe-based compounds in **Figure S4b** also confirms the oxidation state of intercalated Fe atoms to be +2.

### *Bandgap and Electronic Properties*

The electronic properties of the bulk $Fe_{0.17}ZrSe_2$ crystal were characterized using scanning tunnelling spectroscopy (STS) and diffuse reflectance spectroscopy (DFS) at room temperature (**Figure 3**). STS measurements were performed on a freshly cleaved bulk sample in ambient conditions. The sample was mounted into the STS sample holder using carbon tape to attach the top of the sample to an STS probe (see Supporting Information, Section S9 **Figure S6a** for details). At a fixed tip–sample separation, the tunneling current between the tip and the $Fe_{0.17}ZrSe_2$ crystal was monitored while the bias voltage (V) was swept over a given range. In this measurement, $V$ modulates the position of the fermi level within the electronic bands, which is manifest as a variation in the tunneling current, $I$, and therefore also a modulation in differential conductance, $dI/dV$ (the instantaneous slope of the $I$–$V$ trace). When the Fermi level lies within the bandgap, $I$ and $dI/dV$ approach zero, and non-zero $I$ and $dI/dV$ values are expected when the Fermi level lies within the valance or conduction band. **Figure 3a** shows $dI/dV$ tunneling spectra of $ZrSe_2$ and $Fe_{0.17}ZrSe_2$ determined from this STS measurement (**Figure S6b-e**). The flat region of $dI/dV \cong 0$ near the origin corresponds to the band gap,[61] revealing a band gap of approximately 0.43 eV. The comparison between STS for $ZrSe_2$ and $Fe_{0.17}ZrSe_2$ shows that the intercalation of Fe into $ZrSe_2$ substantially decreases the size of the band gap. Likewise, analysis of the DFS Tauc plot of $Fe_{0.17}ZrSe_2$ yields a band gap of ~0.44 eV (**Figure 3b**) (see Supporting Information Section S10 for details), which is consistent with the results of STS. Furthermore, the photoluminescence spectrum (PL) for $Fe_{0.17}ZrSe_2$, as shown in **Figure S7**, exhibits the absence of peaks in the photon energy range of 0.41 – 0.60 eV from 77 to 300 K, consistent with an indirect band-gap semiconductor (see Supporting Information Section S10 for details).

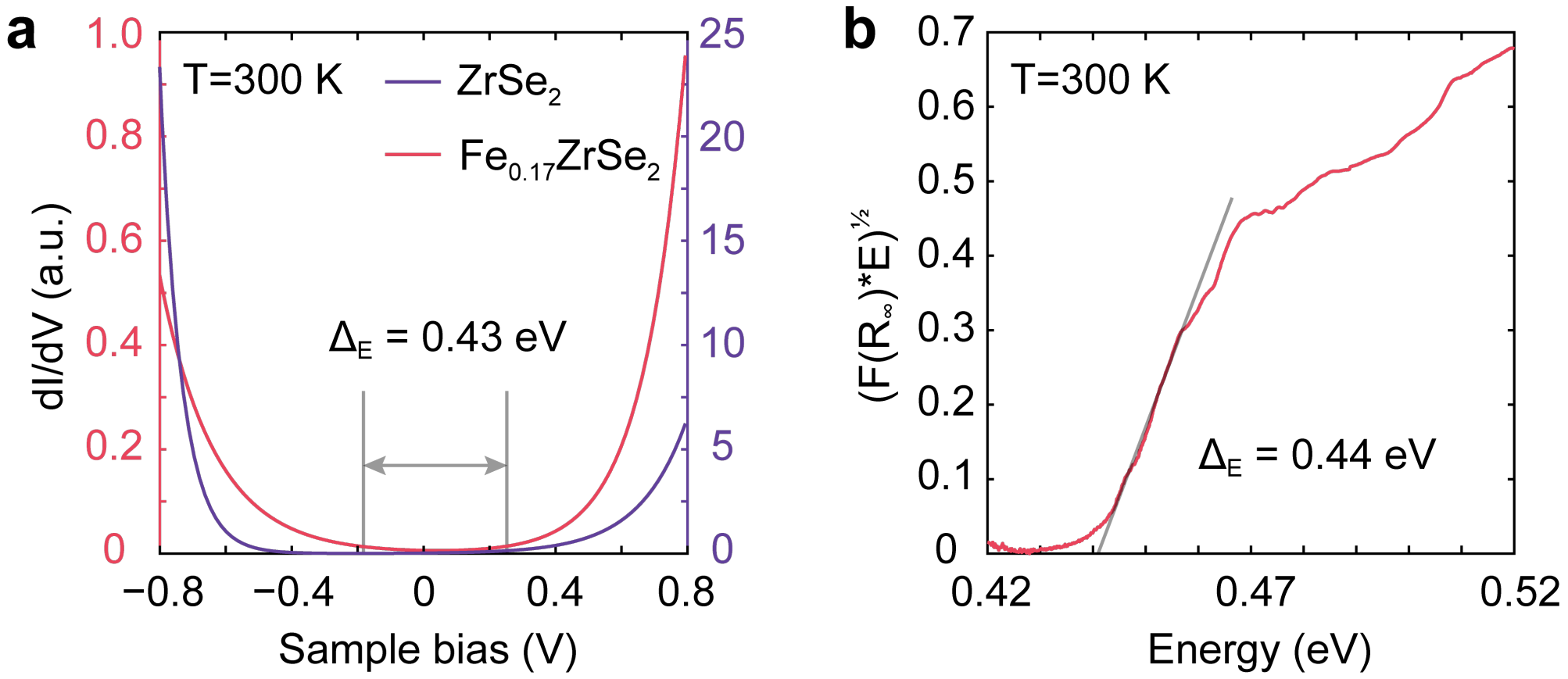

**Figure 3. Bandgap characterization of $Fe_{0.17}ZrSe_2$.** (a) Scanning tunneling spectra of $ZrSe_2$ crystal (purple line) and a $Fe_{0.17}ZrSe_2$ crystal (red line) at room temperature. (b) The Tauc plot of the $Fe_{0.17}ZrSe_2$ crystal. The linear part of the plot is extrapolated to the x-axis. The value of the absorption spectra $F(R_\infty)$ was transformed from the raw diffuse reflectance spectra by applying the Kubelka–Munk function.

### *Variable Temperature Magnetic Measurements*

**Figure 4a** shows temperature-dependent magnetic susceptibility data for $Fe_{0.17}ZrSe_2$ acquired with a magnetic field of 2000 Oe applied along (purple) or perpendicular (red) to the $c$-axis of the crystal. These data show that the in-plane magnetic susceptibility for $Fe_{0.17}ZrSe_2$ is substantially weaker than the out-of-plane magnetic susceptibility, revealing strong magnetocrystalline anisotropy (MCA), consistent with the revious reports.[39] This MCA may arise from the trigonally distorted pseudo-octahedra of the majority of $Fe^{2+}$ centers. We observe a ~3.4% trigonal compression from the size of Fe octahedral site extracted from SCXRD, which results in the qualitative $d$-orbital splitting depicted in **Figure S8**. The unevenly occupied $e_g$ ($d_{xy}$, $d_{x^2-y^2}$) set of octahedral high-spin $Fe^{2+}$ would result in unquenched orbital angular momentum and thus to MCA.

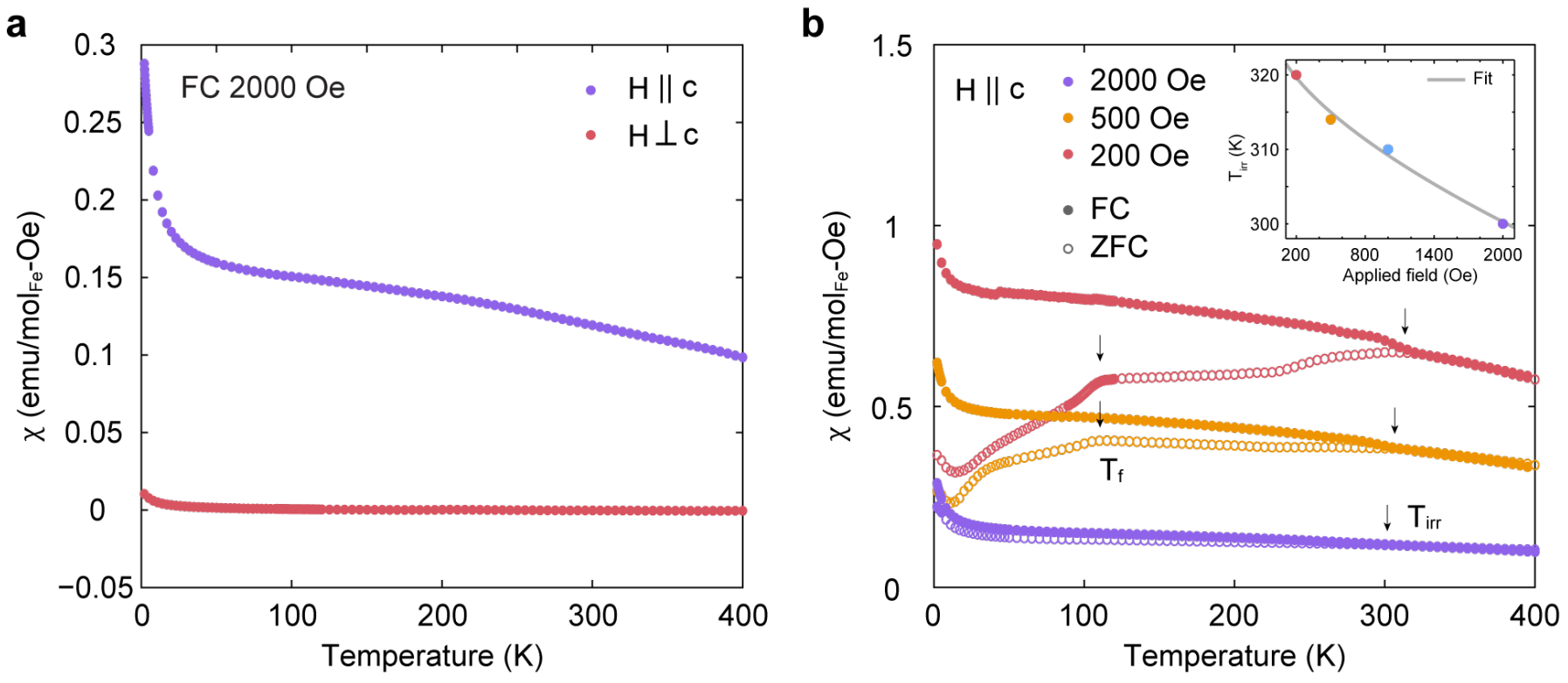

**Figure 4. Temperature dependence of magnetic susceptibility of $Fe_{0.17}ZrSe_2$.** (a) Temperature dependence of out-of-plane (purple dotted line) and in-plane (red dotted line) DC magnetic susceptibility of $Fe_{0.17}ZrSe_2$ single crystals measured with field cooling ($\mu_0 H_{FC}$ = 2000 Oe). (b) Temperature dependence out-of-plane zero-field-cooled (ZFC) and field-cooled (FC) DC magnetic susceptibility for an $Fe_{0.17}ZrSe_2$ single crystal under applied magnetic fields ranging from 200 to 2000 Oe. The ZFC and FC data at a given field are represented by open and closed symbols, respectively. Inset shows the temperature of bifurcation, $T_{irr}$, of ZFC and FC traces for different applied fields (solid dots), fit to the equation $T_{irr}(H) = T_{irr}(0)(1 - AH^n)$.

As a function of temperature, the magnetization of $Fe_{0.17}ZrSe_2$ between 350 and 400 K follows the Curie–Weiss law with $\chi = C/(T - \theta_{CW})$ (**Figure S9**), where $C$ and $\theta_{CW}$ are the Curie constant and Curie–Weiss temperature, respectively. We find $\theta_{CW}$ = –223 K, consistent with the predominance of antiferromagnetic exchange interactions in the high temperature regime. $C$ is determined to be 3.20 emu K/mol$_{Fe}$·Oe and the calculated effective magnetic moment $\mu_{eff}$ of 5.06 $\mu_B$ is close to the theoretical value of 4.90 $\mu_B$ for high spin $Fe^{2+}$ ions, assuming $g = 2$ and $S = 2$ (see Supplemental Information for details). **Figure 4b** depicts the zero-field cooled (ZFC) and field cooled (FC) dc magnetization under various applied fields ranging from 200 to 2000 Oe. These ZFC and FC curves bifurcate around 300 K, arising from irreversibility induced under different cooling protocols. Such bifurcation has been reported in spin-glass and superparamagnetic systems.[39,62-65] To investigate the nature of this state, we determined the irreversible temperatures $T_{irr}$ at each applied field by identifying the bifurcation point of the ZFC and FC curves. In the low-field region, the $T_{irr}$ can be described as a function of the applied magnetic field, $H$, by the following expression:

$$T_{irr}(H) = T_{irr}(0)(1 - AH^n),$$

where $A$ is a constant and $T_{irr}(0)$ is the limit of the irreversible temperature in the absence of a magnetic field. Theoretical models for spin-glass systems predict that in the limit of weak magnetic fields, the spin freezing temperature with weak irreversibility follows the Gabay–Toulouse (GT) line ($T_{irr} \propto H^2$), while the Almeida-Thouless (AT) line ($T_{irr} \propto H^{2/3}$) is followed by systems with strong irreversibility characteristics.[66,67] The experimentally observed field dependence of $T_{irr}$ along with the fit to the above equation are shown in the **inset of Figure 4b**. The value of $n$ obtained from the fit is 0.63, which reveals that $T_{irr}$ follows an approximately $H^{2/3}$ behavior that is consistent with the strong irreversibility of the glassy system.

In **Figure 4b**, we observe that with decreasing temperature below $T_{irr}$, the ZFC curves display a plateau in the measured moment followed by a downturn, whereas the FC curves show continuously increasing moments. ZFC curves in applied fields of 200 and 500 Oe reveal a broad peak near 110 K, and the ZFC trace measured in a field of 500 Oe also shows a transition near 40 K. The upturns observed below 14 K are likely a Curie tail from a paramagnetic impurity. Ac heat capacity measurements of $Fe_{0.17}ZrSe_2$ in the absence of magnetic field (**Figure S11**) show that no distinct thermodynamic phase transitions are present. This absence of a response in heat capacity measurements is consistent with glassy order.[68] The magnetization measurements with applied field perpendicular to the *c*-axis of the crystal also show a bifurcation between the ZFC and FC traces (**Figure S12**), revealing that the spin glass behavior is also anisotropic.

### *Relaxation of magnetization for the spin-glass phases of $Fe_{0.17}ZrSe2$*

A multivalley energy landscape arising from spin frustration emerges when cooling a spin glass through its freezing temperature.[62,64,69] The non-ergodicity of the glass system in turn engenders complex magnetic relaxation processes that are apparent in isothermal remanent magnetization (IRM) and thermoremanent magnetization (TRM) measurements as a slow relaxation of the magnetization over time (**Figure 5**). For the investigation of the slow relaxation dynamics of $Fe_{0.17}ZrSe_2$, IRM and TRM measurements were performed following the protocols outlined in **Figures S13a** and **S13b**, respectively. In short, IRM measurements were carried out by cooling samples down to the target temperature without application of a field, applying a field for a given waiting time ($t_W$), removing the field, and measuring the magnetization as a function of time. TRM measurements were carried out by cooling down to target temperature in the presence of a field, holding at this applied field for $t_W$, removing the field, and then monitoring the change in magnetization as a function of time. See Supplemental Information for additional methodological details. **Figures 5a,b** show that the magnetization relaxes in time upon removal of the applied field, a characteristic of glassy systems. Moreover, the relaxation behavior for both IRM and TRM were best fit using the typical stretched exponential decay function of

$$M_R(t) = M_0 + A \exp[-(t/\tau)^{1-n}]$$

where $M_0$ is the remanent magnetization at $t = 0$ (that is, the intrinsic component of the moment), $A$ is the peak beyond equilibrium values related to glassy component of the magnetization, $\tau$ is the characteristic average relaxation time, and $n$ is the time stretch component. The obtained $\tau$ values at 2 K are around 17 and 24 minutes and the $n$ is found to be approximately 0.58 (**Table 1**). Both values fall in the typical range of glassy systems.[69,70] Moreover, the relaxation times and time stretch components of IRM and TRM measurements are comparable, indicating a common relaxation mechanism in both routines.

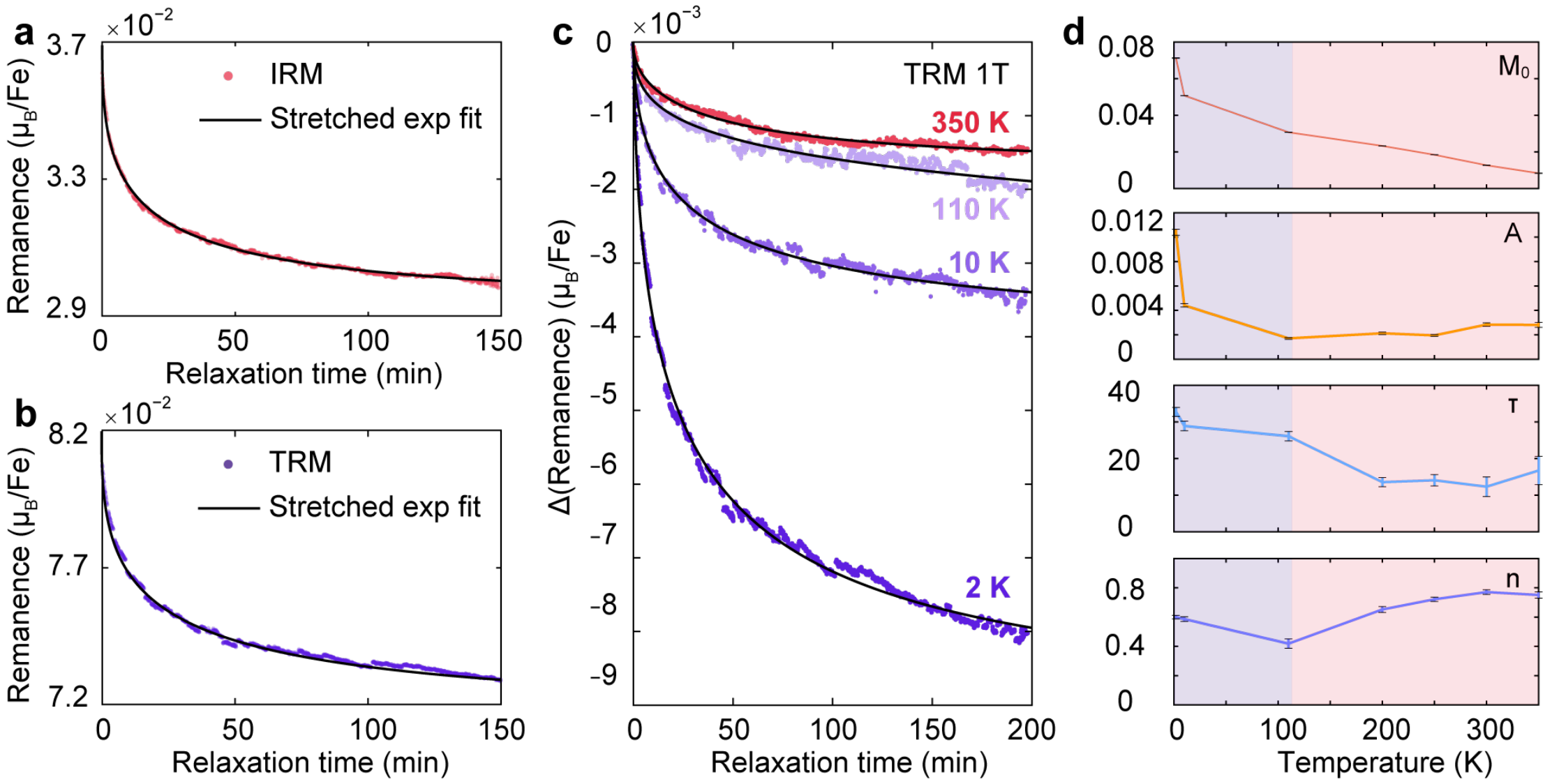

**Figure 5. Slow magnetic relaxation behavior of $Fe_{0.17}ZrSe_2$.** (a,b) IRM (red dots) and TRM (purple dots)data collected from the protocol described below with 1 hour wait time, respectively, and their corresponding fits (black lines). The material was first fast-cooled from 400 K to 60 K by 10 K/min and then slow-cooled from 60 K to 2 K by 1 K/min under a 1 T under zero field (a) or applied field (b) and held in applied field 1 T for a designated wait time, $t_w$, at 2 K, the field was then removed and the TRM or IRM data were collected. The appearance of relaxation dynamics is correlated with the glassy state. Both IRM and TRM show similar dynamics, which indicates a common relaxation mechanism in both routines. (c) TRM measurements (dots) performed at various temperatures after the samples were field cooled in a field of 1 T and their corresponding fits (black lines). The remanences at t = 0 min for all the TRM curves are normalized to 0 $\mu_B$/Fe. All the fits in the figure were performed using a typical stretched exponential decay function $M_R(t) = M_0 + A\exp[-(t/\tau)^{1-n}]$. (d) The extracted parameters from the fitting were plotted as a function of temperature. The fittings for the temperatures under 110 K were shaded with purple whereas the fitting for the temperatures higher than 110 K were shaded with red.

*Table 1. Magnetization relaxation parameters from isothermal remanent magnetization (IRM) and thermoremanent magnetization (TRM) measurements at 2 K.*

| | $M_0$ ($\mu_B$/Fe) | $A$ ($\mu_B$/Fe) | $\tau$ (min) | $n$ |
|---|---|---|---|---|
| **TRM** | 0.0717 | 0.0104 | 23.7 | 0.590 |
| **IRM** | 0.0295 | 0.00745 | 16.6 | 0.571 |

To investigate the temperature-dependent dynamics of the glassy state, TRM measurements were conducted at selected temperatures, as shown in **Figure 5c** and **Figure S14**. The corresponding fitting parameters are presented in **Figure 5d** and **Table S4**. These measurements show that the slow relaxation behavior persists well above room temperature. Notably, $A$ and $\tau$ appear temperature dependent for temperatures between 2 K and 110 K and both parameters decrease with increasing temperature (**Figure 5d**), as expected for a system trapped in metastable states separated by finite energetic barriers. Above 110 K, these parameters appear largely independent of temperature, even as $M_0$ decreases steadily (yet producing a finite $M_0$ up to 350 K). The onset of the long, temperature-dependent $\tau$ values is consistent with the $T_f$ value of 110 K, while the persistent slow relaxation above 110 K may be indicative of a disparate glassy state that is characterized by the bifurcation in FC/ZFC traces and $T_{irr}$ value of **Figure 4b**.

Furthermore, the decay curves in **Figure S15** indicate that longer $t_W$ values are associated with higher magnetization values, which is another characteristic feature of a glassy magnetic system.[62,69]

## *Magnetic hysteresis and exchange bias*

Measurements of magnetization as a function of applied field were performed at 2 K after zero-field cooling the sample. As depicted in **Figures 6a,b**, the magnetization of the system exhibits a hysteresis loop with a coercive field, $\mu_0H_c$, of 866 Oe. At the highest measured applied field of 12 T, the hysteresis loop does not fully saturate, and the moment is measured to be 0.6 $\mu_B$/Fe (substantially lower than the $\mu_{eff}$ of 5.06 measured from C–W fits (**Figure S9**). Variable-field magnetization data were also collected at 2 K after cooling the sample under applied fields of +12 T and –12 T (**Figures 6a,b**), revealing hysteresis loops with distinct lateral shifts from near zero to ∓166 Oe, respectively, accompanied by a shift along the vertical axis. Upon cooling in a positive (negative) field, the loop exhibited a bias in the negative (positive) direction, which is a signature of exchange bias.[2,17] The magnitude of the exchange bias field, $H_{EB}$, was calculated as:

$$\mu_0 H_{EB} = (\mu_0 H_{int1} + \mu_0 H_{int2})/2$$

where $\mu_0 H_{int1}$ and $\mu_0 H_{int2}$ are the intercepts on horizontal axes. The magnitude of $\mu_0 H_c$ is defined as the half-width of the hysteresis loop at the average of the vertical intercepts, as depicted in the inset of **Figure 6a**.

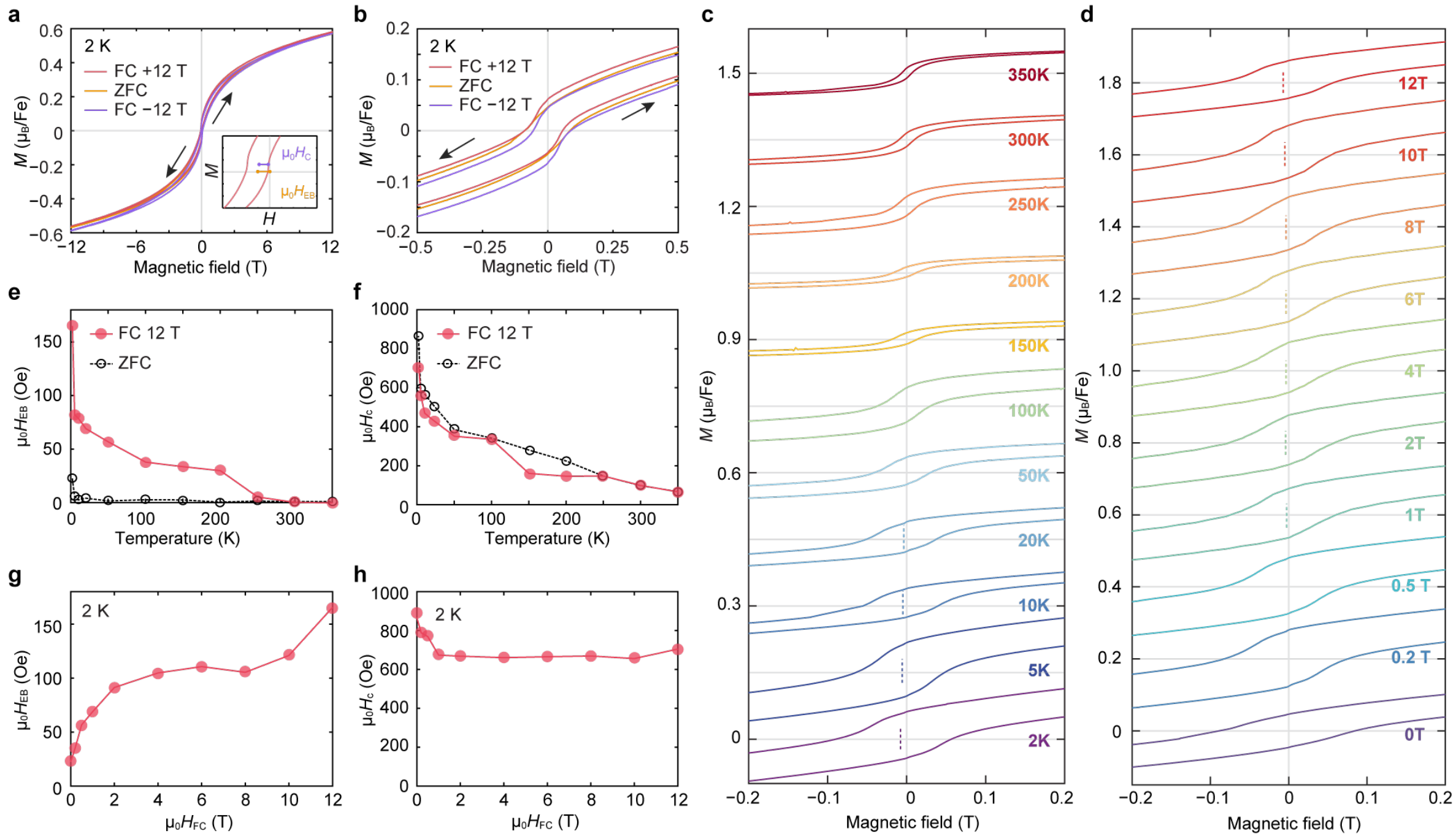

**Figure 6. Exchange bias.** (a) Magnetic hysteresis loops measured after the samples are cooled under 12 T (red curve), −12 T (purple curve), and zero field (orange curve) from 400 K to 2K. The inset in (a) demonstrates the definition of exchange bias field $\mu_0H_{EB}$ (purple), and coercive field $\mu_0H_C$ (orange). $\mu_0H_{EB}$ was extracted from each loop by taking the average of the x intercepts, $\mu_0H_{EB} = (\mu_0H_{int1} + \mu_0H_{int2})/2$; $\mu_0H_C$ was calculated from the half-width of the hysteresis loop at the average of the y intercepts. (b) Expanded view of the region between −0.5 and 0.5 T in (a). The shifts of the hysteresis loops collected under applied fields are symmetrical relative to the zero-field loop. (c) Magnetization versus magnetic field measurements performed on $Fe_{0.17}ZrSe_2$ at various temperatures. Each loop is offset on the y-axis by 0.15 $\mu_B$/Fe. (d,e) Temperature dependence of the extracted exchange bias field, $\mu_0H_{EB}$ (d) and coercive fields, $\mu_0H_C$ (e) after cooling in a 12 T (black curve) and 0 T (grey curve) from 400 K. (f) Magnetization versus magnetic field measurements performed on $Fe_{0.17}ZrSe_2$ under various cooling fields. Each loop is offset on the y-axis by 0.2 $\mu_B$/Fe. (g,h) Cooling field dependence of $\mu_0H_{EB}$ (g) and $\mu_0H_C$ (h) after cooling from 400 K to 2 K. All the hysteresis loops were collected with the magnetic field applied along the c-axis of the samples.

Plots of $M(H)$ over a range of temperatures are shown in **Figure 6c** and the associated temperature-dependent evolution of exchange bias field $\mu_0 H_{EB}$ and coercive field $\mu_0 H_c$ for $\mu_0 H_{FC}$ = 12 T (compared to the corresponding ZFC case) are presented in **Figures 6e,f**. The value of $\mu_0 H_{EB}$ decreases with increasing temperature, and a finite $\mu_0 H_{EB}$ persists up to at least 250 K. Likewise, $\mu_0 H_c$ decreases with increasing temperature but the hysteresis loop is not completely closed at 350 K.

The exchange bias behavior at 2 K was also probed at different cooling fields as shown in **Figures 6d,g–h**. Substantial increases in $\mu_0 H_{EB}$ are observed with increasing cooling fields between 0 to 2 T and 8 to 12 T, but $\mu_0 H_{EB}$ appears unchanged by increasing cooling fields from 2 to 8 T. The value of $\mu_0 H_c$ decreases slightly with increasing cooling field for $\mu_0 H_{FC} \leq 1$ T, and remains effectively invariant with cooling field beyond 1 T.

## Discussion

### *Crystal field stabilization energy and steric effects on occupation of intercalated Fe*

The vdW interfaces of 2*H* and 1*T* polytype TMDs create both octahedral and tetrahedral sites for intercalated ions (**Figure 1**). However, the experimentally determined crystal structures of Fe- intercalated TMDs compounds show that intercalants occupy the octahedral sites exclusively, except for $Fe_xZrSe_2$.[37] The preference for octahedral over tetrahedral sites in Fe-intercalated TMD compounds can be related to the increased coordination number of high-spin $Fe^{2+}$ in octahedral complexes, as well as the negative octahedral site preference energy (OSPE) arising from the difference in crystal field stabilization energies (CFSE) between octahedral and tetrahedral complexes.[37,71,72] The small OSPE in $d^6$ complexes suggests that geometric effects could play a role in determining the stability of complexes. An analysis of Se–Se distances in structures of Fe-intercalated TMDs reported so far (**Figure S19**) shows how geometric effects may explain the coexistence of $Fe^{2+}$ in tetrahedral and octahedral sites in $Fe_xZrSe_2$. The Se–Se distances in the Fe octahedron of $Fe_xMSe_2$ are comparable to analogous Se–Se distance in $Fe_7Se_8$ (in which Fe is exclusively octahedrally coordinated), and the Fe tetrahedron in $Fe_xZrSe_2$ is the closest in size to that found in FeSe (in which Fe is exclusively tetrahedrally coordinated) by comparison to other host lattice TMDs.

This simple analysis might explain why the tetrahedral intercalant site in $1T$-$ZrSe_2$ may more readily accommodate Fe than the corresponding sites in other TMDs, providing a possible explanation for the distribution of Fe intercalants in both tetrahedral and octahedral sites that are observed by SCXRD, HAADF-STEM, and EXAFS.

### *Spin glass behavior*

Structural and spectroscopic characterization (**Figure 2**) demonstrates that $Fe_{0.17}ZrSe_2$ is a crystallographic single-phase material with $Fe^{2+}$ ions distributed in a disordered fashion between the layers of $1T$–$ZrSe_2$, and the material remains semiconducting (**Figure 3**). Multiple measurements of this material display signatures of spin glass behavior up to room temperature. Magnetic susceptibility data (**Figure 4**) display a bifurcation in FC and ZFC measurements at a temperature, $T_{irr}$ (around 300 K), that depends on magnetic field—a phenomenon typically attributed to the freezing of moments.[73,74] This freezing occurs when the crystalline anisotropy of small domains overcomes thermal fluctuations, which leads spins to freeze in random directions. In the low-field region, $T_{irr}$ decreases with increasing applied magnetic field and the dependence of $T_{irr}$ with field (**Figure 4b inset**) shows that the system follows the Almeida–Thouless (AT) line with $T_{irr} \propto H^{2/3}$. This trend is consistent with $Fe_{0.17}ZrSe_2$ representing a glassy system with strong irreversibility characteristics. A negative Curie−Weiss temperature ($\theta_{CW}$ = −223 K) in the apparently paramagnetic regime (above 350 K), is consistent with the predominance of antiferromagnetic exchange interactions in $Fe_{0.17}ZrSe_2$ crystals. Indeed, a downturn at in magnetic susceptibility near 110 K in ZFC measurements is observed under both 200 Oe and 500 Oe applied fields (**Figure 4b**), which could be suggestive of a weak AFM transition, though no predominant phase transition is observed as evinced by specific heat capacity measurements (**Figure S11**). While the ZFC feature at 110 K does indicate a change in magnetic phase, such change in phase could be attributed to an AFM phase coexisting with a spin glass phase either in the high temperature regime or in the low temperature regime. It could also simply correspond to two different spin glass phases, where the secondary low temperature phase is reentrant.[75]

Measurements of magnetization over time (**Figures 5** and **S14**) reveal slow spin relaxation dynamics in $Fe_{0.17}ZrSe_2$ crystals, another characteristic feature of glassy magnetic materials. Below 110 K, the magnetization relaxation is strongly temperature dependent, with time constants varying from 33 to 26 min between 2 and 110 K. However, from 110 K to 350 K, the time constant of the system is largely invariant with temperature (*ca*. 14 min). These relaxation data suggest two regimes of glassy behavior, with the coexistence of a putative AFM-like phase with phase transition temperature about 110 K (**Figure 4b**) serving to further frustrate the spin glass relaxation and significantly increase relaxation times. Such complex magnetic phases, involving the potential coexistence of spin glass and antiferromagnetic phases, might arise from the structural disorder induced by the intercalated Fe atoms occupying both tetrahedral and octahedral sites embedded between 1$T$–$ZrSe_2$ layers, and the competing exchange interactions between magnetic sites. The coexistence of spin-glass and antiferromagnetic phases has been reported in other Fe intercalated TMDs systems $Fe_{0.30}NbS_2$ and $Fe_{0.14}NbSe_2$.[24,76] However, the behavior of $Fe_{0.17}ZrSe_2$ contrasts favorably with that of these other compounds, where Fe atoms occupy only the octahedral interstitial sites in the van der Waals gap of the host lattice. Although $Fe_{0.30}NbS_2$ and $Fe_{0.14}NbSe_2$ also exhibit a glassy behavior, the associated spin-glass phases are observed at $T < 40$ K and $T < 10$ K, respectively, considerably lower temperatures than the room temperature glassy behavior of $Fe_{0.17}ZrSe_2$ observed here. This may again be attributed to the occupancy of Fe in both octahedral and tetrahedral sites that introduces a more global intercalant disorder, that may in turn produce an elevated freezing temperature.

### *Exchange bias effect*

Measurements of magnetization as a function of magnetic field (**Figure 6**) display magnetic hysteresis with a finite coercivity, $\mu_0 H_c$, a behavior that is commonly associated with ferromagnetic systems. However, it is important to note that the opening of hysteresis loops has also been reported in several spin glass materials.[23,25,77] Furthermore, the hysteresis loop measured under ZFC at 2 K does not saturate up to 12 T and the small effective moment, which can arise from a frustrated, multi-degenerate ground state of the system,[66,69] is also consistent with spin glass behavior in $Fe_{0.17}ZrSe_2$. **Figure 6** also shows that when cooled in an applied field, $Fe_{0.17}ZrSe_2$ exbibits a robust exchange bias. Both $\mu_0 H_c$ and the exchange bias field, $\mu_0 H_{EB}$, decrease upon increasing temperature, but a finite $\mu_0 H_c$ is observed up to 350 K and the onset of exchange bias (between 250 K and 300 K) coincides with $T_{irr}$ around 300 K. Once onset, the magnitude of the exchange bias appears independent with decreasing temperature until around 100 K, below which a large increase in $\mu_0 H_{EB}$ is observed.

Again, these behaviors point to two magnetic regimes of this predominant spin glass, separated by the possible onset of a weak, coexisting antiferromagnetic phase at $T_f = 110$ K. Between 2 K and $T_f$, increasing temperature results in thermal fluctuations of the magnetization of both antiferromagnetic and spin glass phases, weakening the coupling between the AFM and SG, which leads to a sharp decrease in both $\mu_0 H_{EB}$ and $\mu_0 H_c$.[78] The coexistence of spin-glass with other magnetic phases has been found in multiple systems besides Fe intercalated TMDs,[79-81] and the interaction between spin-glass and antiferromagnetic phases in those systems have been extensively studied using the random-field Ising model.[82] In artificial bilayer heterostructures of antiferromagnetic and spin glass materials, the interfacial pinning effect of antiferromagnetic domains on spins at the surface of frozen glassy states has been invoked as the origin of an observed exchange bias effect and the interplay between antiferromagnetic order and random fields associated with spin disorder can have a significant impact on the occurrence of spin flips in applied magnetic fields.[83] Since it is the interplay between two magnetic phases that is generally thought to be the key factor giving rise to exchange bias in metallic multilayers, it is reasonable to expect the coexistence of spin-glass and antiferromagnetic phases in single crystallographic phase materials will also exhibit exchange bias behaviors. Nevertheless, these data show that exchange bias can be observed even when the overwhelmingly dominant component is the spin glass.

The increase in $\mu_0 H_{EB}$ with cooling field, which we observe in the low-temperature, putative SG/AFM regime of $Fe_{0.17}ZrSe_2$ (**Figure 6g**), has also been observed in conventional FM/AFM multilayer systems. This effect is explained by considering that the cooling field would act on the AFM component to give rise to an additional induced magnetization. Correspondingly, the magnetic domain size would increase, and both the exchange coupling intensity of the interfacial domains and the unidirectional anisotropy would be strengthened,[84,85] resulting in a stronger pinning interaction between AFM and SG phases that enhances $\mu_0 H_{EB}$.[86,87] Likewise, the slight decrease in $\mu_0 H_c$ with cooling field that we observe (**Figure 6h**) can be explained by considering that larger cooling fields would increase the polarization of the spin glass[87] weakly ordering this phase and reducing the pinning within the SG domain, leading to the suppression of $\mu_0 H_c$.

To assess the potential applicability of the exchange bias phenomenon in $Fe_{0.17}ZrSe_2$, we contextualize this system within a framework of representative exchange bias systems, as summarized in **Table S7**. The observed exchange bias in $Fe_{0.17}ZrSe_2$, measuring 166 Oe at 2 K, aligns with values found in various canonical heterostructure systems.[2] While certain systems can achieve larger exchange bias values, even up to 3 T,[24] rendering them suitable for constructing robust biased permanent magnets, a smaller robust bias in the range of hundreds of Oe is more advantageous for spintronics applications, such as creating a platform for magnetic RAM devices or the electrical manipulation of magnetism, due to its enhanced energy efficiency.

## Conclusions

Combined compositional, structural, and spectroscopic characterization reveals that $Fe_{0.17}ZrSe_2$ crystals exhibit high crystallinity of the 1*T*-$ZrSe_2$ host lattice, while Fe atoms occupy both tetrahedral and octahedral sites, distributed in a disordered fashion between TMD layers. Magnetometry measurements of $Fe_{0.17}ZrSe_2$ demonstrate magnetic irreversibility under zero-field cooling (ZFC), slow relaxation dynamics, the absence of saturation, and a small effective moment under high applied magnetic fields. These findings unveil a frustrated, and multi-degenerate ground state in $Fe_{0.17}ZrSe_2$ single crystals that persists up to room temperature. Interestingly, a robust exchange bias is observed below 250 K, and the coexistence of spin glass phase and a nominal antiferromagnetic phase at temperatures below 110 K appears to contribute to an even more pronounced exchange bias at low temperatures.

Our results highlight the potential of intercalated TMDs like $Fe_{0.17}ZrSe_2$ for spintronics technologies that can operate near room temperature. The spin glass-derived exchange bias observed here persists much higher than that found in other intercalated TMDs like $Fe_{0.33\pm\delta}NbS_2$ and $Fe_{0.14}NbSe_2$, which only display this behavior below 40 K. Importantly, the semiconducting nature of $Fe_{0.17}ZrSe_2$ raises the prospect of strongly modulating or switching these magnetic phases with an electrical bias or optical excitation. More generally, exploring exchange bias in systems like $Fe_{0.17}ZrSe_2$ offers valuable insights into the role of disorder and frustration in magnetic materials more widely, including prototypical metallic magnetic multilayers. This work also opens doors to new design strategies for magnetic film interfaces that leverage materials with inhomogeneous microscopic environments to induce frustration and support exchange bias.

## Associated Content

Supporting Information

Materials and methods, experimental procedures and setup for synthesis and characterization, characterization data, and additional discussion.

## Acknowledgments

The authors thank Dr. N. Settineri of the UC Berkeley CHEXRAY facility for assistance with crystal selection, single crystal diffraction data collection and reduction. This material is based upon work supported by the Air Force Office of Scientific Research under AFOSR Award no. FA9550-20-1-0007. C.J.K. acknowledges support from the NIH (NRSA fellowship award #F32GM142218). R.A.M. is supported as part of the Center for Molecular Quantum Transduction (CMQT), an Energy Frontier Research Center funded by the U.S. Department of Energy, Office of Science, Basic Energy Sciences, under Award No. DE-SC0021314. S.H. acknowledges support from the Blavatnik Innovation Fellowship. L.S.X. acknowledges support from the Arnold and Mabel Beckman Foundation (Award No. 51532) for a postdoctoral fellowship. The XAS data were collected at the Stanford Synchrotron Radiation Lightsource (SSRL) BL 4-3, under the Chemical and Materials Sciences to Advance Clean Energy Technologies and Low-Carbon Manufacturing program funded by the U.S. Department of Energy, Office of Science, Office of Basic Energy Sciences, Chemical Sciences, Geosciences, and Biosciences Division, under contract no. DE-AC0205CH11231 (JY). Use of the SSRL, SLAC National Accelerator Laboratory, is supported by the U.S. Department of Energy, Office of Science, Office of Basic Energy Sciences under Contract No. DE-AC02-76SF00515. Experimental and theoretical work at the Molecular Foundry, LBNL was supported by the Office of Science, Office of Basic Energy Sciences, of the U.S. Department of Energy under Contract No. DE-AC02-05CH11231. Confocal Raman spectroscopy and photoluminescence spectroscopy was supported by a DURIP grant through the Office of Naval Research under Award no. N00014-20-1-2599 (D.K.B.). Other instrumentation used in this work was supported by grants from the W.M. Keck Foundation (Award # 993922), the Canadian Institute for Advanced Research (CIFAR–Azrieli Global Scholar, Award # GS21-011), the Gordon and Betty Moore Foundation EPiQS Initiative (Award #10637), and the 3M Foundation through the 3M Non-Tenured Faculty Award (#67507585). The CHEXRAY facility at UC Berkeley is supported by NIH Shared Instrumentation Grant S10-RR027172.

## References

1. Spaldin, N. A. Ferromagnetic domains. In *Magnetic Materials: Fundamentals and Applications*, 2 ed.; Cambridge University Press: Cambridge, 2010; pp 79–95.
2. Nogues, J.; Schuller, I. K. Exchange bias. *J. Magn. Magn. Mater.* **1999**, *192*, 203–232.
3. Meiklejohn, W. H.; Bean, C. P. New Magnetic Anisotropy. *Phys. Rev.* **1956**, *102*, 1413–1414.
4. Kools, J. C. S. Exchange-biased spin-valves for magnetic storage. *IEEE Trans. Magn.* **1996**, *32*, 3165–3184.
5. Gider, S.; Runge, B. U.; Marley, A. C.; Parkin, S. S. P. The magnetic stability of spin-dependent tunneling devices. *Science* **1998**, *281*, 797–799.
6. Skumryev, V.; Stoyanov, S.; Zhang, Y.; Hadjipanayis, G.; Givord, D.; Nogues, J. Beating the superparamagnetic limit with exchange bias. *Nature* **2003**, *423*, 850–853.
7. Parkin, S. S. P.; Roche, K. P.; Samant, M. G.; Rice, P. M.; Beyers, R. B.; Scheuerlein, R. E.; O'Sullivan, E. J.; Brown, S. L.; Bucchigano, J.; Abraham, D. W.; Lu, Y.; Rooks, M.; Trouilloud, P. L.; Wanner, R. A.; Gallagher, W. J. Exchange-biased magnetic tunnel junctions and application to nonvolatile magnetic random access memory *J. Appl. Phys.* **1999**, *85*, 5828–5833.
8. Mauri, D.; Siegmann, H. C.; Bagus, P. S.; Kay, E. Simple-Model for Thin Ferromagnetic-Films Exchange Coupled to an Antiferromagnetic Substrate. *J. Appl. Phys.* **1987**, *62*, 3047–3049.
9. Malozemoff, A. P. Random-Field Model of Exchange-Anisotropy at Rough Ferromagnetic-Antiferromagnetic Interfaces. *Phys. Rev. B* **1987**, *35*, 3679–3682.
10. Koon, N. C. Calculations of exchange bias in thin films with ferromagnetic/antiferromagnetic interfaces. *Phys. Rev. Lett.* **1997**, *78*, 4865–4868.
11. Schulthess, T. C.; Butler, W. H. Consequences of spin-flop coupling in exchange biased films. *Phys. Rev. Lett.* **1998**, *81*, 4516–4519.
12. O'Grady, K.; Fernandez-Outon, L. E.; Vallejo-Fernandez, G. A new paradigm for exchange bias in polycrystalline thin films. *J. Magn. Magn. Mater.* **2010**, *322*, 883–899.
13. Radu, F.; Abrudan, R.; Radu, I.; Schmitz, D.; Zabel, H. Perpendicular exchange bias in ferrimagnetic spin valves. *Nat. Commun.* **2012**, *3*, 715.
14. Cain, W. C.; Kryder, M. H. Investigation of the Exchange Mechanism in NiFe-TbCo Bilayers. *J. Appl. Phys.* **1990**, *67*, 5722–5724.
15. Vanderzaag, P. J.; Wolf, R. M.; Ball, A. R.; Bordel, C.; Feiner, L. F.; Jungblut, R. A Study of the Magnitude of Exchange Biasing in [111] $Fe_3O_4$/CoO Bilayers. *J. Magn. Magn. Mater.* **1995**, *148*, 346–348.
16. Spaldin, N. A. Exchange bias. In *Magnetic Materials: Fundamentals and Applications*, 2 ed.; Cambridge University Press: Cambridge, 2010; pp 169–174.
17. Kiwi, M. Exchange bias theory. *J. Magn. Magn. Mater.* **2001**, *234*, 584–595.
18. Ali, M.; Adie, P.; Marrows, C. H.; Greig, D.; Hickey, B. J.; Stamps, R. L. Exchange bias using a spin glass. *Nat. Mater.* **2007**, *6*, 70–75.
19. Malozemoff, A. P. Mechanisms of Exchange-Anisotropy. *J. Appl. Phys.* **1988**, *63*, 3874–3879.
20. de Almeida, J. R. L.; Rezende, S. M. Microscopic model for exchange anisotropy. *Phys. Rev. B* **2002**, *65*, 092412.
21. Ali, M.; Marrows, C. H.; Hickey, B. J. Controlled enhancement or suppression of exchange biasing using impurity delta layers. *Phys. Rev. B* **2008**, *77*, 134401
22. Ding, J. F.; Lebedev, O. I.; Turner, S.; Tian, Y. F.; Hu, W. J.; Seo, J. W.; Panagopoulos, C.; Prellier, W.; Van Tendeloo, G.; Wu, T. Interfacial spin glass state and exchange bias in manganite bilayers with competing magnetic orders. *Phys. Rev. B* **2013**, *87*, 054428.
23. Monod, P.; Prejean, J. J.; Tissier, B. Magnetic Hysteresis of CuMn in the Spin-Glass State. *J. Appl. Phys.* **1979**, *50*, 7324–7329.
24. Maniv, E.; Murphy, R. A.; Haley, S. C.; Doyle, S.; John, C.; Maniv, A.; Ramakrishna, S. K.; Tang, Y. L.; Ercius, P.; Ramesh, R.; Reyes, A. P.; Long, J. R.; Analytis, J. G. Exchange bias due to coupling between coexisting antiferromagnetic and spin-glass orders. *Nat. Phys.* **2021**, *17*, 525–530.

25. Murphy, R. A.; Darago, L. E.; Ziebel, M. E.; Peterson, E. A.; Zaia, E. W.; Mara, M. W.; Lussier, D.; Velasquez, E. O.; Shuh, D. K.; Urban, J. J.; Neaton, J. B.; Long, J. R. Exchange Bias in a Layered Metal-Organic Topological Spin Glass. *ACS Cent. Sci.* **2021**, *7*, 1317–1326.
26. Lee, M. Disordered exchange is biased. *Nat. Phys.* **2021**, *17*, 434–435.
27. Tan, S. J. R.; Abdelwahab, I.; Ding, Z. J.; Zhao, X. X.; Yang, T. S.; Loke, G. Z. J.; Lin, H.; Verzhbitskiy, I.; Poh, S. M.; Xu, H.; Nai, C. T.; Zhou, W.; Eda, G.; Jia, B. H.; Loh, K. P. Chemical Stabilization of 1T' Phase Transition Metal Dichalcogenides with Giant Optical Kerr Nonlinearity. *J. Am. Chem. Soc.* **2017**, *139*, 2504–2511.
28. Wang, N. Z.; Shi, M. Z.; Shang, C.; Meng, F. B.; Ma, L. K.; Luo, X. G.; Chen, X. H. Tunable superconductivity by electrochemical intercalation in $TaS_2$. *New J. Phys.* **2018**, *20*, 023014.
29. Friend, R. H.; Beal, A. R.; Yoffe, A. D. Electrical and magnetic properties of some first row transition metal intercalates of niobium disulphide. *Philos. Mag.* **1977**, *35*, 1269–1287.
30. Rao, G. V. S.; Shafer, M. W. Intercalation in Layered Transition Metal Dichalcogenides. In *Intercalated Layered Materials*, Lévy, F., Ed. Springer Netherlands: Dordrecht, 1979; pp 99–199.
31. Husremovic, S.; Groschner, C. K.; Inzani, K.; Craig, I. M.; Bustillo, K. C.; Ercius, P.; Kazmierczak, N. P.; Syndikus, J.; Van Winkle, M.; Aloni, S.; Taniguchi, T.; Watanabe, K.; Griffin, S. M.; Bediako, D. K. Hard Ferromagnetism Down to the Thinnest Limit of Iron-Intercalated Tantalum Disulfide. *J. Am. Chem. Soc.* **2022**, 12167–12176.
32. Morosan, E.; Zandbergen, H. W.; Li, L.; Lee, M.; Checkelsky, J. G.; Heinrich, M.; Siegrist, T.; Ong, N. P.; Cava, R. J. Sharp switching of the magnetization in $Fe_{1/4}TaS_2$. *Phys. Rev. B* **2007**, *75*, 104401.
33. Checkelsky, J. G.; Lee, M.; Morosan, E.; Cava, R. J.; Ong, N. P. Anomalous Hall effect and magnetoresistance in the layered ferromagnet $Fe_{1/4}TaS_2$: The inelastic regime. *Phys. Rev. B* **2008**, *77*, 014433.
34. Nair, N. L.; Maniv, E.; John, C.; Doyle, S.; Orenstein, J.; Analytis, J. G. Electrical switching in a magnetically intercalated transition metal dichalcogenide. *Nat. Mater.* **2020**, *19*, 153–157.
35. Maniv, E.; Nair, N. L.; Haley, S. C.; Doyle, S.; John, C.; Cabrini, S.; Maniv, A.; Ramakrishna, S. K.; Tang, Y. L.; Ercius, P.; Ramesh, R.; Tserkovnyak, Y.; Reyes, A. P.; Analytis, J. G. Antiferromagnetic switching driven by the collective dynamics of a coexisting spin glass. *Sci. Adv.* **2021**, *7*, eabd8452.
36. Kuroiwa, Y.; Nishimura, M.; Nakajima, R.; Abe, H.; Noda, Y. Short-Range Order and Long-Range Order of Fe Atoms in a Spin-Glass Phase and a Cluster-Glass Phase of Intercalation Compounds $Fe_xTiS_2$. *J. Phys. Soc. Jpn.* **1994**, *63*, 4278–4281.
37. Xie, L. S.; Husremovic, S.; Gonzalez, O.; Craig, I. M.; Bediako, D. K. Structure and Magnetism of Iron- and Chromium-Intercalated Niobium and Tantalum Disulfides. *J. Am. Chem. Soc.* **2022**, *144*, 9525–9542.
38. Buhannic, M. A.; Colombet, P.; Danot, M. Spin-Glass Behavior of Iron-Intercalated Zirconium Disulfide. *Solid State Commun.* **1986**, *59*, 77–79.
39. Buhannic, M. A.; Danot, M.; Colombet, P.; Dordor, P.; Fillion, G. Thermopower and Low-DC-Field Magnetization Study of the Layered $Fe_xZrSe_2$ Compounds - Anderson-Type Localization and Anisotropic Spin-Glass Behavior. *Phys. Rev. B* **1986**, *34*, 4790–4795.
40. Mizokawa, T.; Fujimori, A. p-d exchange interaction for 3d transition-metal impurities in II-VI semiconductors. *Phys. Rev. B* **1997**, *56*, 6669–6672.
41. Dietl, T.; Ohno, H.; Matsukura, F.; Cibert, J.; Ferrand, D. Zener model description of ferromagnetism in zinc-blende magnetic semiconductors. *Science* **2000**, *287*, 1019–1022.
42. Ruderman, M. A.; Kittel, C. Indirect Exchange Coupling of Nuclear Magnetic Moments by Conduction Electrons. *Phys. Rev.* **1954**, *96*, 99–102.
43. Kasuya, T. A Theory of Metallic Ferro- and Antiferromagnetism on Zener's Model. *Prog. Theor. Phys.* **1956**, *16*, 45–57.
44. Yosida, K. Magnetic Properties of Cu-Mn Alloys. *Phys. Rev.* **1957**, *106*, 893–898.
45. Hillenius, S. J.; Coleman, R. V. Magnetic-Susceptibility of Iron-Doped 2H-$NbSe_2$. *Phys. Rev. B* **1979**, *20*, 4569–4576.

46. Garvin, J. F.; Morris, R. C. Transport-Properties and Magnetic-Ordering in Iron-Doped $NbSe_2$. *Phys. Rev. B* **1980**, *21*, 2905–2914.
47. Whitney, D. A.; Fleming, R. M.; Coleman, R. V. Magnetotransport and Superconductivity in Dilute Fe Alloys of $NbSe_2$, $TaSe_2$, and $TaS_2$. *Phys. Rev. B* **1977**, *15*, 3405–3423.
48. Telford, E. J.; Dismukes, A. H.; Dudley, R. L.; Wiscons, R. A.; Lee, K.; Chica, D. G.; Ziebel, M. E.; Han, M. G.; Yu, J.; Shabani, S.; Scheie, A.; Watanabe, K.; Taniguchi, T.; Xiao, D.; Zhu, Y. M.; Pasupathy, A. N.; Nuckolls, C.; Zhu, X. Y.; Dean, C. R.; Roy, X. Coupling between magnetic order and charge transport in a two-dimensional magnetic semiconductor. *Nat. Mater.* **2022**, *21*, 754–760.
49. Bediako, D. K.; Rezaee, M.; Yoo, H.; Larson, D. T.; Zhao, S. Y. F.; Taniguchi, T.; Watanabe, K.; Brower-Thomas, T. L.; Kaxiras, E.; Kim, P. Heterointerface effects in the electrointercalation of van der Waals heterostructures. *Nature* **2018**, *558*, 425–429.
50. Yamada, Y.; Ueno, K.; Fukumura, T.; Yuan, H. T.; Shimotani, H.; Iwasa, Y.; Gu, L.; Tsukimoto, S.; Ikuhara, Y.; Kawasaki, M. Electrically Induced Ferromagnetism at Room Temperature in Cobalt-Doped Titanium Dioxide. *Science* **2011**, *332*, 1065–1067.
51. Zhang, J. S.; Yang, A. K.; Wu, X.; van de Groep, J.; Tang, P. Z.; Li, S. R.; Liu, B. F.; Shi, F. F.; Wan, J. Y.; Li, Q. T.; Sun, Y. M.; Lu, Z. Y.; Zheng, X. L.; Zhou, G. M.; Wu, C. L.; Zhang, S. C.; Brongersma, M. L.; Li, J.; Cui, Y. Reversible and selective ion intercalation through the top surface of few-layer $MoS_2$. *Nat. Commun.* **2018**, *9*, 5289.
52. Wiedemeier, H.; Goldman, H. Mass-Transport and Crystal-Growth of the Mixed $ZrS_2$-$ZrSe_2$ System. *J. Less-Common Met.* **1986**, *116*, 389–399.
53. Mao, Q. H.; Geng, X. D.; Yang, J. F.; Li, R. X.; Hao, H. S.; Yang, J. H.; Wang, H. D.; Xu, B. J.; Fang, M. H. Synthesis and semiconducting behavior of Zr (Cu, Fe) $Se_2$-delta single crystals. *Physica B Condens. Matter* **2019**, *567*, 1–4.
54. Lebail, P.; Colombet, P.; Rouxel, J. Synthesis and Properties of New Intercalates $Eu_xZrSe_{1.95}$. *Solid State Ionics* **1989**, *34*, 127–134.
55. Dahn, J. R.; Mckinnon, W. R.; Levyclement, C. Lithium Intercalation in $Li_xZrSe_2$. *Solid State Commun.* **1985**, *54*, 245–248.
56. Manas-Valero, S.; Garcia-Lopez, V.; Cantarero, A.; Galbiati, M. Raman Spectra of $ZrS_2$ and $ZrSe_2$ from Bulk to Atomically Thin Layers. *Appl. Sci. (Basel)* **2016**, *6*, 264.
57. Koyano, M.; Watanabe, H.; Yamamura, Y.; Tsuji, T.; Katayama, S. Magnetic and Raman scattering studies on intercalation compounds $Fe_xNbS_2$. *Mol. Cryst. Liq. Cryst.* **2000**, *341*, 837–842.
58. Mignuzzi, S.; Pollard, A. J.; Bonini, N.; Brennan, B.; Gilmore, I. S.; Pimenta, M. A.; Richards, D.; Roy, D. Effect of disorder on Raman scattering of single-layer $MoS_2$. *Phys. Rev. B* **2015**, *91*, 195411.
59. Wang, H. Z.; Zhao, L. Y.; Zhang, H.; Liu, Y. S.; Yang, L.; Li, F.; Liu, W. H.; Dong, X. T.; Li, X. K.; Li, Z. H.; Qi, X. D.; Wu, L. Y.; Xu, Y. F.; Wang, Y. Q.; Wang, K. K.; Yang, H. C.; Li, Q.; Yan, S. S.; Zhang, X. G.; Li, F.; Li, H. S. Revealing the multiple cathodic and anodic involved charge storage mechanism in an $FeSe_2$ cathode for aluminium-ion batteries by in situ magnetometry. *Energy Environ. Sci.* **2022**, *15*, 311–319.
60. Yang, S. H.; Lee, Y. J.; Kang, H.; Park, S. K.; Kang, Y. C. Carbon-Coated Three-Dimensional MXene/Iron Selenide Ball with Core-Shell Structure for High-Performance Potassium-Ion Batteries. *Nano-Micro Letters* **2022**, *14*, 1–17.
61. Whitman, L. J.; Stroscio, J. A.; Dragoset, R. A.; Celotta, R. J. Geometric and Electronic-Properties of Cs Structures on III-V (110) Surfaces - from 1d and 2d Insulators to 3d Metals. *Phys. Rev. Lett.* **1991**, *66*, 1338–1341.
62. Mydosh, J. A. Spin glasses: redux: an updated experimental/materials survey. *Rep. Prog. Phys.* **2015**, *78*, 052501.
63. Korenblit, I. Y.; Shender, E. F. Spin-Glasses and Nonergodicity. *Sov. Phys. Usp.* **1989**, *157*, 267–310.
64. Binder, K.; Kob, W. *Glassy materials and disordered solids: an introduction to their statistical mechanics*; Rev. ed.; World Scientific, Hackensack, NJ ; London ; Singapore, 2011.
65. Bedanta, S.; Kleemann, W. Supermagnetism. *J. Phys. D: Appl. Phys.* **2009**, *42*, 052501.

66. Kroder, J.; Gooth, J.; Schnelle, W.; Fecher, G. H.; Felser, C. Observation of spin glass behavior in chiral $Mn_{48}Fe_{34}Si_{18}$ with a β-Mn related structure. *AIP Adv.* **2019**, *9*, 055327.
67. Pakhira, S.; Mazumdar, C.; Ranganathan, R.; Giri, S.; Avdeev, M. Large magnetic cooling power involving frustrated antiferromagnetic spin-glass state in $R_2NiSi_3$ (R = Gd,Er). *Phys. Rev. B* **2016**, *94*, 104414.
68. Ramirez, A. P. Strongly Geometrically Frustrated Magnets. *Annu. Rev. Mater. Sci.* **1994**, *24*, 453–480.
69. Mydosh, J. A. *Spin glasses: an experimental introduction*; 1st ed.; Taylor & Francis, London ; Washington, DC, 1993.
70. Mitchler, P. D.; Roshko, R. M.; Ruan, W. Non-equilibrium relaxation dynamics in the spin glass and ferromagnetic phases of CrFe. *Philos. Mag. B* **1993**, *68*, 539–550.
71. Burns, R. G.; Fyfe, W. S. Site of Preference Energy and Selective Uptake of Transition-Metal Ions from Magma. *Science* **1964**, *144*, 1001–1003.
72. Cotton, F. A.; Wilkinson, G. *Advanced Inorganic Chemistry*; 2nd ed.; John Wiley and Sons, Inc., 1966.
73. Gruyters, M. Spin-glass-like behavior in CoO nanoparticles and the origin of exchange bias in layered CoO/Ferromagnet structures. *Phys. Rev. Lett.* **2005**, *95*, 077204.
74. Chen, Q.; Zhang, Z. J. Size-dependent superparamagnetic properties of $MgFe_2O_4$ spinel ferrite nanocrystallites. *Appl. Phys. Lett.* **1998**, *73*, 3156–3158.
75. Kumar, A.; Kaushik, S. D.; Siruguri, V.; Pandey, D. Evidence for two spin-glass transitions with magnetoelastic and magnetoelectric couplings in the multiferroic $(Bi_{1-x}Ba_x)(Fe_{1-x}Ti_x)O_3$ system. *Phys. Rev. B* **2018**, *97*, 104402.
76. Erodici, M. P.; Mai, T. T.; Xie, L. S.; Li, S.; Fender, S. S.; Husremović, S.; Gonzalez, O.; Hight Walker, A. R.; Bediako, D. K. Bridging Structure, Magnetism, and Disorder in Iron-Intercalated Niobium Diselenide, $FexNbSe_2$, below x = 0.25. *J. Phys. Chem. C* **2023**, *127*, 9787–9795.
77. Srivastava, J. K.; Hammann, J.; Asai, K.; Katsumata, K. Magnetic Hysteresis Behavior of Anisotropic Spin-Glass $Fe_2TiO_5$. *Phys. Lett. A* **1990**, *149*, 485–487.
78. Hu, J. G.; Jin, G. J.; Hu, A.; Ma, Y. Q. Temperature dependence of exchange bias and coercivity in ferromagnetic/antiferromagnetic bilayers. *Eur. Phys. J. B* **2004**, *40*, 265–271.
79. Wong, P.; Vonmolnar, S.; Palstra, T. T. M.; Mydosh, J. A.; Yoshizawa, H.; Shapiro, S. M.; Ito, A. Coexistence of Spin-Glass and Antiferromagnetic Orders in the Ising System $Fe_{0.55}Mg_{0.45}Cl_2$. *Phys. Rev. Lett.* **1985**, *55*, 2043–2046.
80. Chillal, S.; Thede, M.; Litterst, F. J.; Gvasaliya, S. N.; Shaplygina, T. A.; Lushnikov, S. G.; Zheludev, A. Microscopic coexistence of antiferromagnetic and spin-glass states. *Phys. Rev. B* **2013**, *87*, 220403.
81. Kleemann, W.; Shvartsman, V. V.; Borisov, P.; Kania, A. Coexistence of Antiferromagnetic and Spin Cluster Glass Order in the Magnetoelectric Relaxor Multiferroic $PbFe_{0.5}Nb_{0.5}O_3$. *Phys. Rev. Lett.* **2010**, *105*, 257202.
82. Young, A. P. *Spin Glasses and Random Fields*; ed.; World Scientific Publishing Co. Pte. Ltd., 1998.
83. Liao, X.; Wei, S.; Wang, Y.; Wang, D.; Wu, K.; Liang, H.; Yang, S.; Svedlindh, P.; Zeng, Y.-J. Large Exchange Bias Triggered by Transition Zone of Spin Glass. *J. Adv. Phys.* **2023**, *2*, 2200043.
84. Shen, H.; Feng, P.; Jiang, G. C.; Wu, A. H. Tunable exchange bias in hard/soft $SmMnO_3$/α-$Mn_2O_3$ composite with AFM structures. *Physica B Condens. Matter* **2020**, *585*, 412084.
85. Borgohain, C.; Mishra, D.; Sarma, K. C.; Phukan, P. Exchange bias effect in $CoFe_2O_4$-$Cr_2O_3$ nanocomposite embedded in $SiO_2$ matrix. *J. Appl. Phys.* **2012**, *112*, 113905.
86. Nowak, U.; Usadel, K. D.; Keller, J.; Miltenyi, P.; Beschoten, B.; Guntherodt, G. Domain state model for exchange bias. I. Theory. *Phys. Rev. B* **2002**, *66*, 014430.
87. Usadel, K. D.; Nowak, U. Exchange bias for a ferromagnetic film coupled to a spin glass. *Phys. Rev. B* **2009**, *80*, 014418.

**TOC graphic:**

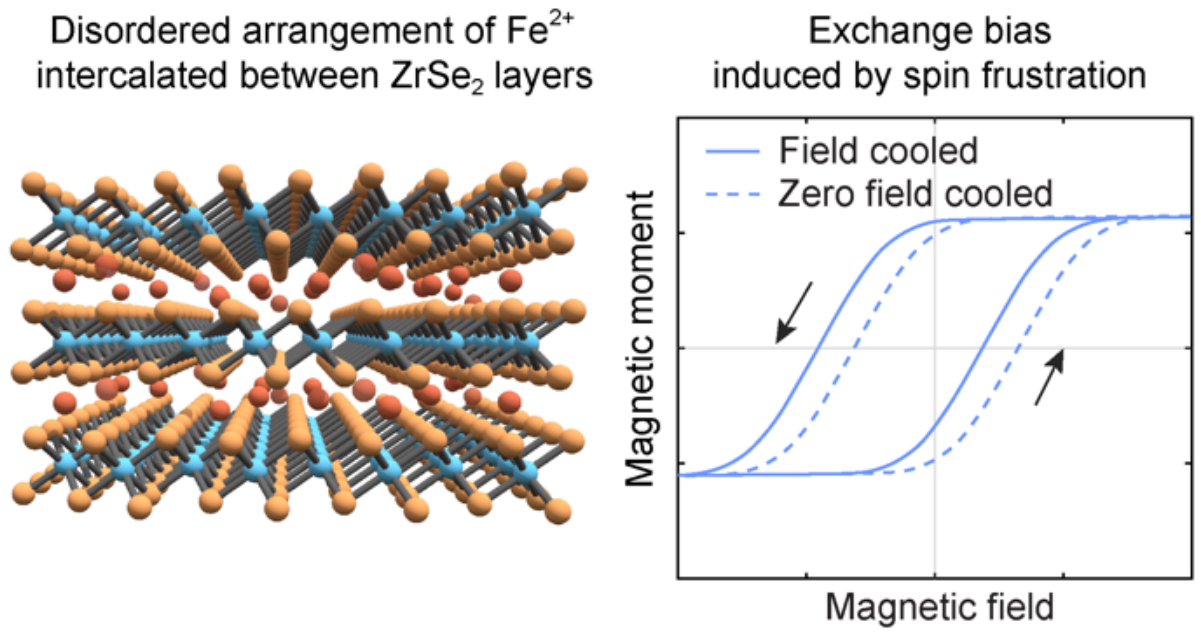

Disordered arrangement of Fe2+
intercalated between ZrSe2 layers
Exchange bias
induced by spin frustration
Field cooled
Zero field cooled
Magnetic moment
Magnetic field

*Supporting Information for*

# Near room-temperature intrinsic exchange bias in an Fe intercalated $ZrSe_2$ spin glass

Zhizhi Kong[1], Corey J. Kaminsky[2], Catherine K. Groschner[1], Ryan A. Murphy[1], Yun Yu[1], Samra Husremović[1], Lilia S. Xie[1], Matthew P. Erodici[1], R. Soyoung Kim[3], Junko Yano[2], and D. Kwabena Bediako[1,3]*

[1] *Department of Chemistry, University of California, Berkeley, California 94720, United States*

[2]*Molecular Biophysics and Integrated Bioimaging Division, Lawrence Berkeley National Laboratory, Berkeley, CA 94720, USA*

[3]*Chemical Sciences Division, Lawrence Berkeley National Laboratory, Berkeley, CA 94720, USA*

**Correspondence to: bediako@berkeley.edu*

**Table of Contents**

## S1. Chemicals

Unless otherwise stated, all materials and reagents were used as received.

The following chemicals were used for the chemical vapor transport (CVT) growth of crystalline iron intercalated zirconium diselenide ($Fe_xZrSe_2$). Iron powder (spherical, <10 μm, purity >99.9%, metal basis), and selenium powder (200 mesh, purity 99.999%, metals basis) were purchased from Alfa Aesar (Ward Hill, MA. USA). Zirconium pellets (catalog # ZR39X88-10G, purity >99.9%) was purchased from R.D. Mathis Co. (Long Beach, CA, USA). Iodine (catalog # 229695-20G, purity >99.999%, trace metals basis) was purchased from Sigma Aldrich Co., LLC. (St. Louis, MO, USA). $ZrSe_2$ single crystals were purchased from HQ graphene (Groningen, The Netherlands). The as-grown $Fe_xZrSe_2$ flakes were washed using toluene in an Ar-filled glove box. Glassware was oven-dried at a temperature of 150 °C for ≥ 4 h, and allowed to cool in an evacuated glove box antechamber prior to use. Anhydrous toluene was stored over 3- or 4-Å molecular sieves prior to use.

## S2. Chemical vapor transport growth of $Fe_xZrSe_2$

$Fe_xZrSe_2$ was synthesized following procedures reported previously.[5] A mixture of 70.9 mg (0.778 mmol) of zirconium pellet, 123.0 mg (1.557 mmol) selenium powder, 24 mg (0.429 mmol) Fe powder along with transport agent 40.2 mg (0.158 mmol) of $I_2$ was loaded into 30 cm long quartz ampoule in atmospheric conditions. The ampoule was then evacuated to $1 \times 10^{-2}$ Torr. The contents of ampoule were then cooled using liquid nitrogen to minimize the vaporization of $I_2$. When the pressure dropped below $1 \times 10^{-4}$ Torr, the ampoule was sealed, and placed in a 3-zone furnace. The temperature settings for hot zone and cold zone are shown in **Figure S1a.** The heating and cooling rate was around 55 °C/h. **Figure S1c** shows that high-quality hexagonal crystals with diameters of several millimeters were obtained. SEM-EDS measurements (**Figure S2**) were used to determine elemental composition as Section S1.2

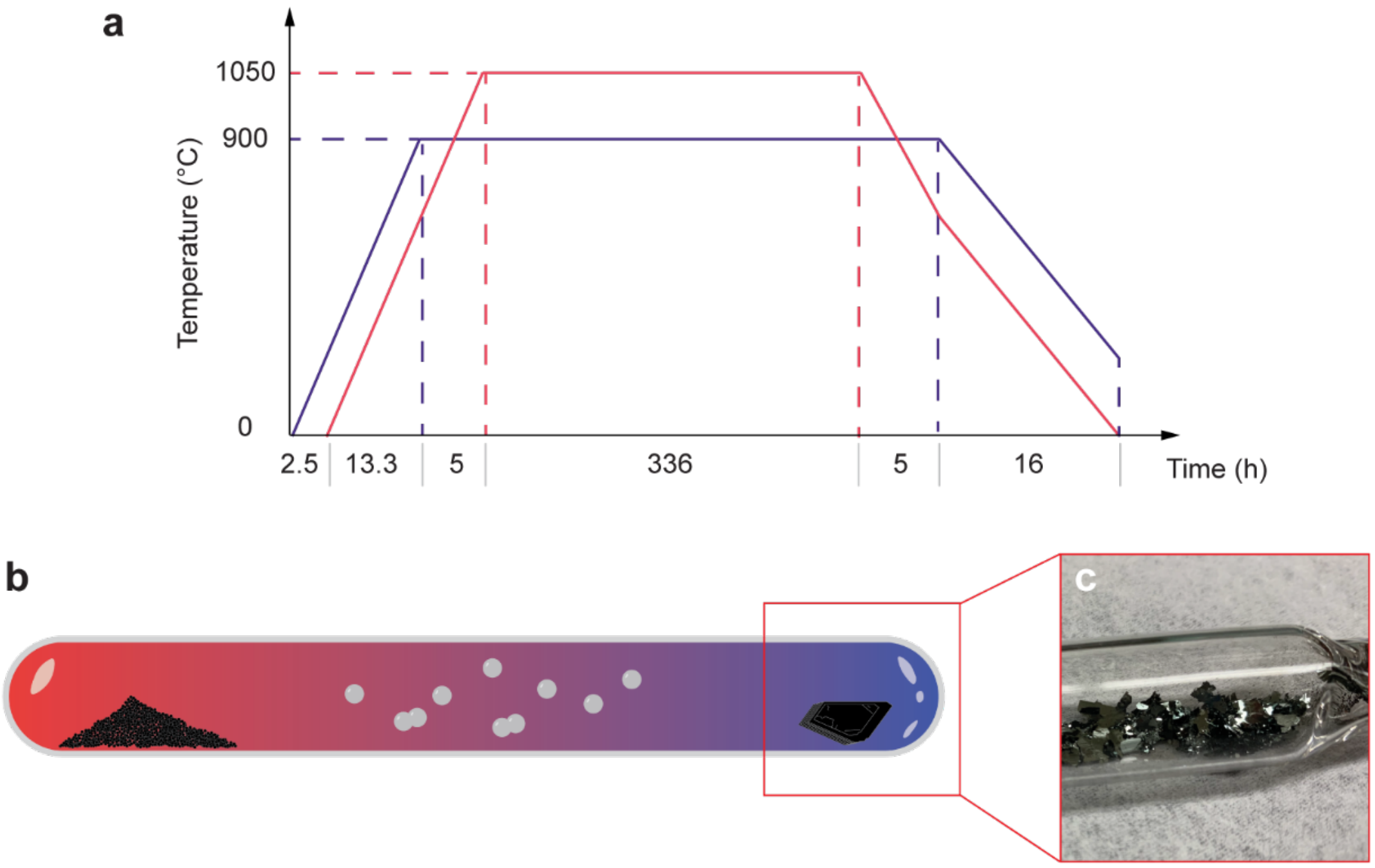

**Figure S1.** (a) Temperature settings of source zone (red line) and growth zone (purple line) for $Fe_xZrSe_2$ single crystal growth. (b) Schematic of a typical CVT experiment for bulk crystal growth of TMDs. Source materials, Zr pellets, Fe powder, and Se powder as well as transport agent $I_2$ are placed in one side of ampoule, which is evacuated, sealed, and positioned in a 3-zone furnace with a temperature gradient. (b) The image of as-synthesized iron intercalated $ZrSe_2$ crystals.

## S3. Scanning electron microscopy and energy dispersive x-ray spectroscopy

Energy Dispersive X-ray Spectroscopy (EDS) measurements were performed on the crystal to extract the stoichiometry of iron intercalated between the $ZrSe_2$ layers. Scanning electron microscopy images were recorded on a Thermo Scientific Scios 2 DualBeam scanning electron microscope (SEM). Energy dispersive x-ray spectroscopy (EDS) of the $Fe_xZrSe_2$ crystals was performed with an Oxford Symmetry EDX detector using 20 keV accelerating voltage. Elemental compositions were determined by integrating under the characteristic spectrum peaks for each element using the AZtecLive Platform (Oxford Instruments NanoAnalysis).

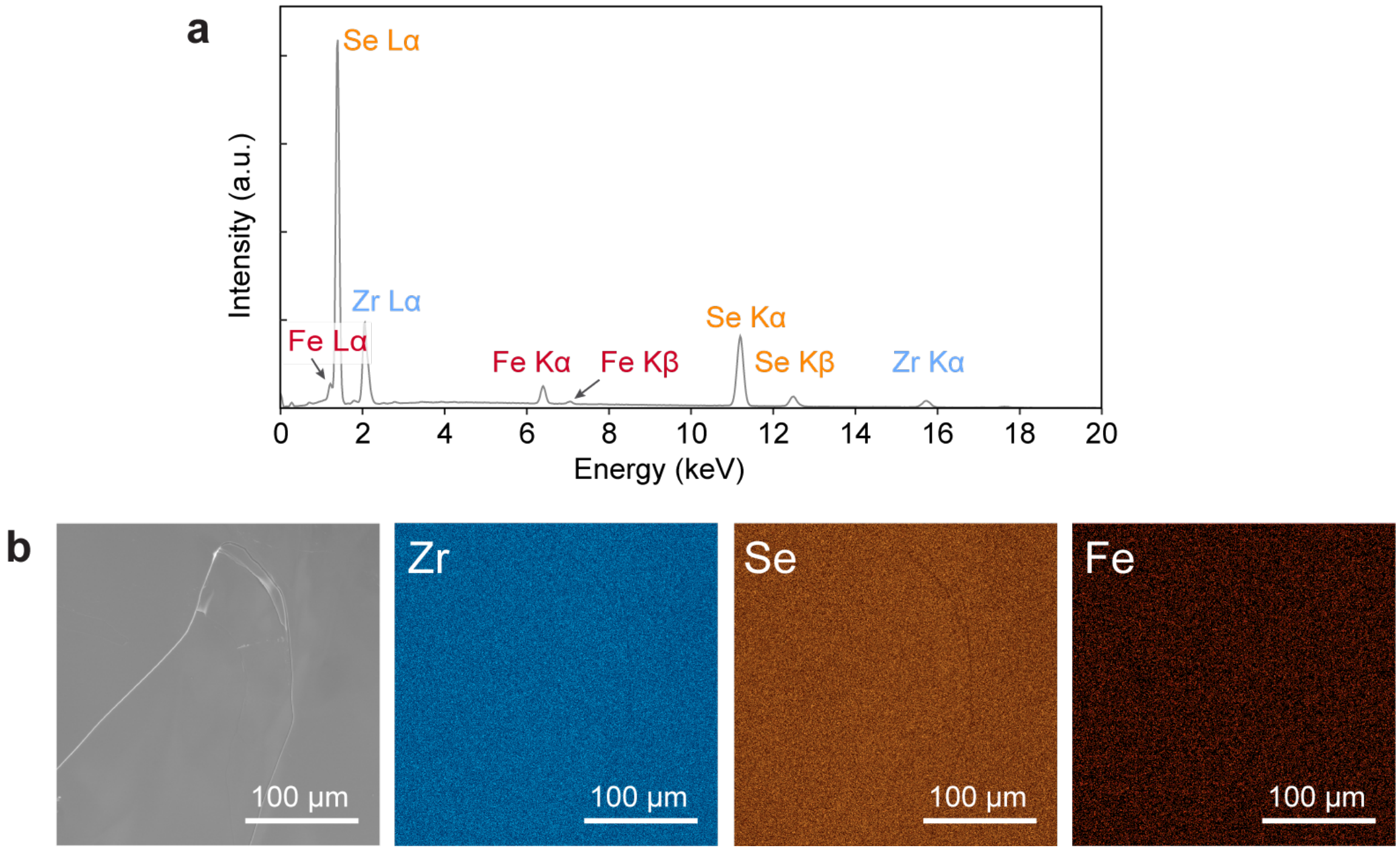

**Figure S2.** EDS measurement of the Fe-intercalated $ZrSe_2$ is presented. (a) The dispersion spectroscopy give an atomic ratio (Se:Zr:Fe) of 2.02:1:0.17. (b) SEM image and SEM-EDX mapping for elements of Zr, Se, and Fe for an as-grown $Fe_{0.17}ZrSe_2$ flake. The elemental maps reveal a uniform spatial distribution of Fe, Zr, and Se over a large area of approximately 90,000 $\mu m^2$.

**S4. Transmission electron microscopy and selected area electron diffraction**

TEM samples for imaging along the c-axis:

A single crystal of $Fe_{0.17}ZrSe_2$ was mechanically exfoliated with Blue Tape (Ultron Systems Inc.) and polydimethylsiloxane (PDMS) films (Gel-Pak) and transferred onto a NORCADA Location Tagged Micro-Porous TEM grid with 0.50 mm × 0.50 mm, 200 nm thick $SiN_x$ membrane with 2-μm holes. Exfoliated flakes with thicknesses < 50 nm were identified using optical microscopy and atomic force microscopy. The TEM grid was cleaned in a vacuum annealer at $2.1\times10^{-6}$ torr, 350 °C for 30 min immediately prior to stacking the sample onto the TEM grid.

Selected area electron diffraction (SAED) was performed with FEI TitanX microscope at the National Center for Electron Microscopy (NCEM) at the Molecular Foundry, Lawrence Berkeley National Laboratory. The SAED patterns were obtained at 60 kV applied voltage, collected along the $[10\bar{1}0]$ zone axis of the $Fe_{0.17}ZrSe_2$ flakes. The obtained patterns correspond to a 720 nm sample region, defined using a 40 μm diameter SAED aperture.

Cross-sectional samples for imaging along the *ab*-plane:

The cross-sectional TEM samples were prepared following procedures previously reported.[6] In brief, cross-sectional TEM samples were prepared with Thermo Scientific Scios 2 focused ion beam (FIB) scanning electron microscopy (FIB-SEM) systems. $Fe_{0.17}ZrSe_2$ flakes were exfoliated using Blue Tape and PDMS film onto a $SiO_2$/Si substrate. A 200 nm coating of Pt or C was deposited over the target $Fe_{0.17}ZrSe_2$ flake using an electron beam at 5 kV and 1.6 nA. This was followed by a deposition of 2.5 μm Pt or C layer using a gallium-ion beam at 30 kV and 0.1 nA. The initial bulk-out was performed with a 30 kV Ga beam and 3 nA current, while the bulk-out cleaning was performed with a 1 nA current. For the initial thinning, a 30 kV Ga beam was used, while the current was increasingly lowered as the sample became thinner. Initial thinning started with a 1 nA current, while the final current for the rough thinning was typically 0.3 nA. After the sample was approximately 150 nm thick, the FIB beam was switched to the 5 kV and 40 pA settings. The final polishing was performed at 2 kV and 40 pA to reduce the ion beam damage.

High-angle annular dark-field scanning transmission electron microscopy (HAADF-STEM) of cross-sectional TEM samples was performed on transmission electron aberration-corrected microscope (TEAM0.5) microscope at NCEM. The microscope was operated at 300 kV with a probe convergence angle of 30 mrad.

A single crystal of $Fe_{0.17}ZrSe_2$ was mechanically exfoliated with blue tape and transferred onto PDMS stamp/glass slide. Exfoliated flakes with thicknesses < 50 nm were identified using optical microscopy. The target flake was dropped onto a TEM grid with 200-nm-thick amorphous $SiN_x$ membrane. Only diffraction peaks corresponding to the host lattice 1*T*-$ZrSe_2$ were observed while the diffraction peaks corresponding to ordered superlattices were not.

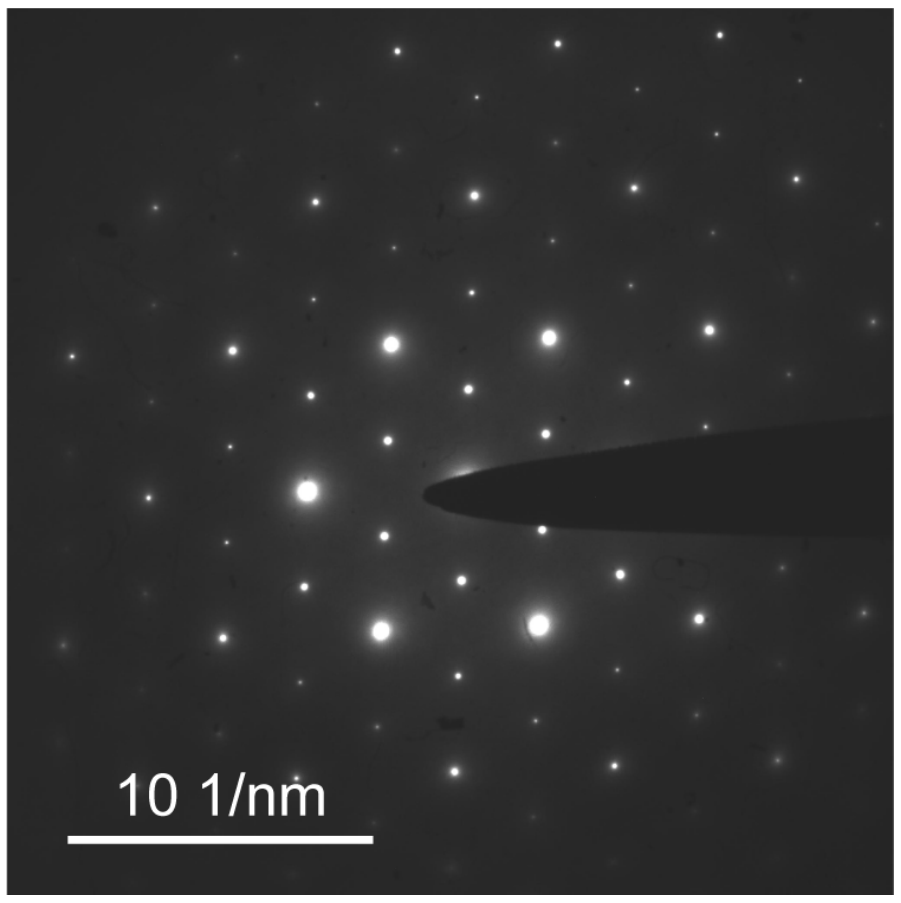

**Figure S3.** SAED pattern of a $Fe_{0.17}ZrSe_2$ flake.

## S5. Single crystal x-ray diffraction

Single crystal x-ray diffraction (SCXRD) was obtained at UC Berkeley CHEXRAY crystallographic facility. SCXRD analysis was performed on a single crystal coated with Paratone-N oil and mounted on a MiTeGen MicroMount. Data were collected on a Rigaku XtaLab P200 (ChexTWOFACE) equipped with a MicroMax 007HF rotating anode and a Pilatus 200K hybrid pixel array detector using Mo Kα radiation (λ = 0.71073 Å). The temperature of the crystal was maintained at 100(2) K throughout collection. Data collection, refinement, and reduction were performed with CrysAlisPro (version 1.171.41.112a, Rigaku Corporation, Oxford, UK). A multi-scan absorption correction was applied using the SCALE3 ABSPACK scaling algorithm within CrysAlisPro. The structure was solved using direct methods with SHELXS[7,8] and refined with SHELXL (version 2014/07) ,[9,10] with refinement of $F^2$ on all data by full-matrix least squares, operated in the OLEX2[11] interface. All atoms were refined anisotropically. The 3D molecular structure figures were visualized with VESTA version 3.5.7 software,[12] while schematics were drawn in Autodesk 3ds Max and Adobe Illustrator version 27.3.1.

**Table S1.** Crystal data and structure refinement for $Fe_{0.17}ZrSe_2$

| | | |
|---|---|---|
| Empirical formula | Fe0.17Se2Zr | |
| Formula weight (g/mol) | 258.63 | |
| Temperature (K) | 100 | |
| Crystal system | Trigonal | |
| Space group | P-3m1 | |
| Radiation | Mo Kα (λ = 0.71073) | |
| Unit cell dimensions | a = b = 3.7564(2) Å | $\alpha = \beta = 90°$ |
| | c = 6.1078(4) Å | $\gamma = 120°$ |
| Volume ($Å^{-3}$) | 74.636(10) | |
| $Z$ | 1 | |
| Density (calculated) ($g/cm^3$) | 5.754 | |
| Absorption coefficient ($mm^{-1}$) | 28.516 | |
| $F$ (000) | 112.0 | |
| Crystal size ($mm^3$) | 0.075 × 0.05 × 0.003 | |
| $\theta$ (°) | 3.335 to 30.987 | |
| Index ranges | $-5 \le h \le 5, -5 \le k \le 5, -8 \le l \le 8$ | |
| Reflections collected | 3773 | |
| Independent reflections | 114 | |
| Completeness to $\theta_{full}$ | 1.000 | |
| Absorption correction | Semi-empirical from equivalents | |
| Refinement method | Full-matrix least-squares on $F^2$ | |
| Data / restraints / parameters | 114 / 6 / 11 | |
| Goodness-of-fit on $F^2$ | 0.965 | |
| Final $R$ indexes [$I > 2\sigma(I)$] | $R_1 = 0.0336$, $wR_2 = 0.0988$ | |
| Final $R$ indexes [all data] | $R_1 = 0.0353$, $wR_2 = 0.1021$ | |
| Largest diff. peak / hole (e $Å^{-3}$) | 1.58 / −1.97 | |

**S6. Raman spectroscopy**

Raman spectra were acquired with a HORIBA LabRAM Evo Raman spectrometer using a 532 nm wavelength laser source. A 100× (NA = 0.9) objective (M Plan Achromat lens, Olympus Corporation) was used with a laser spot size of ~1 μm and a laser power of ~ 20 μW. Spectra were acquired with a grating of 600 grooves/mm, 1 s acquisition times, and 20 accumulations in ambient conditions (room temperature, 1 atm pressure). Higher laser powers (> 20 μW) and longer acquisition times (> 5 s) were found to lead to sample degradation.

**S7. X-ray absorption spectroscopy**

The XAS samples were prepared by sandwiching as-grown $Fe_{0.17}ZrSe_2$ crystals between two strips of KAPTON tape. XAS data were collected at the Stanford Synchrotron Radiation Lightsource (SSRL) beamline 4-3 at the Fe *K*-edge. Spectra were calibrated against an Fe foil. Samples were placed in a He filled chamber during data collection. No beam damage was observed during data collection.

Data were analyzed in the Demeter 0.9.26 suite of programs using Ifeffit 1.2.12. Spectra were imported to Athena for calibration, merging and spline fitting. The final spectra were imported into Artemis. Input files for path generation in Artemis were created using WebAtoms to convert CIF files into FEFF8 input files. The symmetry of the tetrahedral sites was not perfect, leading to slightly different length Fe–Se scattering paths. No significant difference was found between the choice of these paths. Both $R_{\mathrm{eff}}$ values for these paths are reported below with the unused path $R_{\mathrm{eff}}$ given in parentheses in the summary. **Table S2** is a summary of the paths used in the mode.

**Table S2. Summary of the paths used in the mode**

| Type | R effective | Degeneracy |
|---|---|---|
| Fe-Se Octahedral | 2.623 | 6 |
| Fe-Fe Octahedral | 3.756 | 6 |
| Fe-Zr Octahedral | 3.054 | 2 |
| Fe-Se Tetrahedral | 2.241 (alternative: 2.386) | 4 |
| Fe-Fe Tetrahedral | 2.831 | 3 |
| Fe-Zr Tetrahedral | 3.050 | 3 |

### S7.1. X-ray absorption near edge structure

**Figure S4a** shows the Fe K-edge XANES spectrum for $Fe_{0.17}ZrSe_2$. We attribute the pre-peak A (~7112 eV) to the local Fe tetrahedral ligand field that explicitly allows dipole transitions into 3*d* related states.[13] The intensity is low due to the low occupation of tetrahedral Fe. The feature B in the rising main absorption edge (the maximum point in the derivative absorption spectrum ~7118 eV) appears due to the 1s → 4p transition.[14] We assign the peak-like feature C (~7121 eV) to the 1s → 4p states admixed with the *d* states of the chalcogen atoms.[2] The D peak (~7125 eV) of 1s → 4p main transition is allowed by electric dipole matrix approximation.[15]

**Figure S4b** shows the comparison of Fe *K*-edge XANES spectra for $Fe_{0.17}ZrSe_2$ and the standard compound $Fe^{2+}$ (FeO), $Fe^{3+}$ ($LaSrFeO_4$), and $Fe^{4+}$ ($SrFeO_{3-x}$)[1] and for the reported spectrum for FeSe.[2] The rising main absorption edge for $Fe_{0.17}ZrSe_2$ spectrum is observed close to the those of $Fe^{2+}$ (FeSe) and $Fe^{2+}$ (FeO), consistent with an oxidation state of +2, consistent with XPS result shown in the main text.

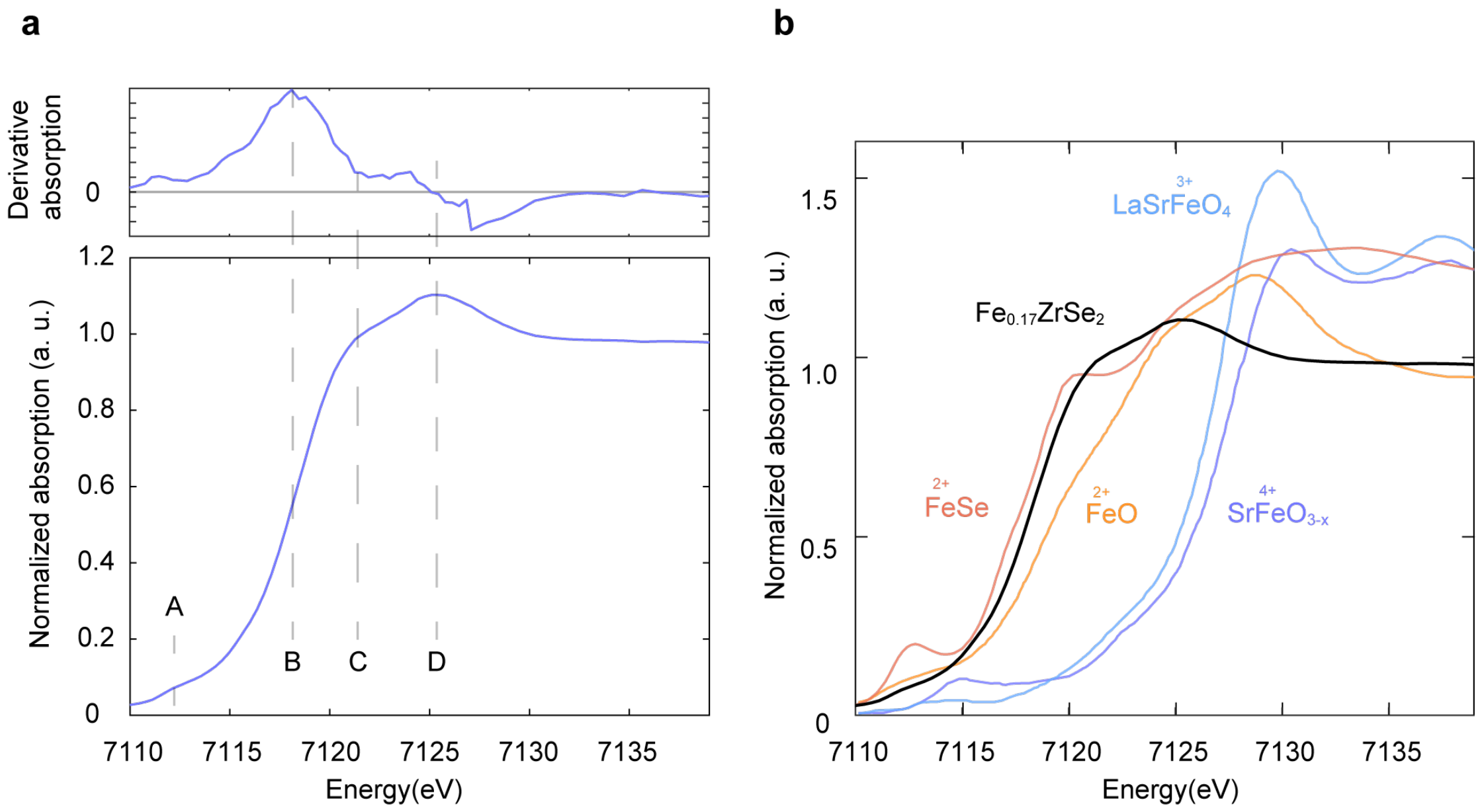

**Figure S4.** Fe K-edge X-ray absorption near edge structure (XANES) for $Fe_{0.17}ZrSe_2$. (a) Normalized Fe K-edge XANES spectrum for $Fe_{0.17}ZrSe_2$ and the corresponding first-order derivative absorption spectrum. (b) Fe K-edge XANES spectra for the standard compound $Fe^{2+}$ (FeO), $Fe^{3+}$ ($LaSrFeO_4$), and $Fe^{4+}$ ($SrFeO^{3-}$) and for the reported FeSe, compared to that of $Fe_{0.17}ZrSe_2$. $Fe^{2+}$ (FeO), $Fe^{3+}$ ($LaSrFeO_4$), and $Fe^{4+}$ ($SrFeO^{3-}$) spectra are adapted from ref. 1 FeSe spectrum is adapted from ref. 2

## S7.2. Extended x-ray absorption fine structure fitting

EXAFS fits were performed against data in $k$-space with $k$, $k^2$ and $k^3$ weighted data for $k$ between 3 and 12 $Å^{-1}$ with a Hanning window and in $R$-space for $R$ between 1.6-4.2 Å. Each fit had 14.5 independent points with 9 variables. For all fits, the amplitude was set to 0.8, though qualitatively similar results were obtained with amplitude set to 0.7. The amplitude factor in the fits for the octahedral and tetrahedral paths were weighted by a factor $x$ or $1–x$, respectively, where $x = 1$ corresponds to 100% octahedral sites and $x = 0$ corresponds to 100% tetrahedral sites. In a given fit, $x$ was constant and a series of fits with the same parameters were performed where $x$ was varied in steps of 0.05 or 0.1 from 0 to 1. Though the fit at 0.95 has the lowest overall value for the statistical assessments of the fit (i.e., reduced Red. $\chi^2$ and R-factor) (**Figure S5**), this fit also contains an unreasonably large value for the change in path length for tetrahedral Fe–Fe scattering path, $\Delta R_{Td, Fe}$, and a negative value for the Debye-Waller analog for the tetrahedral Fe–Se path, $\sigma^2_{Td, Se}$. Thus, we identified $x = 0.9$ as the best fit. These fitting results are summarized in **Table S3.**

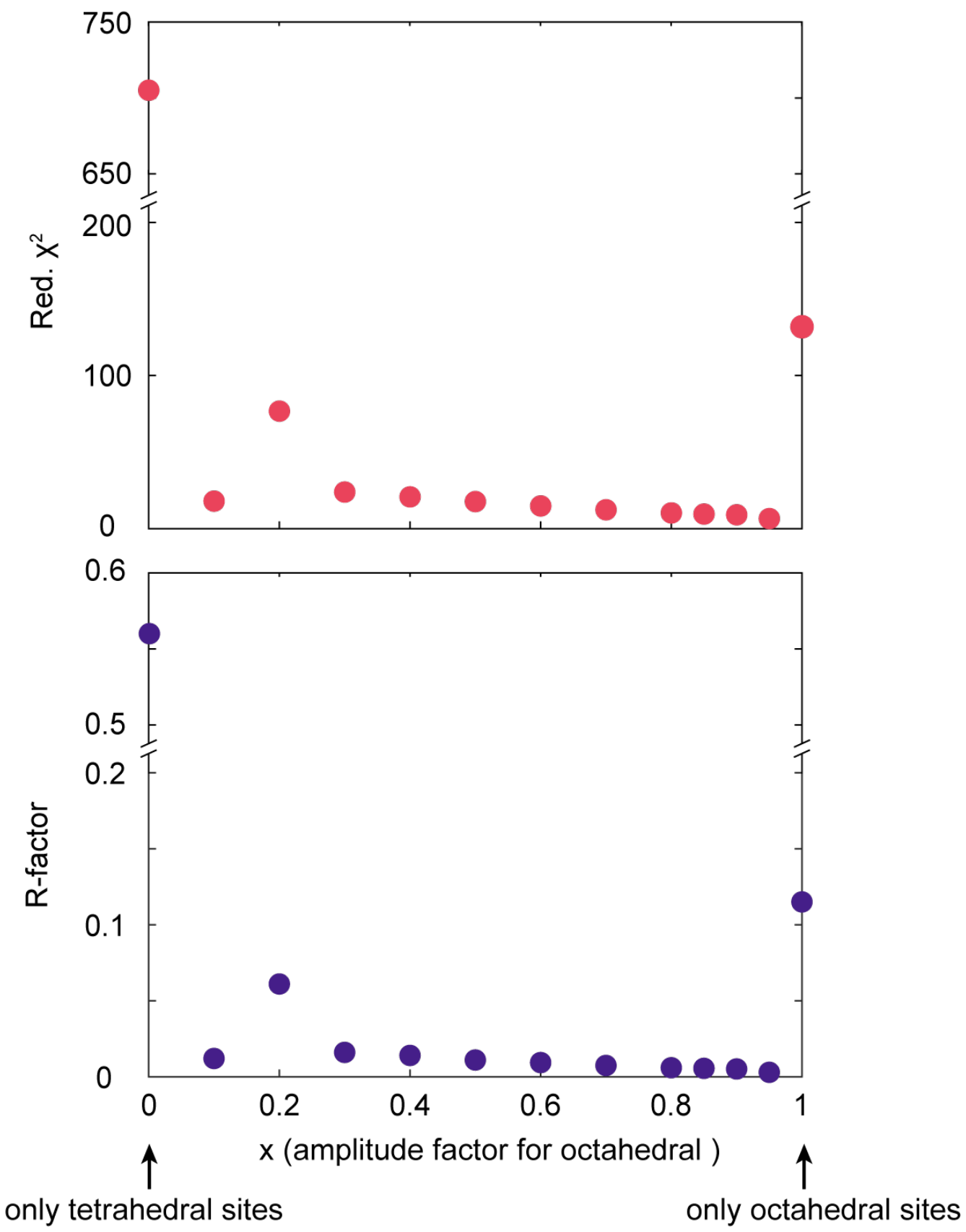

**Figure S5.** EXAFS fits against data in $k$-space with $k$, $k^2$ and $k^3$ weighted data. The statistical assessments of the fit, i.e., reduced Red. $\chi^2$ and R-factor were plotted against the amplitude factor in the fits for the octahedral and tetrahedral paths. The octahedral or tetrahedral sites were weighted by a factor $x$ or $1–x$.

**Table S3. The fitting results of different ratios of octa-Fe to tetra-Fe**

| $x$ | $E^0$ | $\Delta R_{Oh, Se}$ | $\sigma^2_{Oh, Se}$ | $\Delta R_{Oh, Fe}$ | $\sigma^2_{Oh, Fe}$ | $\Delta R_{Td, Se}$ | $\sigma^2_{Td, Se}$ | $\Delta R_{Td, Fe}$ | $\sigma^2_{Td, Fe}$ | Red. $\chi^2$ | R-factor |
|---|---|---|---|---|---|---|---|---|---|---|---|
| 1[a] | 6.82 | 0.027 | 0.014 | 0.064 | 0.0378 | 0[b] | 0.003[b] | 0[b] | 0.003[b] | 131 | 0.115 |
| 0.95 | 4.0 ± 0.7 | -0.003 ± 0.005 | 0.0146 ± 0.0003 | 0.04 ± 0.02 | 0.033 ± 0.002 | 0.091 ± 0.004 | -0.0011 ± 0.0004 | 0.67 ± 0.02 | 0.001 ± 0.002 | 4.13 | 0.0029 |
| 0.9 | 4.8 ± 1.0 | -0.007 ± 0.009 | 0.0120 ± 0.0007 | 0.04 ± 0.02 | 0.036 ± 0.004 | 0.11 ± 0.01 | 0.0016 ± 0.0007 | 0.03 ± 0.02 | 0.001 ± 0.002 | 6.64 | 0.0051 |
| 0.85 | 5.0 ± 1.1 | -0.01 ± 0.01 | 0.0105 ± 0.0007 | 0.04 ± 0.02 | 0.035 ± 0.004 | 0.13 ± 0.01 | 0.0034 ± 0.0008 | 0.02 ± 0.02 | 0.002 ± 0.002 | 7.13 | 0.0055 |
| 0.8 | 5.5 ± 1.2 | -0.00(7) ± 0.01 | 0.0093 ± 0.0007 | 0.04 ± 0.03 | 0.034 ± 0.004 | 0.14 ± 0.01 | 0.0049 ± 0.0009 | 0.02 ± 0.02 | 0.002 ± 0.002 | 7.88 | 0.0059 |
| 0.7 | 6.5 ± 1.4 | -0.00(3) ± 0.01 | 0.0077 ± 0.0008 | 0.06 ± 0.03 | 0.032 ± 0.005 | 0.16 ± 0.02 | 0.007 ± 0.001 | 0.02 ± 0.02 | 0.004 ± 0.002 | 9.96 | 0.0074 |
| 0.6 | 7.5 ± 1.6 | 0.0(0) ± 0.01 | 0.0067 ± 0.0008 | 0.07 ± 0.04 | 0.029 ± 0.005 | 0.18 ± 0.02 | 0.010 ± 0.002 | 0.01 ± 0.02 | 0.005 ± 0.002 | 12.5 | 0.0093 |
| 0.5 | 8.5 ± 1.9 | 0.00(5) ± 0.01 | 0.0061 ± 0.0009 | 0.08 ± 0.04 | 0.027 ± 0.006 | 0.20 ± 0.03 | 0.013 ± 0.022 | 0.01 ± 0.02 | 0.007 ± 0.002 | 15.4 | 0.011 |
| 0.4 | 9.4 ± 2.1 | 0.01 ± 0.01 | 0.0055 ± 0.0009 | 0.10 ± 0.04 | 0.024 ± 0.006 | 0.21 ± 0.03 | 0.017 ± 0.003 | 0.00 ± 0.02 | 0.008 ± 0.003 | 18.5 | 0.014 |
| 0.3 | 10 ± 2 | 0.01 ± 0.1 | 0.0048 ± 0.0009 | 0.11 ± 0.05 | 0.019 ± 0.006 | 0.22 ± 0.04 | 0.022 ± 0.005 | -0.00(5) ± 0.03 | 0.010 ± 0.003 | 21.7 | 0.016 |
| 0.2 | 4 ± 3 | 1.6 ± 0.1 | 0.01 ± 0.02 | 0.0(3) ± 0.1 | .02 ± 0.01 | 0.09 ± 0.04 | 0.026 ± 0.006 | -0.13 ± 0.02 | 0.0065 ± 0.0009 | 75.2 | 0.061 |
| 0.1 | 10 ± 2 | -0.561 ± 0.009 | 0.004 ± 0.001 | 0.11 ± 0.04 | 0.007 ± 0.004 | 0.036 ± 0.008 | 0.014 ± 0.001 | -0.11 ± 0.02 | 0.010 ± 0.001 | 15.7 | 0.012 |
| 0[a] | 29 | 0[b] | 0.003 [b] | 0 [b] | 0.003 [b] | 0.085 | 0.0192 | 1.27 | 0.103 | 705 | 0.56 |

[a] Error bars of ± 0.000000 in model output for this row. [b] Initial guess value, parameter not fit under these conditions.

## S8. X-ray photoelectron spectroscopy

X-ray photoelectron spectroscopy (XPS) measurements were performed using a Thermo Scientific K-Alpha$^{+}$ (Al Kα radiation, h$\nu$ = 1486.6 eV) (Thermo Fisher Scientific Inc) equipped with an electron flood gun. XPS data were analyzed using Thermo Scientific Avantage Data System software (version 5.9914), and a Smart background was applied prior to peak deconvolution and integration.

## S9. Scanning tunnelling spectroscopy

Scanning tunneling spectroscopy (STS) measurements were performed on freshly cleaved bulk $Fe_{0.17}ZrSe_2$ crystals using a Park Systems NX10 STM module operated in ambient conditions. Pt–Ir tips were fabricated via electrochemical etching of 0.25 mm Pt–Ir wires in 1.5 M $CaCl_2$ solutions. The STS $I$–$V$ curve was acquired by turning off the feedback loop, holding the tip a fixed distance above the surface, and sweeping the voltage from +0.8 to −0.8 V and then back to +0.8 V with duration of 1 s. The d$I$/d$V$ trace was obtained by taking the derivative of the $I$–$V$ curve.

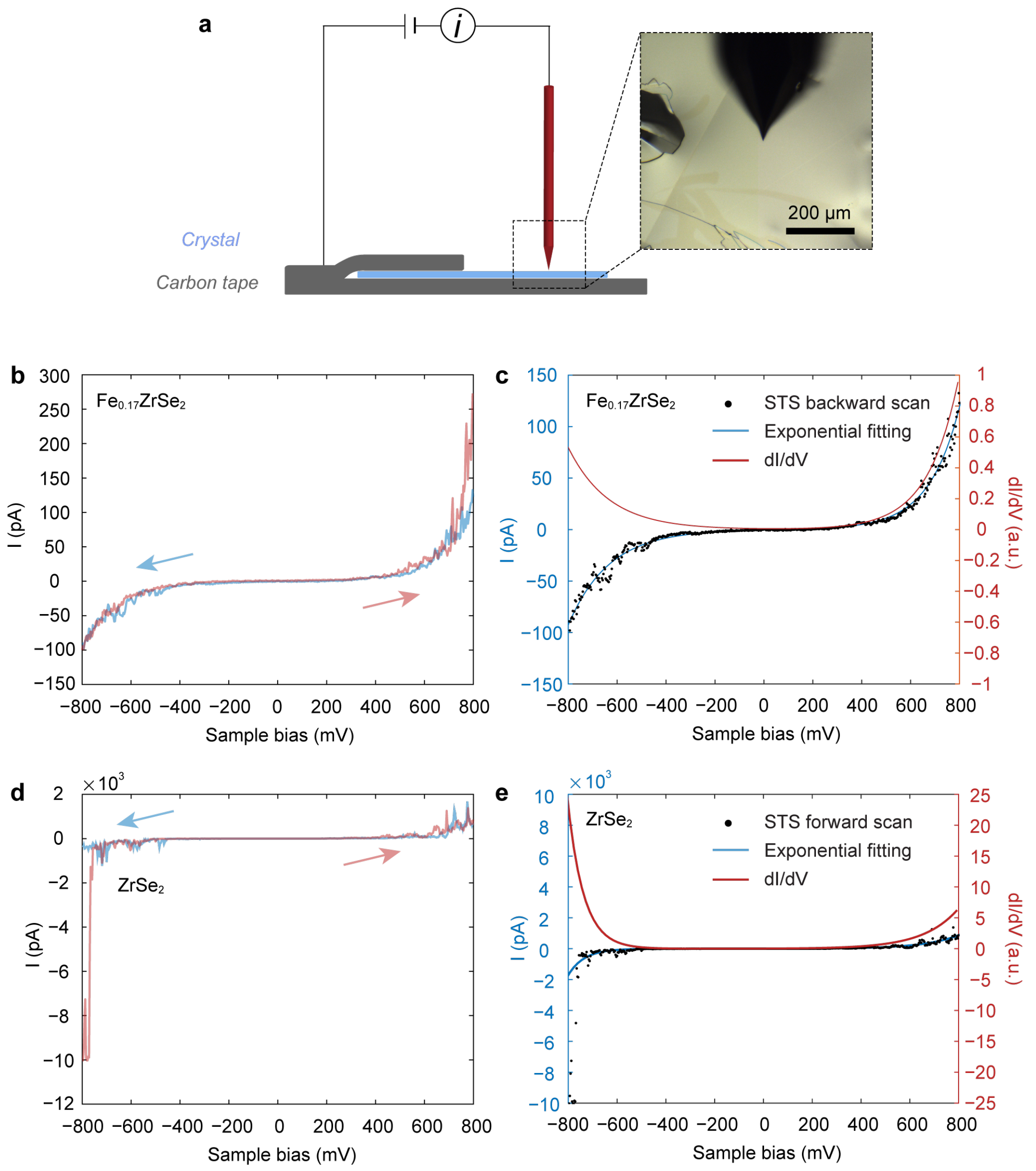

**Figure S6.** (a) Schematic of STS measurement setup. STS were performed on a freshly cleaved bulk sample in ambient conditions. The sample was mounted into the STS sample holder using carbon tape to attach the top of the sample to an STS probe. At a fixed tip–sample separation, the tunneling current was monitored while the bias voltage was swept from +0.8 V to –0.8 V (backward scan), and from –0.8 V to +0.8 V (forward scan). (b) The raw data of measured current of a $Fe_{0.17}ZrSe_2$ crystal with different applied sample bias. The backward scan is marked in a blue line and the forward scan is marked in a red line. (c) To mitigate the impact of raw data noise on the derivative of current versus sample bias, the raw data was fit with a two-term exponential model (blue line). The fitted data was used to obtain the $dI/dV$ curve by taking the derivative of the current with respect to the sample bias (red line). (d) The raw data of measured current of a $ZrSe_2$ crystal with different applied sample bias and (e) the corresponding fitting and $dI/dV$ curve. Data values at the extremes were excluded for the exponential fitting.

**S10. Photoluminescence spectroscopy and UV−vis−NIR diffuse reflectance spectroscopy**

Photoluminescence spectra (PL) were acquired with a HORIBA LabRAM Evo Raman spectrometer using a 532 nm wavelength laser source. For the ambient measurement (room temperature, 1 atm pressure), a 100× (NA = 0.9) objective (M Plan Achromat lens, Olympus Corporation) was used with a laser spot size of ~1 μm and a laser power of ~ 20 μW. Spectra were acquired with a grating of 600 grooves/mm, 3 s acquisition times, and 2 accumulations. Higher laser powers (> 20 μW) and longer acquisition times (> 5 s) were found to lead to sample degradation. For low temperature measurement, a 100× objective was used with a laser spot size of ~5 μm and a laser power of ~ 1 Mw. A sample was loaded in a Microscopy Cryostat (Model No.: CFM-1738-102, Cryo Industries of America, Inc. Manchester, NH, USA), which was attached to a Turbo Pumping station with Diagphram roughing pump (Model: HiCube 80 Eco, DN 63 ISO-K, MVP 015-4, Pfeiffer Vacuum Inc., Nashua, NH, USA) to maintain the pressure of $10^{-2}$ Pa during the measurement. Temperature is adjusted by a Temperature Controller (Model 325, Lake Shore Cryotronics, Inc., Westerville, OH, USA).

Photoluminescence spectra (PL) were collected on pristine $ZrSe_2$ crystals at room temperature in atmosphere (**Figure S7a**) and $Fe_{0.17}ZrSe_2$ in the temperature range from 77 to 300 K under vacuum (**Figure S7b**). No PL peak was observed for $ZrSe_2$, in line with $ZrSe_2$ being an indirect bandgap semiconductor, which is consistent with LAPW band calculations and experimental band structure obtained from ARPES.[16] Upon intercalation, **Figure 2** shows that the $ZrSe_2$ layers themselves do not undergo a phase change from. Accordingly, we suggest that a simple band filling and Fermi level renormalization process takes place upon intercalation. It seems reasonable to speculate therefore, that Fe intercalated $ZrSe_2$ persists as an indirect bandgap semiconductor. This conclusion is consistent with the PL measurements of $Fe_{0.17}ZrSe_2$ that display no peak from 77 to 300 K. Future ARPES measurements and/or theoretical calculations will be helpful in confirming the band structure of $Fe_{0.17}ZrSe_2$.

Diffuse reflectance spectroscopy measurements were performed on polycrystalline $Fe_{0.17}ZrSe_2$ samples, which were diluted in $BaSO_4$ and ground with a mortar and pestle to produce a homogenous powder in Ar-filled glove box. Diffuse reflectance UV-vis-NIR spectra were collected on a CARY 5000 spectrophotometer equipped with a Praying Mantis diffuse reflection accessory (Harrick Scientific Products, Inc.) and interfaced with Varian Win UV software.

The measured diffuse reflectance spectrum of $Fe_{0.17}ZrSe_2$ was transformed to the corresponding absorption spectrum by applying the Kubelka−Munk function:[17]

$$F(R_\infty) = \frac{K}{S} = \frac{(1 - R_\infty)^2}{2R_\infty}$$

where $R_\infty$ is the diffuse reflectance, while K and S are the absorption and scattering coefficients, respectively.

The band-gap energy of semiconductors is usually determined by a Tauc plot,[18] which assumes that the energy-dependent absorption coefficient α could be interpreted by the following equation:

$$(\alpha \cdot h\nu)^{1/\gamma} = A\,(h\nu - E_g)$$

where h is the Planck constant, ν is the photon's frequency, A is a proportionality constant independent of the photon energy, and $E_g$ is the band gap energy. The value of the γ factor is determined by the nature of the electron transition and is equal to 1/2 or 2 for direct or indirect transition band gaps, respectively.[19] According to photoluminescence spectrum (PL) for $Fe_{0.17}ZrSe_2$ in **Figure S7b**, no peak was observed in the energy range of 0.41 ~ 0.60 eV, which indicates that $Fe_{0.17}ZrSe_2$ is an indirect band gap semiconductor. Therefore, the $(F(R_\infty) \cdot E)^{1/2}$ was plotted against photon energy E in the main text. The *x*-intercept of the linear fit of the Tauc plot gives an estimate of the band gap energy, which is around 0.44 eV.

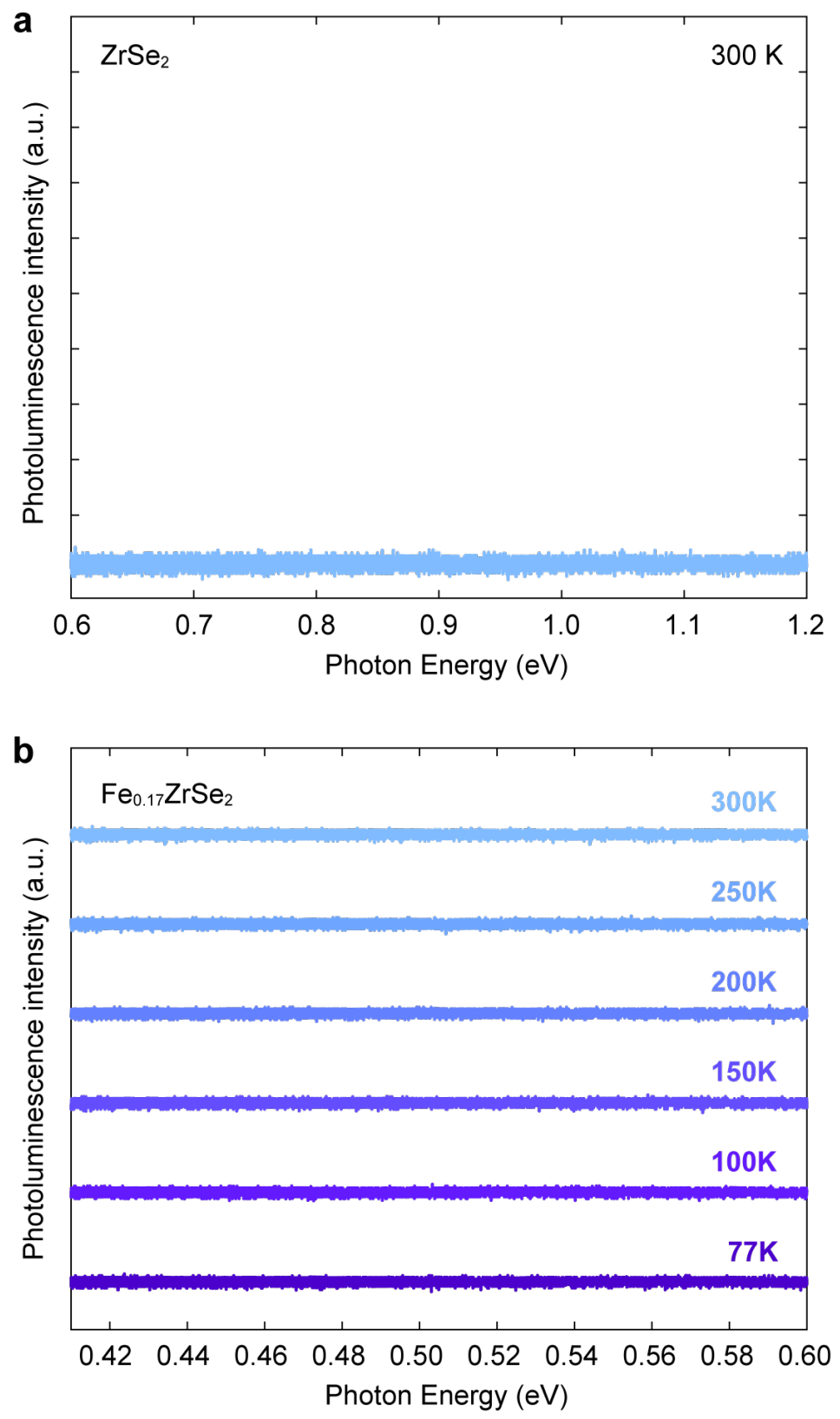

**Figure S7.** PL spectra for (a) pristine $ZrSe_2$ at room temperature subtracted by $SiO_2$/Si background. (b) $Fe_{0.17}ZrSe_2$ at temperature ranging from 77 to 300 K after subtraction of $SiO_2$/Si background.

**S11. Magnetometry measurements**

DC magnetization measurements as a function of temperature and applied field were carried out in a Quantum Design Physical Property Measurement System (PPMS) Dynacool equipped with a 12 T superconducting magnet using the Vibrating Sample Magnetometer (VSM) option, with a detection limit of $10^{-6}$ emu. For out-of-plane measurements, 7.89 mg of $Fe_{0.17}ZrSe_2$ flakes were stacked into polypropylene VSM powder sample holder, which was snapped into the brass half-tube. For in-plane measurements, 2.40 mg of $Fe_{0.17}ZrSe_2$ flakes were stacked onto MPMS 3 Quartz Paddle Sample Holder (C130A) and secured using Kapton tape.

**S12. Magnetocrystalline anisotropy of $Fe_{0.17}ZrSe_2$**

The fractional trigonal distortion is defined as $(a_2–a_1)/a_1$. For the regular octahedron, the fractional trigonal distortion is zero; for the trigonally elongated/compressed Fe environment, the fractional trigonal distortion is greater/smaller than zero. Values of $a_1$ = 3.253 Å and $a_2$ = 3.144 Å for $Fe_{0.17}ZrSe_2$ were exacted from the SCXRD solved structure. The fractional trigonal distortion is –3.35%, indicating that oct-Fe atoms in $Fe_{0.17}ZrSe_2$ locate in trigonally compressed distorted pseudo-octahedral coordination environment. The trigonally distorted pseudo-octahedron coordination environment gives rise to a qualitative $d$-orbital splitting diagram of $e_g$ $(d_{xy}, d_{x^2-y^2})$, $A_{1g}$ $(d_{z^2})$, and $e_g$ $(d_{xz}, d_{yz})$.[3] XPS showed the oxidization state of Fe is +2. The weak crystal field results in a high-spin $d^6$ electron configuration for $Fe^{2+}$ (S = 2). An unevenly occupied $e_g$ set of $Fe^{2+}$ leads to an unquenched orbital angular momentum and large spin orbit coupling.[3]

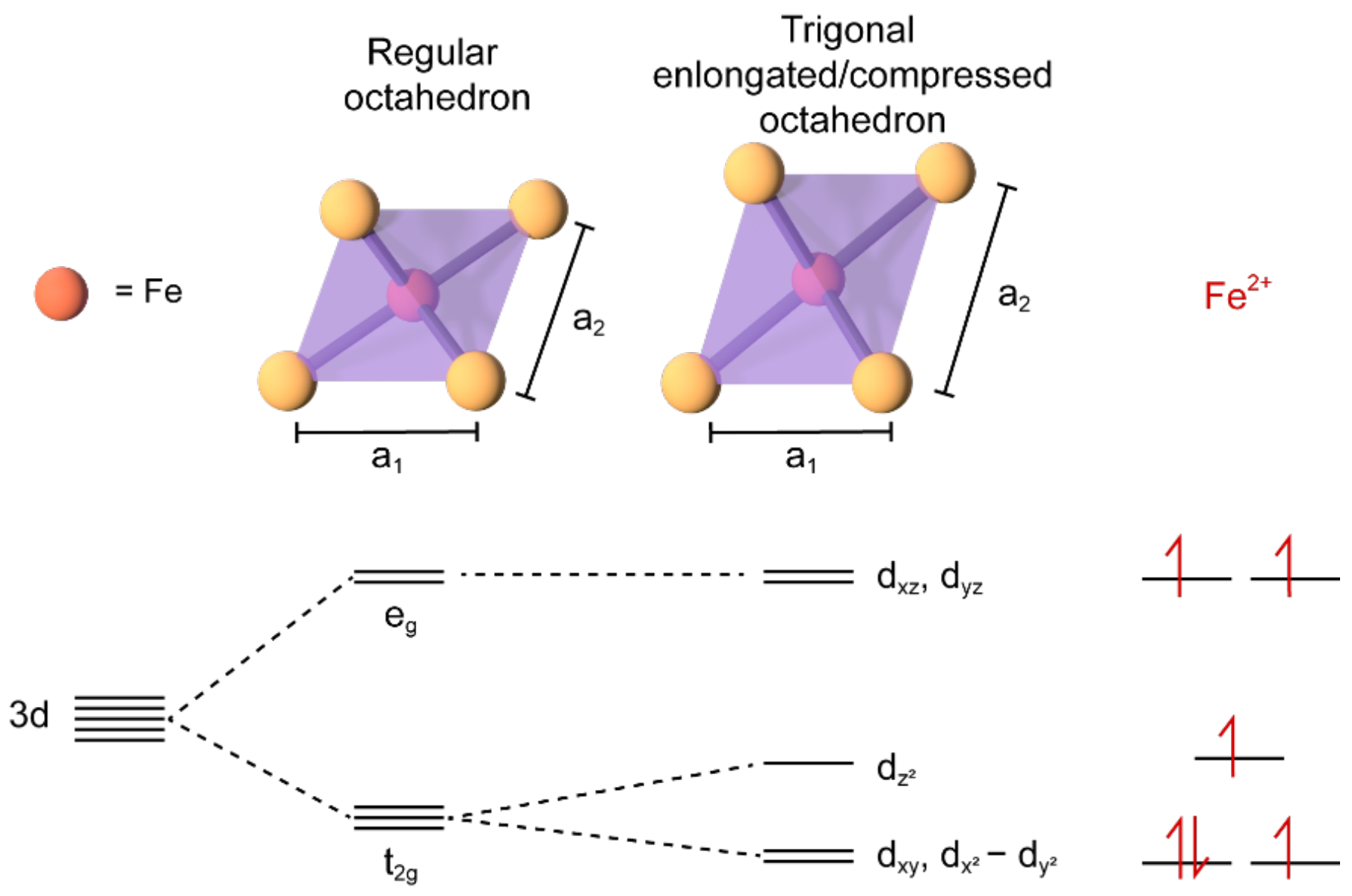

**Figure S8.** Schematic of perfect octahedral (left) and trigonally distorted (right) octahedral Fe environments, as seen along the $[2\bar{1}\bar{1}0]$] zone axis of the $Fe_{0.17}ZrSe_2$ lattice. Qualitative d-orbital splitting diagrams for intercalant high-spin $Fe^{2+}$ in a trigonally distorted pseudo-octahedral coordination environment. Schematic adapted from ref. 3

**S13. Curie–Weiss fit**

Curie–Weiss fits were performed on the susceptibility data for $Fe_{0.17}ZrSe_2$ in the temperature range of 350 – 400 K. (**Figure S9**). The following functions were used to calculated effective moment $\mu_{eff}$:

$$\chi = C/(T - \theta_{CW})$$

$$\mu_{eff} = \sqrt{8C}$$

where $C$ is the Curie constant and $\theta_{CW}$ is the Curie-Weiss temperature. The extracted $C$, $\theta_{CW}$, and $\mu_{eff}$ are shown below in **Figure S9**.

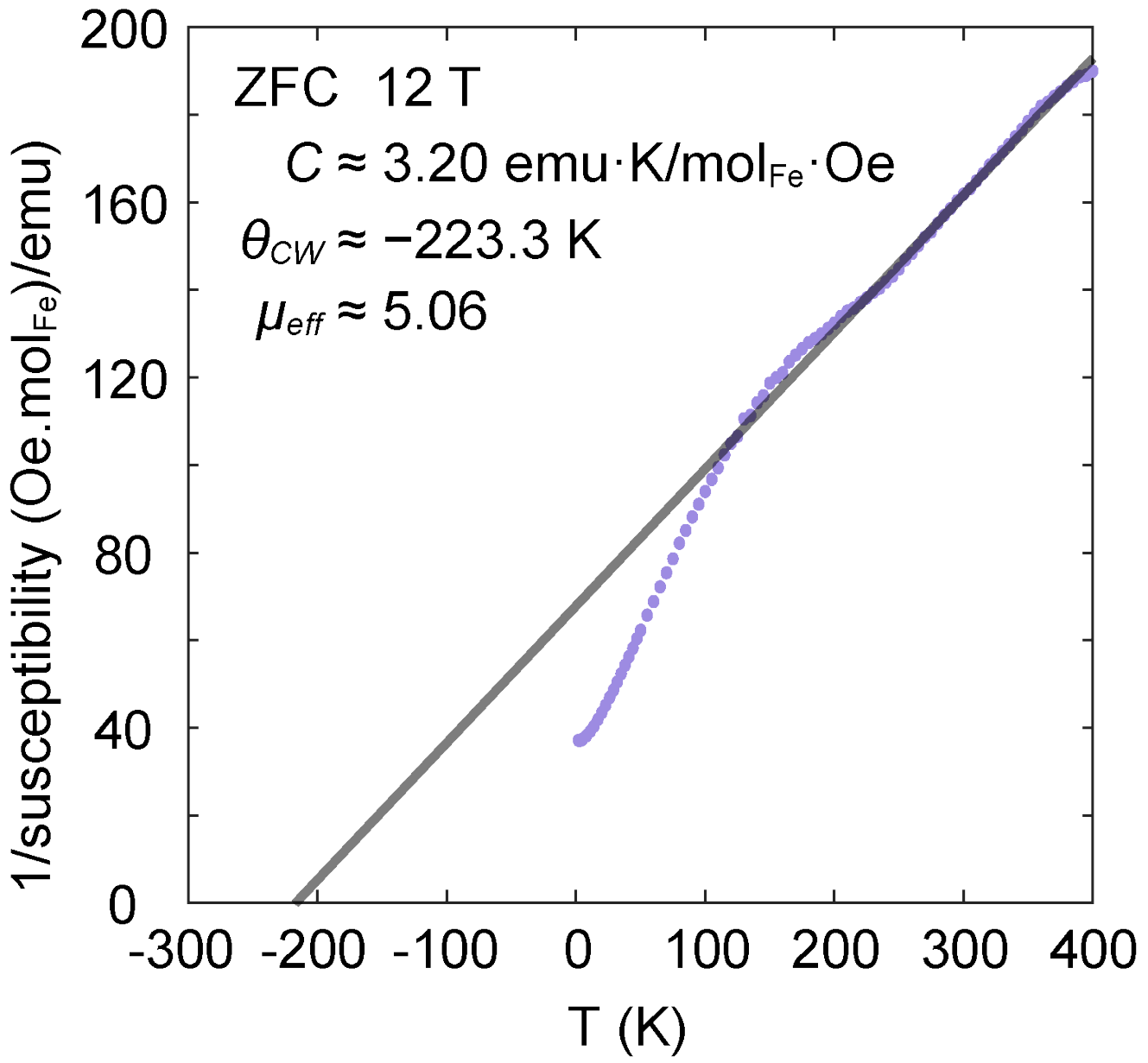

**Figure S9.** Inverse susceptibility as a function of temperature for $Fe_{0.17}ZrSe_2$ (purple dots). The data above 350 K was fitted to the Curie–Weiss model $\chi = C/(T - \theta_{CW})$(black line).

## S14. Out-of-plane magnetization of $Fe_{0.17}ZrSe_2$

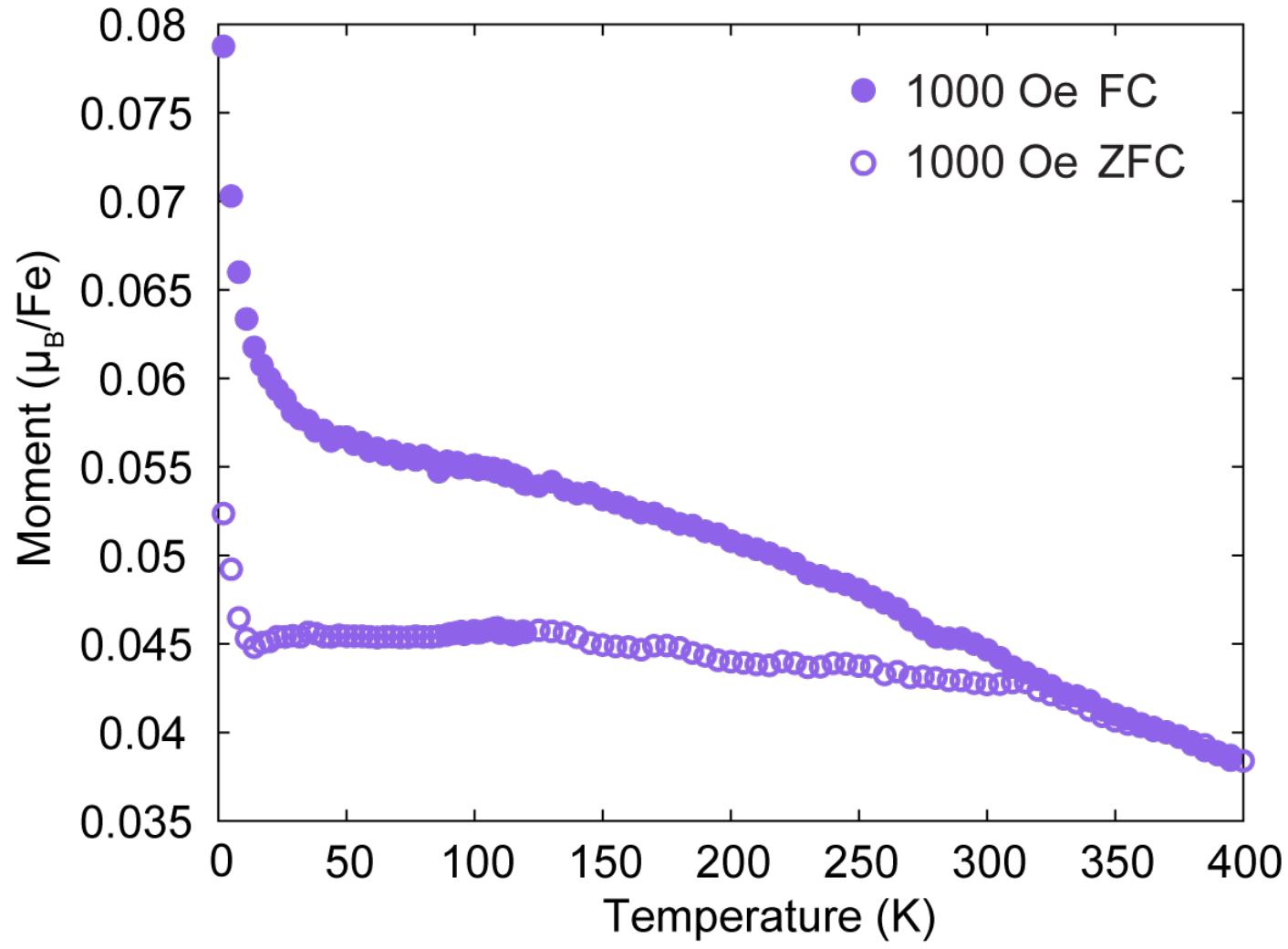

**Figure S10.** Temperature dependence out-of-plane zero-field-cooled (ZFC) (open dots) and field-cooled (FC) (filled dots) DC magnetization for $Fe_{0.17}ZrSe_2$ under applied magnetic fields 1000 Oe.

**S15. Heat capacity**

Heat capacity measurements as a function of temperature were carried out in a PPMS Dynacool using the Heat Capacity Option. 3.24 mg of $Fe_{0.17}ZrSe_2$ crystals were stacked together and compressed into one pellet and mounted onto heat capacity sample puck. The ac heat capacity measurement as a function of temperature was performed on this $Fe_{0.17}ZrSe_2$ pellet in the absence of an external magnetic field (**Figure S11a**). The specific-heat curve of $Fe_{0.17}ZrSe_2$ displays a broad transition but no sharp features, consistent with glassy order. The freezing of spins in different directions and the presence of frustration result in a distribution of energy levels. As a result, the heat capacity shows a broad maximum rather than a sharp peak.[4]

Further, the absence of sharp transitions in the specific heat capacity vs. temperature trace points to the homogeneity of the samples. The magnetometry data points to the coexistence of AFM and spin glass phases in the $Fe_{0.17}ZrSe_2$ below 110 K. If the two phases were segregated in discrete volumes throughout the sample, the specific-heat measurement would show a peak associated with the AFM transition overlaid with a broad glassy background. Therefore, the absence of the sharp transition reveals that there is no phase segregation in the sample.

A key characteristic indicating the presence of a spin glass state in the heat capacity measurement is its linear behavior of specific heat at low temperatures. A plot of the specific heat $C_p$ as a function of $T$ and the corresponding linear fitting for $Fe_{0.17}ZrSe_2$ at low temperature (< 3.5 K) are presented in the **Figure S11b**. A positive curvature takes place to force $C_p$ to zero as temperature approaches 0 K as a consequence of the third law of thermodynamics. The linear temperature dependence of the specific heat in low temperature limit can be understood by the two-level tunnelling model, which is the simplified model from the rough energy landscape and macroscopic degeneracy of ground states in a spin glass. A sketch of a two-level energy diagram is shown in **Figure S11c**. There are two nearly similar energy minima separated with a tunnelling barrier (adapted from ref. 4). Assuming there is constant density of states for the two-level excitations. The quantum-mechanical tunnelling through the barrier between the two levels can result in a small rearrangement of some spins, which is temperature independent. Therefore, a linear proportionality of specific heat at low $T$ naturally arises for the randomly frozen spin-glass.[4]

**Figure S11d** reveals a plot of the specific heat per unit temperature ($C_p/T$) as a function of $T^2$ for $Fe_{0.17}ZrSe_2$ at low temperature (< 10 K), which can be well expressed by using the following equation:[20-21]

$$C_P\ (\mathrm{T}) = \gamma T + \beta T^3 + \delta T^5$$

where $\gamma T$ describes the electronic contribution to the heat capacity and $\gamma$ is the electron specific heat coefficient (Sommerfeld constant), $\beta T^3$ is the phonon contribution and $\beta$ stands for phonon specific heat coefficient, $\delta T^5$ and reflects the deviation term. As a result, the fitted values of parameters $\gamma$, $\beta$, and $\delta$ are 14.7(4) mJ/mol $K^2$, 0.55(3) mJ/mol $K^4$, and 1.2(3) × $10^{-3}$ mJ/mol $K^6$, respectively.

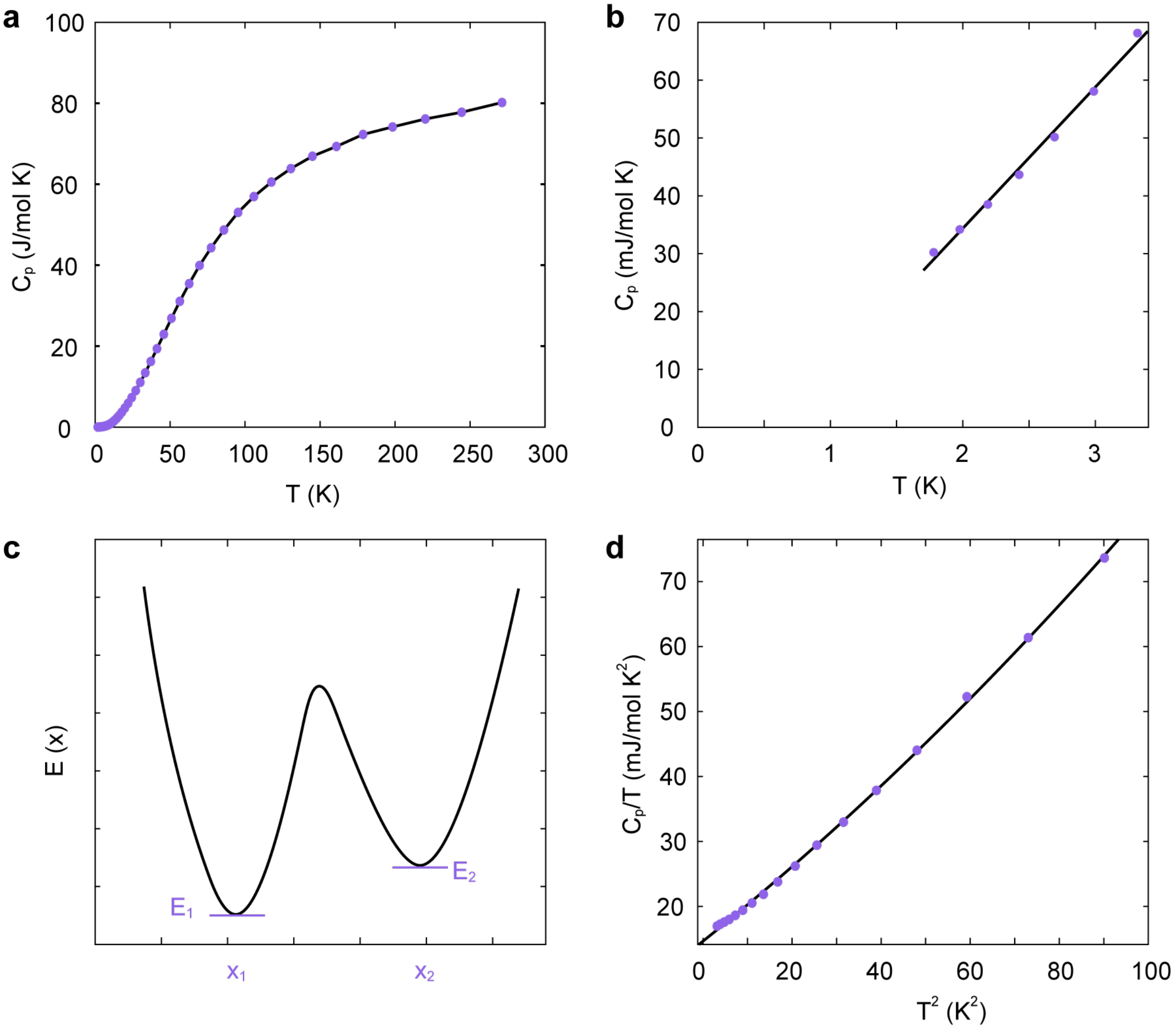

**Figure S11.** (a) Specific heat capacity ($C_p$) for $Fe_{0.17}ZrSe_2$ as a function of temperature (purple dots). Smoothed black line is a guide for the eyes. (b) Low temperature region of (a) (purple dots). The black line is the linear fitting curve. (c) A sketch of a two-level energy diagram (adapted from ref. 4). (d) Specific heat capacity divided by temperature ($C_p$/T) of $Fe_{0.17}ZrSe_2$ as a function of temperature squared ($T^2$) (purple dots). The black line is the curve of data fitted to $C_P$ (T) $= \gamma T + \beta T^3 + \delta T^5$.

## S16. In-plane magnetization of $Fe_{0.17}ZrSe_2$

The in-plane magnetization of the $Fe_{0.17}ZrSe_2$ crystals was measured as a function of temperature with the applied magnetic field parallel to the c-axis. The bifurcation between ZFC and FC curves reveals an anisotropic spin glass behavior along ab-plane, which has been reported before.[22] In Section S8, we discussed the magnetocrystalline anisotropy behavior along the *c*-axis in the $Fe_{0.17}ZrSe_2$ crystal, which might arise from the unquenched angular momentum of pseudo-octahedral $Fe^{2+}$ coordination environment. However, it is worth noting that the *c*-axis cannot be exclusively considered as the easy axis, as spin glass behavior is also observed for fields perpendicular to the *c*-axis, although to a lesser extent. Therefore, the $Fe_xZrSe_2$ spin-glass system cannot be described according to a pure Ising model. Generally, the spin–orbit coupling effects in tetrahedral $Fe^{2+}$ complexes are weaker compared to octahedral coordination environments. However, due to the disordered distribution of Fe at the octahedral and tetrahedral sites, a random distribution of spin directions, potentially canted with respect to the c-axis, is expected to emerge. Therefore, the Ising character of the octahedral $Fe^{2+}$ spins may be significantly altered, and the anisotropic spin glass behavior may arise.[22]

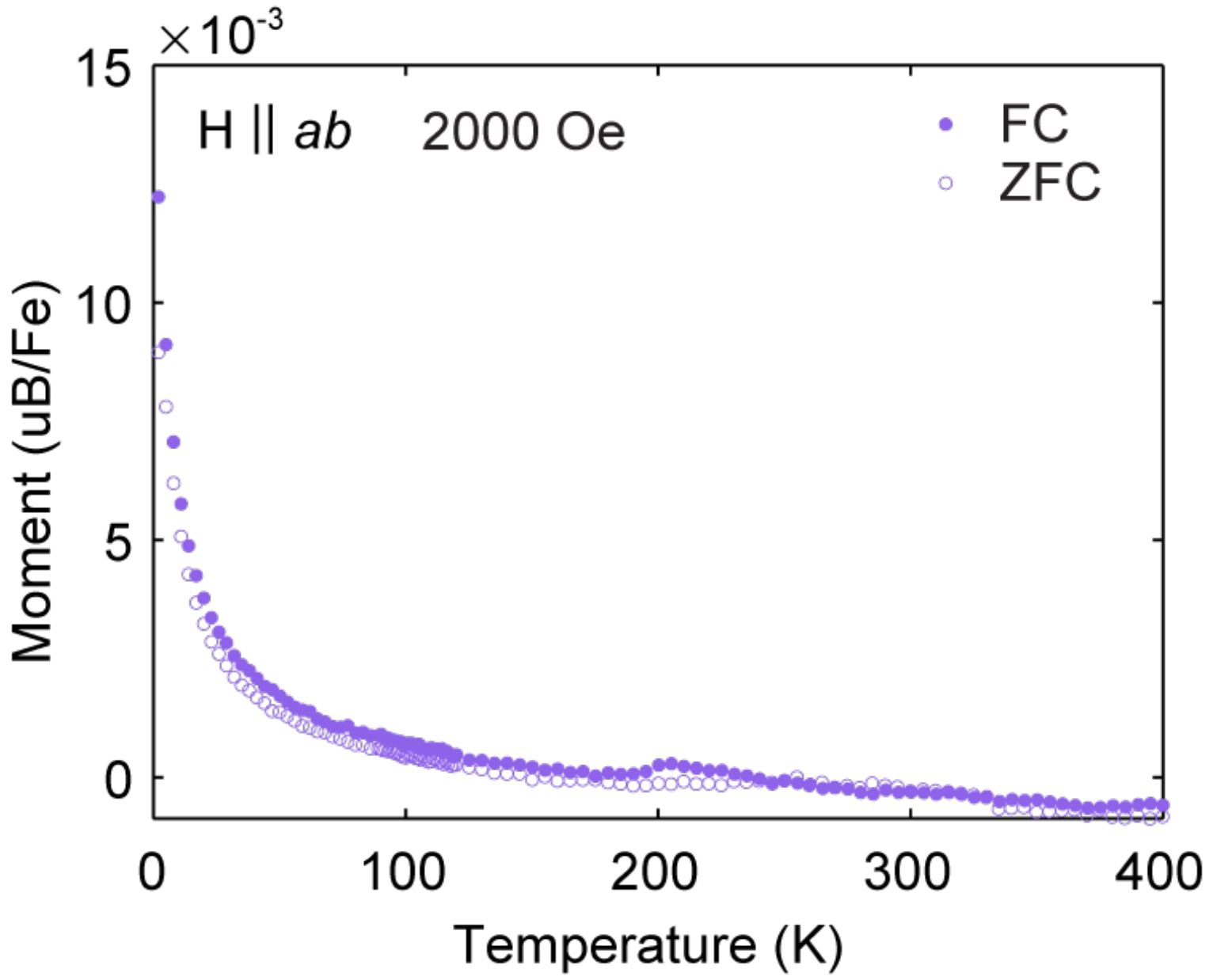

**Figure S12.** Temperature dependence in-plane zero-field-cooled (ZFC) and field-cooled (FC) DC magnetization for $Fe_{0.17}ZrSe_2$ under applied magnetic fields 2000 Oe.

### S17. Relaxation analysis

Both isothermal remanent magnetization (IRM) and thermoremanent magnetization (TRM) measurements were performed on $Fe_{0.17}ZrSe_2$ following the protocols outlined in **Figures S13a** and **S13b**, respectively. TRM measurements were conducted in the Quantum Design PPMS with VSM option using the following protocol: (a) warm the sample to 400 K in zero magnetic field, (b) apply the field along crystallographic $c$-axis to 1 T, (c) fast cool the sample to 60 K above the target temperature at 10 K/min, (d) slow cool to the target temperature at 1 K/min, (e) hold the sample in 1 T field for wait time $t_W$ one hour, and (f) set the field to 0 T and measuring the remanent magnetization in the sample over set time.

IRM measurements were conducted with same (a), (c)–(f) steps. For step (b), apply zero field along crystallographic $c$-axis.

The relaxation measurements for both IRM and TRM were best fit using the $M_R(t) = M_0 + A\exp[-(t/\tau)^{1-n}]$, where $M_0$ is remanence, irreversible part of the change in magnetization, $A$ is peak beyond equilibrium values related to glassy component of the magnetization, $\tau$ is the characteristic relaxation time, and $n$ is the time stretch component.

TRM measurements were conducted at different temperatures to study the effect of temperature on slow dynamics of $Fe_{0.17}ZrSe_2$ crystals. **Figure S14a** shows that the remanence decreases with increasing temperature, which was consistent with FC data. **Figure S14b and Table S4** reveals that in the range of 2 ~ 110 K, the parameters $M_0$, $A$, $\tau$, and $n$ changed drastically with increasing temperature. $\tau$ and $n$ decreased with increasing temperatures, which suggests that energy barriers, which trap the system in its metastable state, are reduced at higher temperatures. When the temperature is higher than 110 K, the parameter $M_0$ still decreased with increasing temperature, indicating the polarization of spins in the $Fe_{0.17}ZrSe_2$ crystals is weaker at higher temperature. However, the $Fe_{0.17}ZrSe_2$ still exhibits slow relaxation behavior for $T > 110$ K and parameters $A$, $\tau$, and $n$ did not show obvious changes with temperature, consistent with the existence of a glassy phase at high temperature. We note that the fitting equation is not useful when $T$ is higher than the freezing temperature, and relaxation curves at high temperature are noisy, making it challenging to extract precise relaxation parameters. Nevertheless, these fits provide some general characterization of the magnetic relaxation at high temperatures.

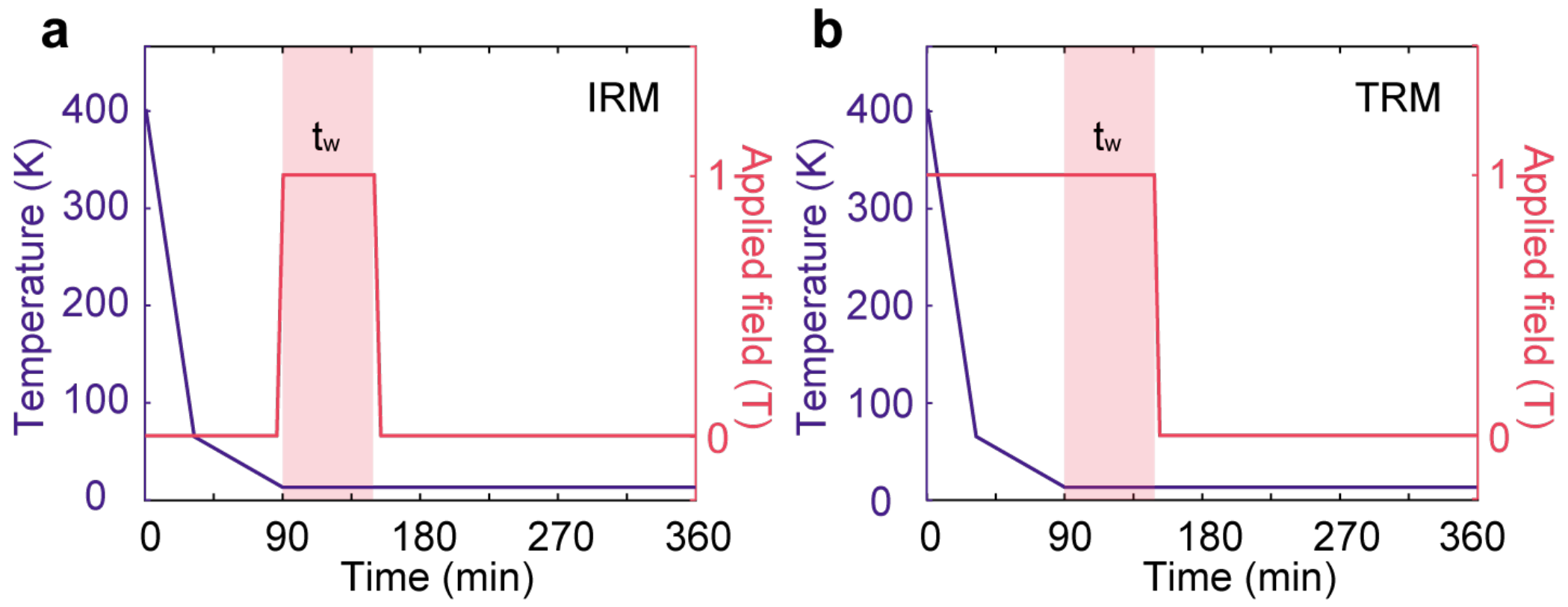

**Figure S13.** Illustration of isothermal remanent magnetization (IRM) measurements (a) and thermoremanent magnetization (TRM) measurements (b). The material was first fast-cooled from 400 K to 60 K by 10 K/min and then slow-cooled from 60 K to 2 K by 1 K/min under a 1 T applied field (a) or under zero field (b) and held in applied field 1 T for a designated wait time, $t_W$, at 2 K, the field was then removed and the TRM or IRM data were collected.

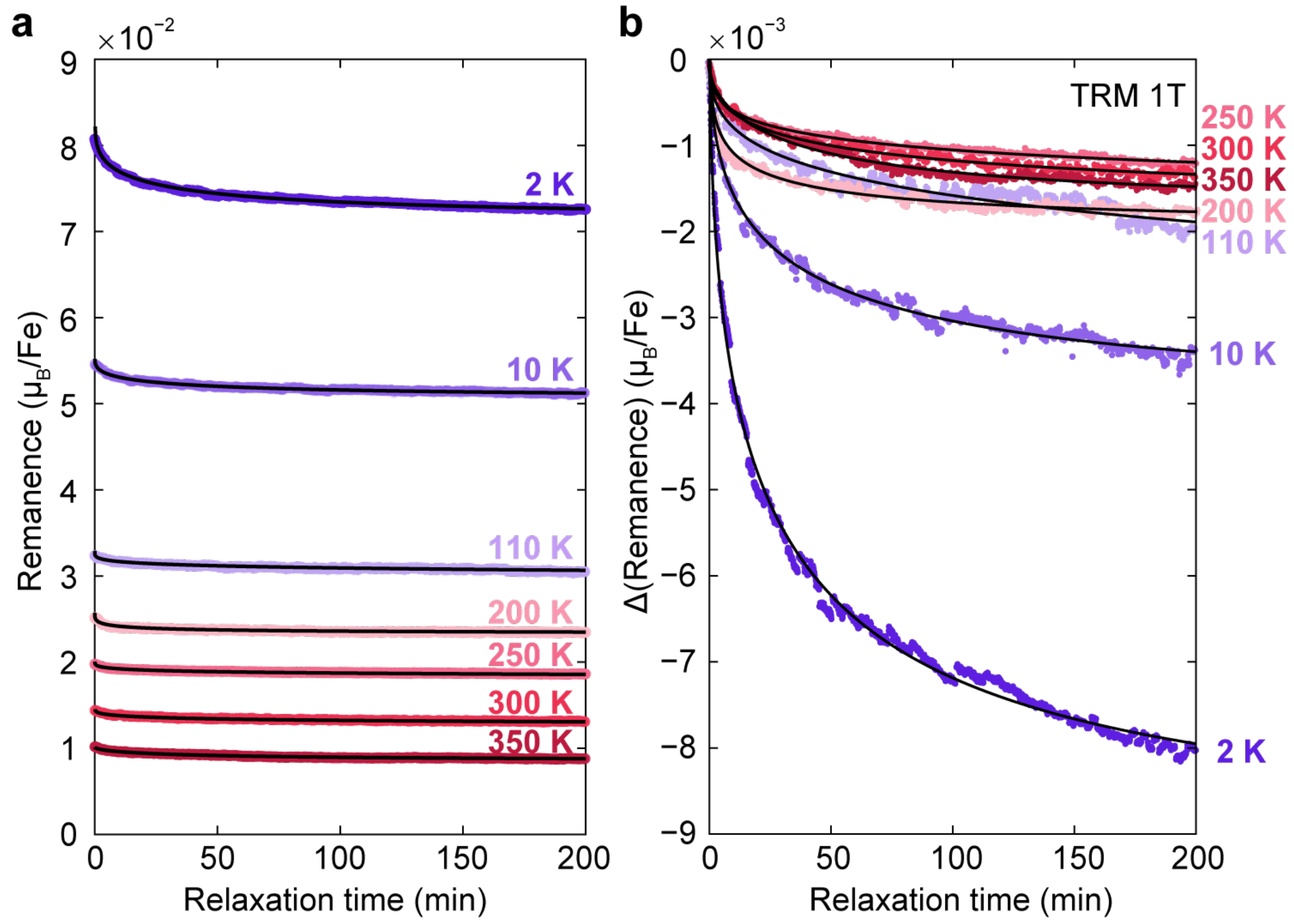

**Figure S14.** (a) The raw TRM curves at different temperature were obtained by FC in a 1 T magnetic field to the target temperature, waiting 60 min, removing this field, and finally measuring the variation in magnetization over 200 min. The TRM data plotted here are after field removal. (b) Change in TRM in (a) determined by subtracting the remanence at $t$ = 0 min for all the TRM data (colored dots) and their corresponding fits (black lines).

**Table S4. Temperature dependence of TRM relaxation fitting results**

$$M_R(t) = M_0 + A \exp[-(t/\tau)^{1-n}]$$

| Temp (K) | $M_0$ ($\mu_B$/Fe) | A ($\mu_B$/Fe) | $\tau$ (min) | n |
|---|---|---|---|---|
| 2 | 0.07127 | 0.0104 | 32.74 | 0.5997 |
| 10 | 0.05078 | 0.004411 | 28.9 | 0.5866 |
| 110 | 0.03072 | 0.001706 | 26.1 | 0.4184 |
| 200 | 0.02328 | 0.002122 | 13.55 | 0.6517 |
| 250 | 0.01843 | 0.001949 | 14.06 | 0.7214 |
| 300 | 0.01271 | 0.002835 | 12.31 | 0.7707 |
| 350 | 0.008366 | 0.002811 | 16.73 | 0.7509 |

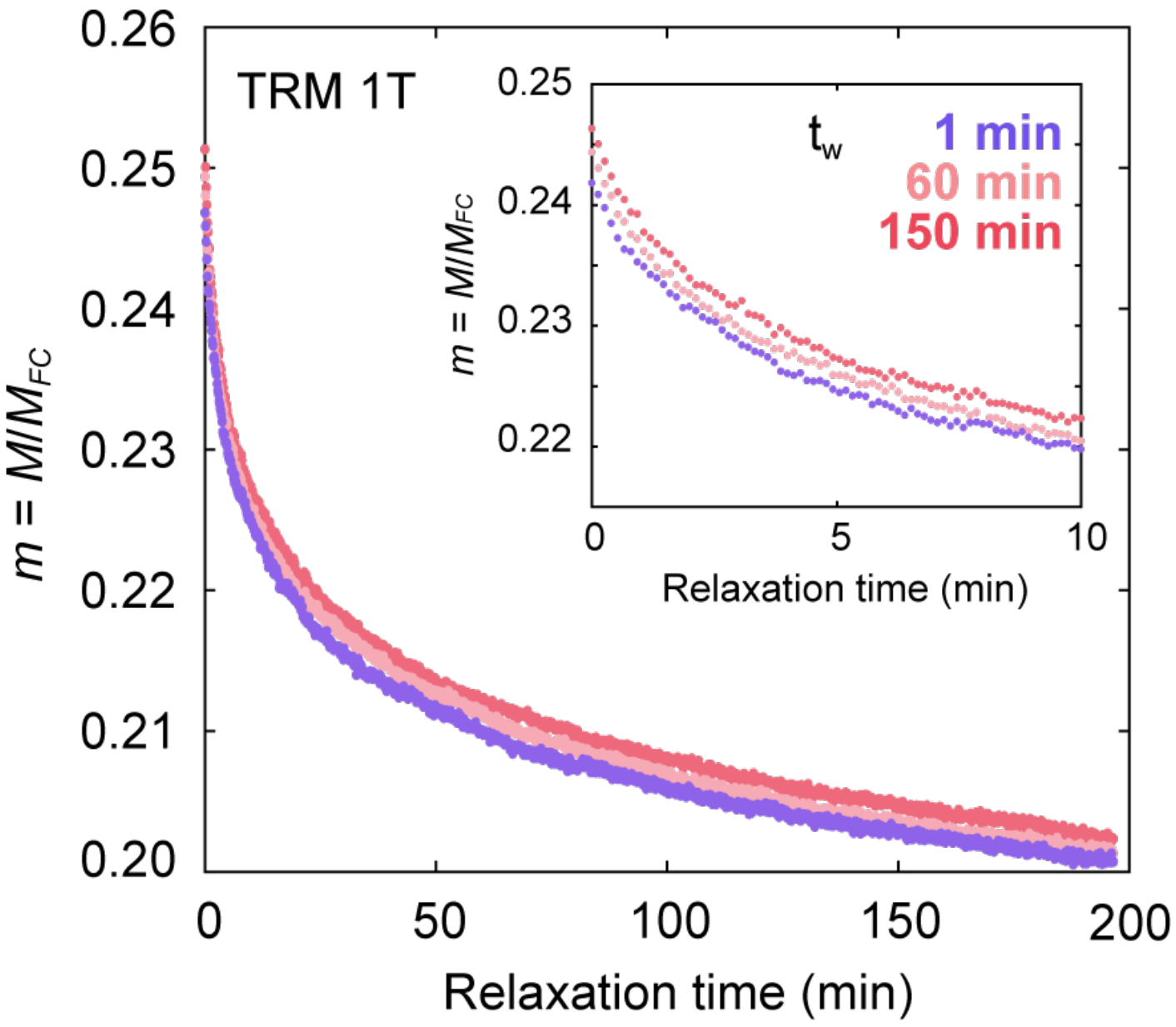

**Figure S15.** TRM data with different wait times. Inset: zoom in on the relaxation time ranging from 0 to 10 min. Magnetization increases with longer wait times, indicating the magnetic aging characteristic of spin glasses.

**S18. Variable temperature magnetization data as a function of field**

The temperature-dependent magnetization versus applied magnetic field scans were performed in the range of −12 T to 12 T. Zoomed in plots of FC magnetization sweep in the field range of −0.2 T to 0.2 T are presented in the main text. The side-by-side comparison of ZFC and FC plots are shown below, with zoomed-in plots in the range of −0.2 T to 0.2 T (**Figure S16**) and full range plots in the range of −12 T to 12 T (**Figure S17**).

**Figure S18** shows the cooling field-dependent magnetization versus field scans performed at 2 K. The zoom-in plots of magnetization sweep in the field range of −0.2 T to 0.2 T were presented in the main text. The sample was field cooled down from 400 K to 2 K under the applied field and then the applied field was set from the cooling field to 12 T. All loops were taken from +12 T to −12 T and back to +12 T.

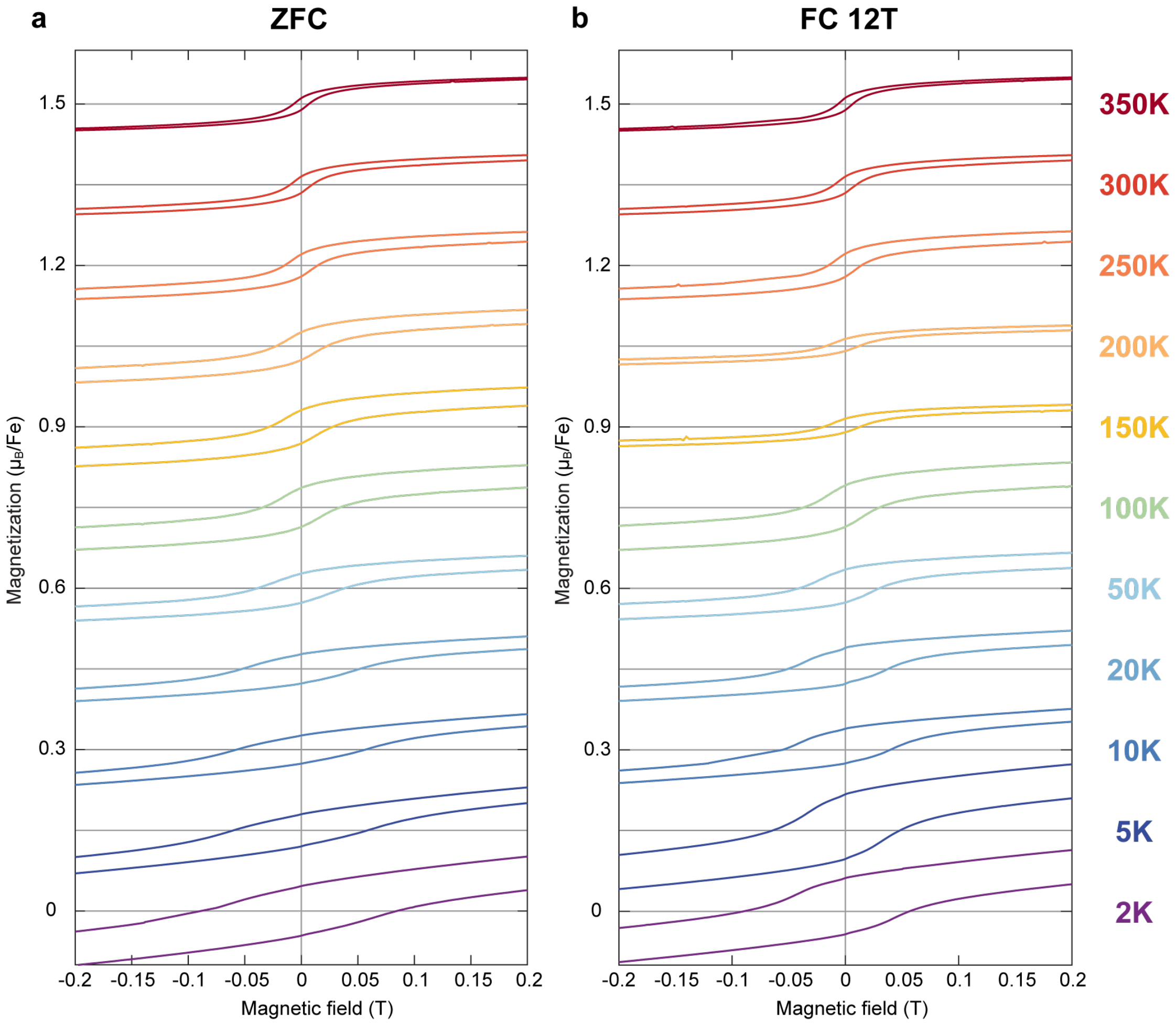

**Figure S16.** Magnetization versus magnetic field measurements at different temperatures. (a) ZFC and (b) FC under a 12 T external field and then sweep field from +12 T to −12 T back to +12 T. The applied field range of −0.2 T to 0.2 T is presented here. Each loop is offset on the y-axis by 0.15 $\mu_B$/Fe. The magnetic field was applied along the c-axis of the samples.

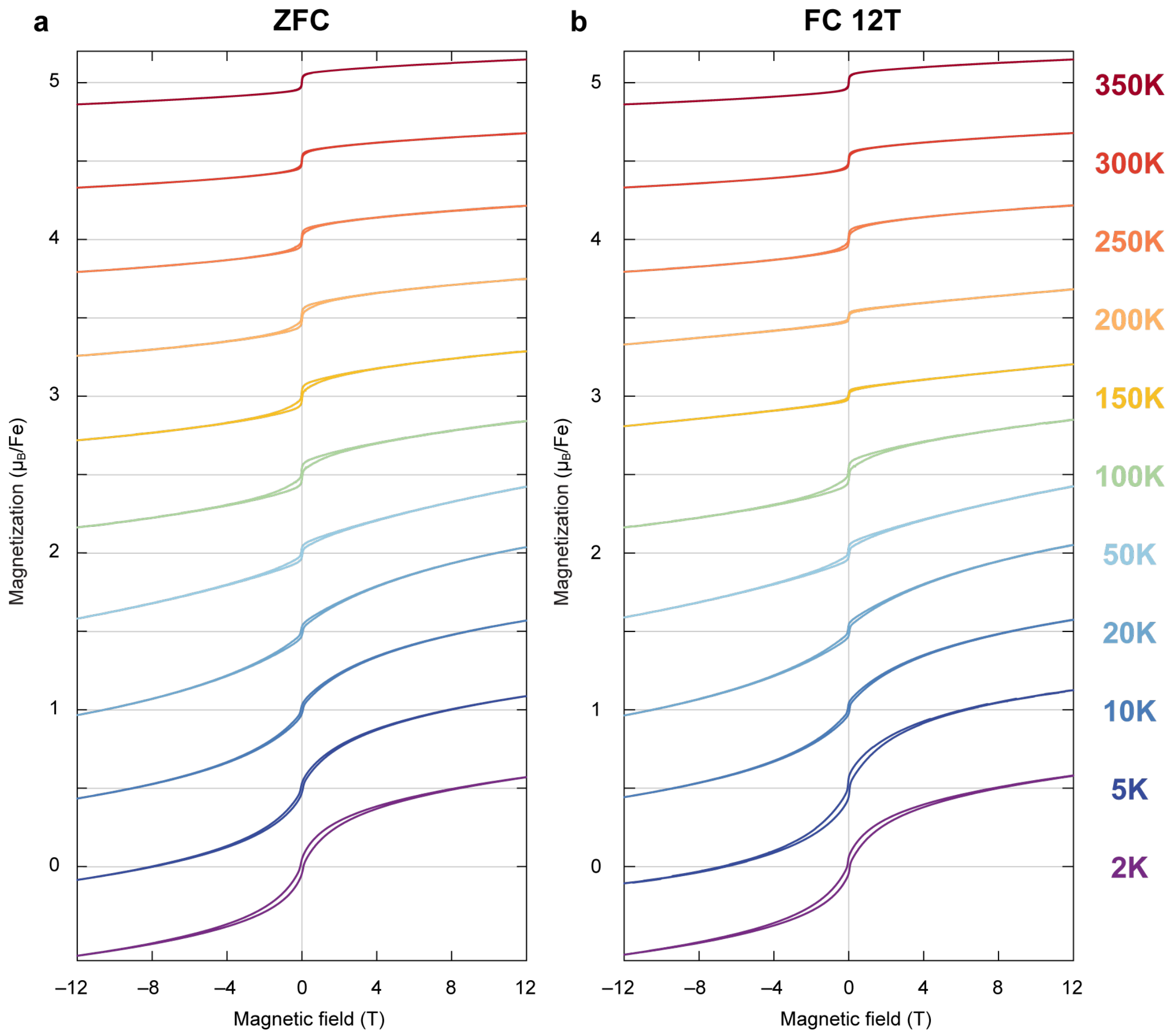

**Figure S17.** Magnetization versus magnetic field measurements at different temperatures. Same as Figure S16 but presented with the full range of applied field. Each loop is offset on the y-axis by 0.5 $\mu_B$/Fe.

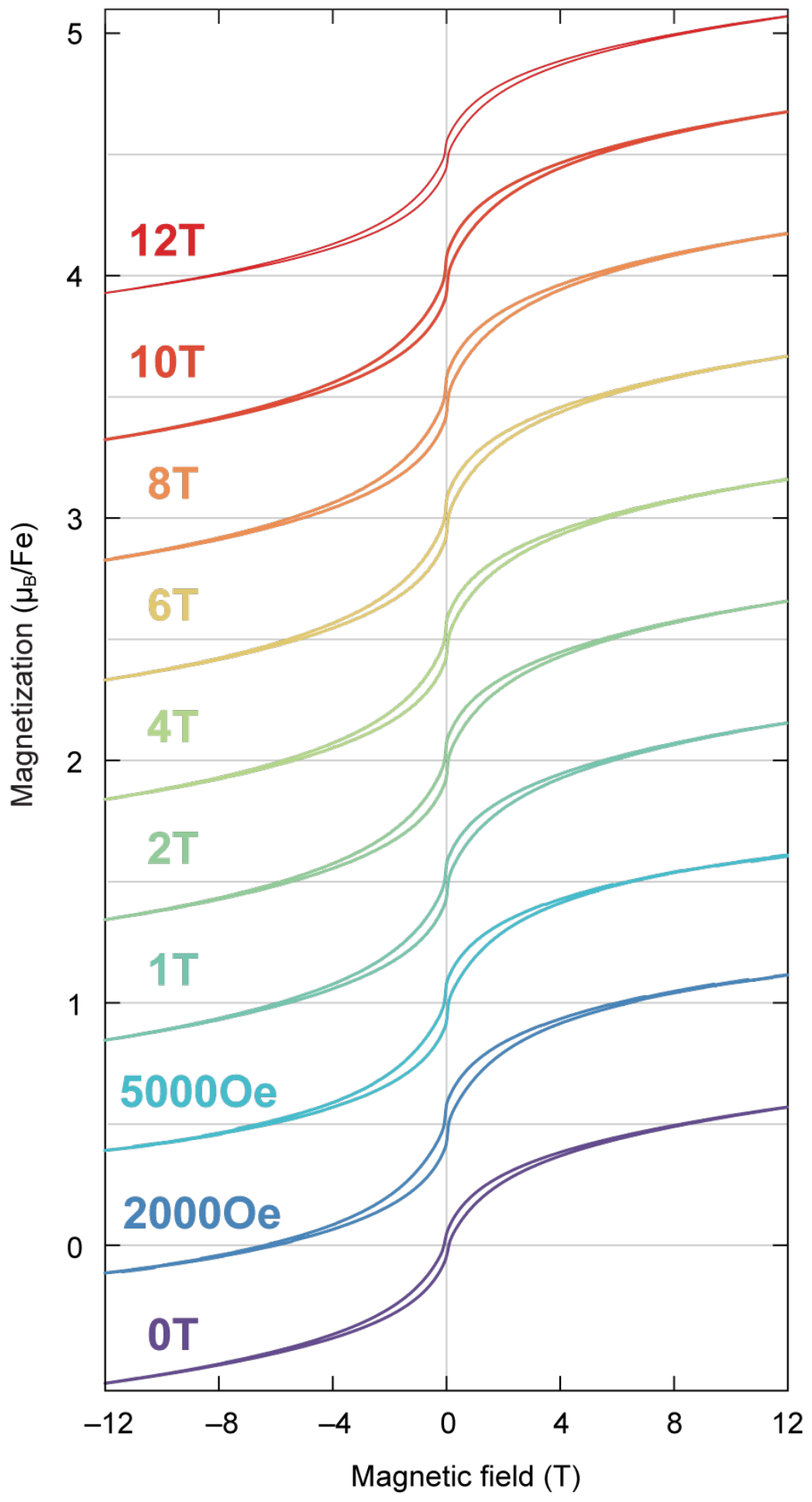

**Figure S18.** Magnetization versus magnetic field measurements at 2K with sample cooled down in different cooling fields. Each loop is offset on the y-axis by 0.5 $\mu_B$/Fe The magnetic field was applied along the c-axis of the samples.

**S19. Steric effect on the occupation of Fe in vdWs gap**

To gain insights into the impact of steric effects on complex stability, we compare the sizes of octahedra and tetrahedra in iron selenide compounds and Fe-intercalated TMDs. We examined the reported crystal structures for $FeSe_x$ compounds and $Fe_xMSe_2$ ($M$ = transition metal) and extracted the Se–Se distances of octahedra (**Table S5**) and tetrahedra (**Table S6**). Compositions without published crystal structures are marked with a “–”. The extracted Se–Se distances of octahedra are plotted in **Figure S19b** and those of tetrahedra are presented in **Figure S19c**.

This simple analysis shows that the sizes of the $FeSe_6$ octahedra in $Fe_xMSe_2$ compounds are within the range of those found in $FeSe_x$ compounds. However, we find that the sizes of the $FeSe_4$ tetrahedra in $Fe_xMSe_2$ compounds are consistently smaller than those found in FeSe, but $Fe_xZrSe_2$, possesses the $FeSe_4$ tetrahedron that is closest to that found in FeSe. This comparison hints to the reason why Zr-based TMDs are uniquely capable of accommodating a significant fraction of Fe intercalated in tetrahedral sites. Further theoretical investigations are required to gain a deeper understanding of the influence of steric effects on the preferred intercalation sites of Fe.

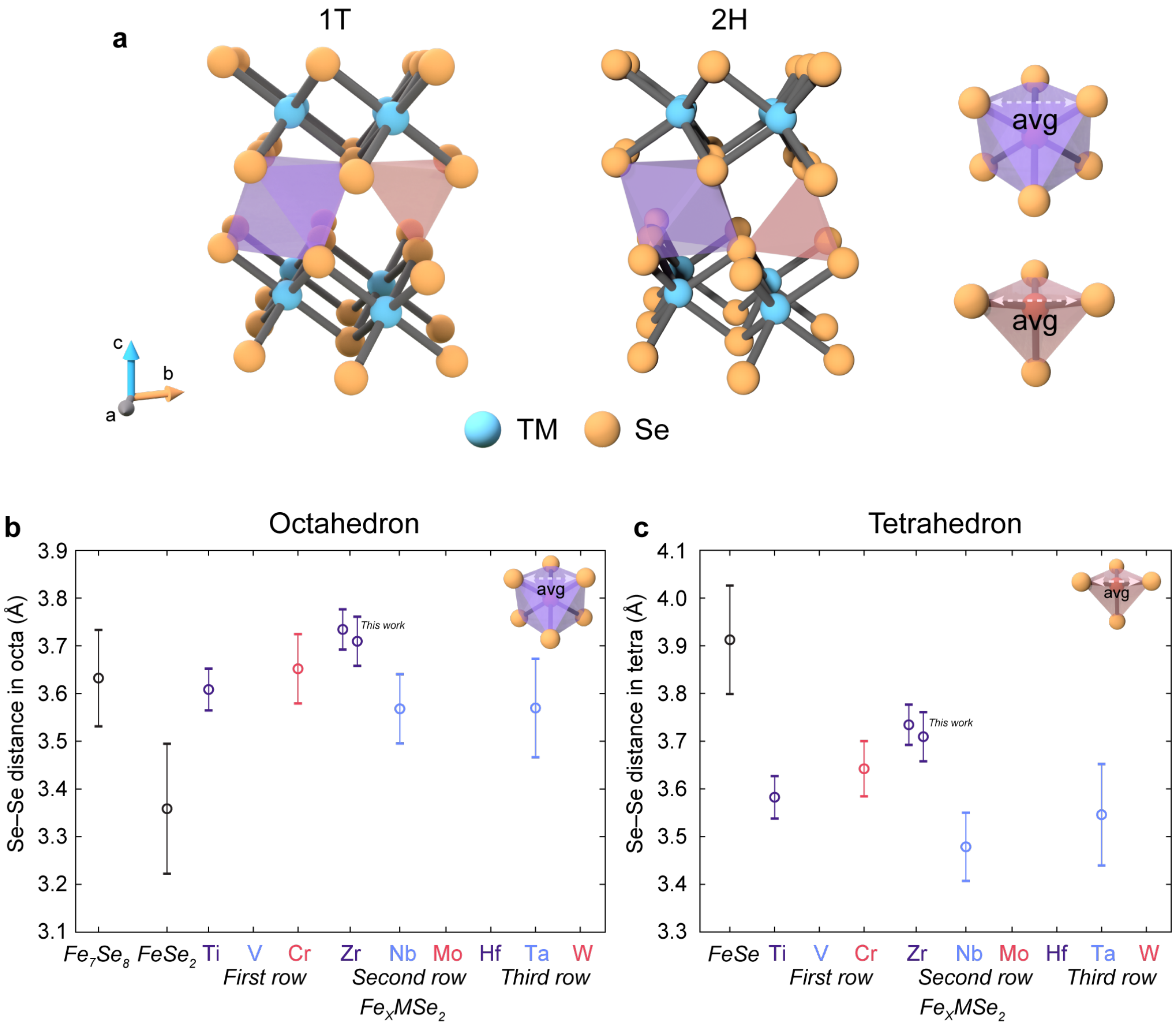

**Figure S19.** The comparison of Se−Se distances for $FeSe_4$ tetrahedra and $FeSe_6$ octahedra in $FeSe_x$ compounds versus in Fe intercalated TMDs.

**Table S5. Se−Se distance for $FeSe_6$ octahedra**

| | Average Se−Se distance (Å) | Standard deviation Se−Se distance (Å) | ref |
|---|---|---|---|
| $FeSe_2$ | 3.36 | 0.14 | 23-28 |
| $Fe_7Se_8$ | 3.63 | 0.10 | 29 |
| | | | |
| $Fe_xTiSe_2$ | 3.61 | 0.04 | 30-33 |
| $Fe_xVSe_2$ | – | – | |
| $Fe_xCrSe_2$ | 3.65 | 0.07 | 34-36 |
| $Fe_xZrSe_2$ | 3.73 | 0.04 | 37-38 |
| $Fe_xNbSe_2$ | 3.57 | 0.07 | 39 |
| $Fe_xMoSe_2$ | – | – | |
| $Fe_xHfSe_2$ | – | – | |
| $Fe_xTaSe_2$ | 3.57 | 0.10 | 40 |
| $Fe_xWSe_2$ | – | – | |
| $Fe_xZrSe_2$ (This work) | 3.71 | 0.05 | |

**Table S6. Se−Se distances in $FeSe_4$ tetrahedra**

| | Average Se−Se distance (Å) | Standard deviation Se−Se distance (Å) | ref |
|---|---|---|---|
| FeSe | 3.91 | 0.11 | 41-48 |
| | | | |
| $Fe_xTiSe_2$ | 3.58 | 0.04 | 30-33 |
| $Fe_xVSe_2$ | × | × | |
| $Fe_xCrSe_2$ | 3.64 | 0.06 | 34-36 |
| $Fe_xZrSe_2$ | 3.73 | 0.04 | 37-38 |
| $Fe_xNbSe_2$ | 3.48 | 0.07 | 39 |
| $Fe_xMoSe_2$ | × | × | |
| $Fe_xHfSe_2$ | × | × | |
| $Fe_xTaSe_2$ | 3.55 | 0.11 | 40 |
| $Fe_xWSe_2$ | × | × | |
| $Fe_xZrSe_2$ (This work) | 3.71 | 0.05 | |

**S20. The contextualization of representative exchange bias systems**

**Table S7. Representative exchange bias systems**

| Structure | Materials | Magnetic phase | largest $H_{EB}$ | Temp (largest $H_{EB}$) | ref |
|---|---|---|---|---|---|
| Intrinsic exchange bias | $YbFe_2O_4$ | AFM/FM two magnetic sublattices | 19 kOe | 5 K | 49 |
| | Aurivillius oxides $Bi_{10}Fe_{6+y}Ti_{3-y}O_{30+\delta}$ | coupling between AFM and spin glass | 38 Oe | 300 K | 50 |
| | $Fe_xNbS_2$ (x = 0.30, 0.35) | coupling between AFM and spin glass | 30 kOe | 1.8 K | 51 |
| | $Mn_3(C_6S_6)$ | Geometric frustrated spin glass | 1625 Oe | 2 K | 52 |
| | $Fe_{0.17}ZrSe_2$ | Structural disorder induced spin glass | 166 Oe | 2 K | This work |
| heterostructure | Co/CoO | Ferromagnet/ Ferromagnet | 9500 Oe | 4-10 K | 53 |
| | Co/CoN | | 3200 Oe | 4-10 K | |
| | Ni/NiO | Ferromagnet/ antiferromagnet | 400 Oe | 4-10 K | |
| | $Fe/Fe_3O_4$ | Ferromagnet/ Ferrimagnet/ | 120 Oe | 4-10 K | |
| | $Fe/Fe_2N$ | Ferromagnet/ Ferromagnet | 300 Oe | 4-10 K | |
| | Co/CuMn | Ferromagnet/spin glass | 85 Oe | 2 K | 54 |

**S21. Reference:**

1. Veith, G. M.; Chen, R. J.; Popov, G.; Croft, M.; Shokh, Y.; Nowik, I.; Greenblatt, M. Electronic, magnetic, and magnetoresistance properties of the n=2 Ruddlesden-Popper phases $Sr_3Fe_{2-x}CO_xO_{7-\delta}$ (0.25 <= x <= 1.75). *J. Solid State Chem.* **2002**, *166*, 292–304.
2. Joseph, B.; Iadecola, A.; Simonelli, L.; Mizuguchi, Y.; Takano, Y.; Mizokawa, T.; Saini, N. L. A study of the electronic structure of $FeSe_{1-x}Te_x$ chalcogenides by Fe and Se K-edge x-ray absorption near edge structure measurements. *J Phys-Condens Mat* **2010**, *22*, 485702.
3. Xie, L. S.; Husremovic, S.; Gonzalez, O.; Craig, I. M.; Bediako, D. K. Structure and Magnetism of Iron- and Chromium-Intercalated Niobium and Tantalum Disulfides. *J. Am. Chem. Soc.* **2022**, *144*, 9525–9542.
4. Mydosh, J. A. *Spin glasses: an experimental introduction*; 1st ed.; Taylor & Francis, London ; Washington, DC, 1993.
5. Ueno, K. Introduction to the Growth of Bulk Single Crystals of Two-Dimensional Transition-Metal Dichalcogenides. *J. Phys. Soc. Jpn.* **2015**, *84*, 121015.
6. Husremović, S.; Groschner, C. K.; Inzani, K.; Craig, I. M.; Bustillo, K. C.; Ercius, P.; Kazmierczak, N. P.; Syndikus, J.; Van Winkle, M.; Aloni, S.; Taniguchi, T.; Watanabe, K.; Griffin, S. M.; Bediako, D. K. Hard Ferromagnetism Down to the Thinnest Limit of Iron-Intercalated Tantalum Disulfide. *J. Am. Chem. Soc.* **2022**, *144*, 12167–12176.
7. Sheldrick, G. M. A short history of SHELX. *Acta Cryst. A* **2008**, *64*, 112–122.
8. Sheldrick, G. M., SHELXS-2014: Program for the Solution of Crystal Structures. University Of Göttingen: 2014.
9. Sheldrick, G. M. Crystal structure refinement with SHELXL. *Acta Cryst. C* **2015**, *71*, 3–8.
10. Sheldrick, G. M., SHELXL-2014: Crystallographic Software Package. Bruker AXS, Inc.: Madison, WI: 2014.
11. Dolomanov, O. V. a. B., Luc J. and Gildea, Richard J. and Howard, Judith A. K. and Puschmann, Horst OLEX2: a complete structure solution, refinement and analysis program. *J. Appl. Crystallogr.* **2009**, *42*, 339–341.
12. Momma, K.; Izumi, F. VESTA 3 for three-dimensional visualization of crystal, volumetric and morphology data. *J. Appl. Crystallogr.* **2011**, *44*, 1272–1276.
13. Wong, J.; Lytle, F. W.; Messmer, R. P.; Maylotte, D. H. K-Edge Absorption-Spectra of Selected Vanadium Compounds. *Phys. Rev. B* **1984**, *30*, 5596–5610.
14. Zhu, J.; Zeng, Z. H.; Li, W. X. K-Edge XANES Investigation of Fe-Based Oxides by Density Functional Theory Calculations. *J. Phys. Chem. C* **2021**, *125*, 26229–26239.
15. Kim, M. G.; Cho, H. S.; Yo, C. H. Fe K-edge X-ray absorption (XANES/EXAFS) spectroscopic study of the nonstoichiometric $SrFe_{1-x}Sn_xO_{3-y}$ system. *J. Phys. Chem. Solids* **1998**, *59*, 1369–1381.
16. Brauer, H. E.; Starnberg, H. I.; Holleboom, L. J.; Hughes, H. P. The Electronic-Structure of $ZrSe_2$ and $Cs_xZrSe_2$ Studied by Angle-Resolved Photoelectron-Spectroscopy. *J Phys-Condens Mat* **1995**, *7*, 7741–7760.
17. Kubelka, P. M., Franz A contribution to the optics of pigments. *Z. Tech. Phys.* **1931**, *12*, 593–599.
18. Tauc, J. Optical properties and electronic structure of amorphous Ge and Si. *Mater. Res. Bull.* **1968**, *3*, 37–46.
19. Lopez, R.; Gomez, R. Band-gap energy estimation from diffuse reflectance measurements on sol-gel and commercial $TiO_2$: a comparative study. *J. Sol-Gel Sci. Technol.* **2012**, *61*, 1–7.
20. Tong, P.; Sun, Y. P.; Zhu, X. B.; Song, W. H. Strong electron-electron correlation in the antiperovskite compound $GaCNi_3$. *Phys. Rev. B* **2006**, *73*, 245106.
21. Goetsch, R. J.; Anand, V. K.; Pandey, A.; Johnston, D. C. Structural, thermal, magnetic, and electronic transport properties of the $LaNi_2(Ge_{1-x}P_x)_2$ system. *Phys. Rev. B* **2012**, *85*, 054517.
22. Buhannic, M. A.; Danot, M.; Colombet, P.; Dordor, P.; Fillion, G. Thermopower and Low-DC-Field Magnetization Study of the Layered $Fe_xZrSe_2$ Compounds - Anderson-Type Localization and Anisotropic Spin-Glass Behavior. *Phys. Rev. B* **1986**, *34*, 4790–4795.

23. Tengner, S. Über Diselenide und Ditelluride von Eisen, Kobalt und Nickel. *Z. Anorg. Allg. Chem.* **1938**, *239*, 126–132.
24. Schuster, W.; Mikler, H.; Komarek, K. L. Transition metal-chalcogen systems, VII.: The iron-selenium phase diagram. *Monatshefte für Chemie / Chemical Monthly* **1979**, *110*, 1153–1170.
25. Pickardt, J.; Reuter, B.; Riedel, E.; Söchtig, J. On the formation of $FeSe_2$ single crystals by chemical transport reactions. *J. Solid State Chem.* **1975**, *15*, 366–368.
26. Buryanova, E. Z. K., A.I. A new mineral—Ferroselite. *Doklady Akademii Nauk SSSR* **1955**, *105*, 812–813.
27. Arne Kjekshus, T. R. Coumpounds with the Marcasite Type Crystal Structure. XI. High Temperature Studies of Chalcogenides. *Acta Cryst. A* **1975**, *29*, 443–452.
28. Andresen, A. K. T. R. A. F. Compounds with the Marcasite Type Crystal Structure. IX. Structural Data for $FeAs_2$, $FeSe_2$, $NiAs_2$, $NiSb_2$, and $CuSe_2$. *Acta Chem. Scand.* **1974**, *28*, 996–1000.
29. Parise, J. B.; Nakano, A.; Tokonami, M.; Morimoto, N. Structure of iron selenide 3*C*-$Fe_7Se_8$. *Acta Cryst. B* **1979**, *35*, 1210–1212.
30. Shkvarina, E. G.; Titov, A. A.; Shkvarin, A. S.; Postnikov, M. S.; Radzivonchik, D. I.; Plaisier, J. R.; Gigli, L.; Gaboardi, M.; Titov, A. N. Thermal disorder in the $Fe_{0.5}TiSe_2$. *J. Alloys Compd.* **2020**, *819*, 153016.
31. Lyding, J. W.; Ratajack, M. T.; Kannewurf, C. R.; Goodman, W. H.; Ibers, J. A.; Marsh, R. E. Structure, Electrical Transport, and Optical-Properties of a New Ordered Iron Intercalated Dichalcogenide, $Fe_{0.34}TiSe_2$. *J. Phys. Chem. Solids* **1982**, *43*, 599–607.
32. Huntley, D. R.; Sienko, M. J.; Hiebl, K. Magnetic-Properties of Iron-Intercalated Titanium Diselenide. *J. Solid State Chem.* **1984**, *52*, 233–243.
33. Calvarin, G.; Gavarri, J. R.; Buhannic, M. A.; Colombet, P.; Danot, M. Crystal and Magnetic-Structures of $Fe_{0.25}TiSe_2$ and $Fe_{0.48}TiSe_2$. *Revue De Physique Appliquee* **1987**, *22*, 1131–1138.
34. Riedel, E.; Al-Juani, A.; Rackwitz, R.; Söchtig, H. Spinelle mit substituierten Nichtmetallteilgittern. VIII. Röntgenographische und elektrische Eigenschaften, Mößbauer- und IR-Spektren des Systems $FeCr_2(S_{1-x}Se_x)_4$. *Z. Anorg. Allg. Chem.* **1981**, *480*, 49–59.
35. Kim, J. H. K. S. J. K. B. W. L. C. S. Neutron and Mössbauer studies of $FeCr_2Se_4$. *J. Appl. Phys.* *99*, 08F714.
36. Morris, B. L.; Russo, P.; Wold, A. Magnetic properties of $ACr_2Se_4$ (A = Fe, Co, Ni) and $NiCr_2S_4$. *J. Phys. Chem. Solids* **1970**, *31*, 635–638.
37. Buhannic, M.-A.; Ahouandjinou, A.; Danot, M.; Rouxel, J. Double coordinence du fer dans la phase $Fe_xZrSe_2$ (0 < x < 0,25): propriétés magnétiques et caractéristiques Mössbauer. *J. Solid State Chem.* **1983**, *49*, 77–84.
38. Ahouandjinou, A. T., Luc; Rouxel, Jean Chimie minérale. Les phases $Fe_xZrSe_2$ (0< x< 0.25), $Co_xZrSe_2$ (0< x< 0.33) et $Ni_xZrSe_2$ (0< x< 0.50). *Comptes Rendus des Seances de l'Academie des Sciences, Serie C: Sciences Chimiques* **1976**, 727–730.
39. Erodici, M. P.; Mai, T. T.; Xie, L. S.; Li, S.; Fender, S. S.; Husremović, S.; Gonzalez, O.; Hight Walker, A. R.; Bediako, D. K. Bridging Structure, Magnetism, and Disorder in Iron-Intercalated Niobium Diselenide, $Fe_xNbSe_2$, below x = 0.25. *J. Phys. Chem. C* **2023**.
40. Mrotzek, B. H. A. Synthese, Struktur und Eigenschaften ternaerer Tantalselenide. *Zeitschrift für Kristallographie. Supplement issue* **1995**, 169.
41. Pomjakushina, E.; Conder, K.; Pomjakushin, V.; Bendele, M.; Khasanov, R. Synthesis, crystal structure, and chemical stability of the superconductor $FeSe_{1-x}$. *Phys. Rev. B* **2009**, *80*, 024517.
42. Li, Z. F.; Ju, J.; Tang, J.; Sato, K.; Watahiki, M.; Tanigaki, K. Structural and superconductivity study on α-$FeSe_x$. *J. Phys. Chem. Solids* **2010**, *71*, 495–498.
43. Horigane, K.; Hiraka, H.; Ohoyama, K. Relationship between Structure and Superconductivity in $FeSe_{1-x}Te_x$. *J. Phys. Soc. Jpn.* **2009**, *78*, 074718.
44. Millican, J. N.; Phelan, D.; Thomas, E. L.; Leao, J. B.; Carpenter, E. Pressure-induced effects on the structure of the FeSe superconductor. *Solid State Commun.* **2009**, *149*, 707–710.

45. Margadonna, S.; Takabayashi, Y.; Ohishi, Y.; Mizuguchi, Y.; Takano, Y.; Kagayama, T.; Nakagawa, T.; Takata, M.; Prassides, K. Pressure evolution of the low-temperature crystal structure and bonding of the superconductor FeSe ($T_C$=37 K). *Phys. Rev. B* **2009**, *80*, 064506.
46. McQueen, T. M.; Huang, Q.; Ksenofontov, V.; Felser, C.; Xu, Q.; Zandbergen, H.; Hor, Y. S.; Allred, J.; Williams, A. J.; Qu, D.; Checkelsky, J.; Ong, N. P.; Cava, R. J. Extreme sensitivity of superconductivity to stoichiometry in $Fe_{1+\delta}Se$. *Phys. Rev. B* **2009**, *79*, 014522.
47. Margadonna, S.; Takabayashi, Y.; McDonald, M. T.; Kasperkiewicz, K.; Mizuguchi, Y.; Takano, Y.; Fitch, A. N.; Suard, E.; Prassides, K. Crystal structure of the new $FeSe_{1-x}$ superconductor. *Chem. Commun.* **2008**, *43*, 5607–5609.
48. Kumar, R. S.; Zhang, Y.; Sinogeikin, S.; Xiao, Y. M.; Kumar, S.; Chow, P.; Cornelius, A. L.; Chen, C. F. Crystal and Electronic Structure of FeSe at High Pressure and Low Temperature. *J. Phys. Chem. B* **2010**, *114*, 12597–12606.
49. Sun, Y.; Cong, J. Z.; Chai, Y. S.; Yan, L. Q.; Zhao, Y. L.; Wang, S. G.; Ning, W.; Zhang, Y. H. Giant exchange bias in a single-phase magnet with two magnetic sublattices. *Appl. Phys. Lett.* **2013**, *102*, 172406.
50. Wang, G. P.; Chen, Z. Z.; He, H. C.; Meng, D. C.; Yang, H.; Mao, X. Y.; Pan, Q.; Chu, B. J.; Zuo, M.; Sun, Z. H.; Peng, R. R.; Fu, Z. P.; Zhai, X. F.; Lu, Y. L. Room Temperature Exchange Bias in Structure-Modulated Single-Phase Multiferroic Materials. *Chem. Mater.* **2018**, *30*, 6156–6163.
51. Maniv, E.; Murphy, R. A.; Haley, S. C.; Doyle, S.; John, C.; Maniv, A.; Ramakrishna, S. K.; Tang, Y. L.; Ercius, P.; Ramesh, R.; Reyes, A. P.; Long, J. R.; Analytis, J. G. Exchange bias due to coupling between coexisting antiferromagnetic and spin-glass orders. *Nat. Phys.* **2021**, *17*, 525–530.
52. Murphy, R. A.; Darago, L. E.; Ziebel, M. E.; Peterson, E. A.; Zaia, E. W.; Mara, M. W.; Lussier, D.; Velasquez, E. O.; Shuh, D. K.; Urban, J. J.; Neaton, J. B.; Long, J. R. Exchange Bias in a Layered Metal-Organic Topological Spin Glass. *ACS Cent. Sci.* **2021**, *7*, 1317–1326.
53. Nogués, J.; Schuller, I. K. Exchange bias. *J. Magn. Magn. Mater.* **1999**, *192*, 203–232.
54. Ali, M.; Adie, P.; Marrows, C. H.; Greig, D.; Hickey, B. J.; Stamps, R. L. Exchange bias using a spin glass. *Nat. Mater.* **2007**, *6*, 70–75.